\documentclass[12pt]{article}
\usepackage{graphicx}
\usepackage{longtable}
\usepackage{amsfonts}
\usepackage{amssymb}
\usepackage{a4wide}

\newtheorem{theo}{Theorem}

\begin{document}

{\renewcommand{\thefootnote}{\fnsymbol{footnote}}
\hfill  AEI--2005--185, IGPG--06/1--6\\
\medskip
\hfill gr--qc/0601085\\
\medskip
\begin{center}
{\LARGE  Loop Quantum Cosmology}\\
\vspace{1.5em}
Martin Bojowald\footnote{e-mail address: {\tt bojowald@gravity.psu.edu}}
\\
\vspace{0.5em}
Institute for Gravitational Physics and Geometry,\\
The Pennsylvania State
University,\\
104 Davey Lab, University Park, PA 16802, USA\\
and\\
Max-Planck-Institute for Gravitational Physics\\
Albert-Einstein-Institute\\Am M\"uhlenberg 1, 14476 Potsdam, Germany\\
\vspace{1.5em}
\end{center}
}

\setcounter{footnote}{0}

\begin{abstract}
  Quantum gravity is expected to be necessary in order to understand
  situations where classical general relativity breaks down. In
  particular in cosmology one has to deal with initial singularities,
  i.e.\ the fact that the backward evolution of a classical space-time
  inevitably comes to an end after a finite amount of proper time.
  This presents a breakdown of the classical picture and requires an
  extended theory for a meaningful description. Since small length
  scales and high curvatures are involved, quantum effects must play a
  role. Not only the singularity itself but also the surrounding
  space-time is then modified. One particular realization is loop
  quantum cosmology, an application of loop quantum gravity to
  homogeneous systems, which removes classical singularities. Its
  implications can be studied at different levels. Main effects are
  introduced into effective classical equations which allow to avoid
  interpretational problems of quantum theory. They give rise to new
  kinds of early universe phenomenology with applications to inflation
  and cyclic models. To resolve classical singularities and to
  understand the structure of geometry around them, the quantum
  description is necessary. Classical evolution is then replaced by a
  difference equation for a wave function which allows to extend
  space-time beyond classical singularities. One main question is how
  these homogeneous scenarios are related to full loop quantum
  gravity, which can be dealt with at the level of distributional
  symmetric states. Finally, the new structure of space-time arising
  in loop quantum gravity and its application to cosmology sheds new
  light on more general issues such as time.
\end{abstract}

\newpage

\section{Introduction}
\label{section:introduction}

\begin{flushright}
\begin{quote}
 {\em Die Grenzen meiner Sprache bedeuten die Grenzen meiner Welt.}

 ({\em The limits of my language mean the limits of my world.})
\end{quote}

 {\sc Ludwig Wittgenstein}

  Tractatus logico-philosophicus
\end{flushright}

While general relativity is very successful in describing the
gravitational interaction and the structure of space and time on large
scales \cite{Will}, quantum gravity is needed for the small-scale
behavior. This is usually relevant when curvature, or in physical
terms energy densities and tidal forces, becomes large. In cosmology
this is the case close to the big bang, and also in the interior of
black holes. We are thus able to learn about gravity on small scales
by looking at the early history of the universe.

Starting with general relativity on large scales and evolving backward
in time, the universe becomes smaller and smaller and quantum effects
eventually become important. That the classical theory by itself
cannot be sufficient to describe the history in a well-defined way is
illustrated by singularity theorems \cite{SingTheo} which also apply
in this case: After a finite time of backward evolution the classical
universe will collapse into a single point and energy densities
diverge. At this point, the theory breaks down and cannot be used to
determine what is happening there. Quantum gravity, with its different
dynamics on small scales, is expected to solve this problem.

The quantum description does not only present a modified dynamical
behavior on small scales but also a new conceptual setting. Rather
than dealing with a classical space-time manifold, we now have
evolution equations for the wave function of a universe.  This opens a
vast number of problems on various levels from mathematical physics to
cosmological observations, and even philosophy.  This review is
intended to give an overview and summary of the current status of
those problems, in particular in the new framework of loop quantum
cosmology.

\section{The viewpoint of loop quantum cosmology}
\label{section:viewpoint}

Loop quantum cosmology is based on quantum Riemannian geometry, or
loop quantum gravity \cite{Rov:Loops,ALRev,ThomasRev,Rov}, which is an
attempt at a non-perturbative and background independent quantization
of general relativity. This means that no assumptions of small fields
or the presence of a classical background metric are made, both of
which is expected to be essential close to classical singularities
where the gravitational field would diverge and space degenerates. In
contrast to other approaches to quantum cosmology there is a direct
link between cosmological models and the full theory
\cite{PhD,SymmRed}, as we will describe later in Sec.~\ref{s:Link}.
With cosmological applications we are thus able to test several
possible constructions and draw conclusions for open issues in the
full theory. At the same time, of course, we can learn about physical
effects which have to be expected from properties of the quantization
and can potentially lead to observable predictions.  Since the full
theory is not completed yet, however, an important issue in this
context is the robustness of those applications to choices in the full
theory and quantization ambiguities.

The full theory itself is, understandably, extremely complex and thus
requires approximation schemes for direct applications. Loop quantum
cosmology is based on symmetry reduction, in the simplest case to
isotropic geometries \cite{IsoCosmo}. This poses the mathematical
problem as to how the quantum representation of a model and its
composite operators can be derived from that of the full theory, and
in which sense this can be regarded as an approximation with suitable
correction terms. Research in this direction currently proceeds by
studying symmetric models with less symmetries and the relations
between them. This allows to see what role anisotropies and
inhomogeneities play in the full theory.

While this work is still in progress, one can obtain full
quantizations of models by using basic features as they can already be
derived from the full theory together with constructions of more
complicated operators in a way analogous to what one does in the full
theory (see Sec.~\ref{s:Analog}). For those complicated operators, the
prime example being the Hamiltonian constraint which dictates the
dynamics of the theory, the link between model and the full theory is
not always clear-cut.  Nevertheless, one can try different versions in
the model in explicit ways and see what implications this has, so
again the robustness issue arises. This has already been applied to
issues such as the semiclassical limit and general properties of
quantum dynamics. Thus, general ideas which are required for this new,
background independent quantization scheme, can be tried in a rather
simple context in explicit ways to see how those constructions work in
practice.

At the same time, there are possible phenomenological consequences in
the physical systems being studied, which is the subject of
Sec.~\ref{s:Effective}. In fact, it turned out, rather surprisingly,
that already very basic effects such as the discreteness of quantum
geometry and other features briefly reviewed in Sec.~\ref{s:LQG}, for
which a reliable derivation from the full theory is available, have
very specific implications in early universe cosmology. While
quantitative aspects depend on quantization ambiguities, there is a
rich source of qualitative effects which work together in a
well-defined and viable picture of the early universe.  In such a way,
as illustrated later, a partial view of the full theory and its
properties emerges also from a physical, not just mathematical
perspective.

With this wide range of problems being investigated we can keep our
eyes open to input from all sides. There are mathematical consistency
conditions in the full theory, some of which are identically satisfied
in the simplest models (such as the isotropic model which has only one
Hamiltonian constraint and thus a trivial constraint algebra). They
are being studied in different, more complicated models and also in the
full theory directly. Since the conditions are not easy to satisfy,
they put stringent bounds on possible ambiguities. From physical
applications, on the other hand, we obtain conceptual and
phenomenological constraints which can be complementary to those
obtained from consistency checks. All this contributes to a test and
better understanding of the background independent framework and its
implications.

Other reviews of loop quantum cosmology at different levels can be
found in
\cite{WS:MB,MexicoVI,Inverted,ICGC,LoopCosRev,IUFM,QuantGeoCos}. For
complementary applications of loop quantum gravity to cosmology see
\cite{Kodama,Kodama2,KodInfl,KodLin,QCInfl,ACosConst}.

\section{Loop quantum gravity}
\label{s:LQG}

Since many reviews of full loop quantum gravity
\cite{Rov:Loops,ThomasRev,ALRev,Rov,NPZRev} as well as shorter
accounts
\cite{AA:Advances,AA:Action,RovelliDialog,SmolinInvite,AP:LoopSpin,Beyond}
are already available, we describe here only those properties which
will be essential later on. Nevertheless, this review is mostly
self-contained; our notation is closest to that in \cite{ALRev}. A
recent bibliography can be found at \cite{LQGBib}.

\subsection{Geometry}

General relativity in its canonical formulation \cite{ADM} describes
the geometry of space-time in terms of fields on spatial slices.
Geometry on such a spatial slice $\Sigma$ is encoded in the spatial
metric $q_{ab}$, which presents the configuration variables. Canonical
momenta are given in terms of extrinsic curvature $K_{ab}$ which is
the derivative of the spatial metric under changing the spatial slice.
Those fields are not arbitrary since they are obtained from a solution
of Einstein's equations by choosing a time coordinate defining the
spatial slices, and space-time geometry is generally covariant. In the
canonical formalism this is expressed by the presence of constraints
on the fields, the diffeomorphism constraint and the Hamiltonian
constraint. The diffeomorphism constraint generates deformations of a
spatial slice or coordinate changes, and when it is satisfied spatial
geometry does not depend on which coordinates we choose on space.
General covariance of space-time geometry also for the time coordinate
is then completed by imposing the Hamiltonian constraint. This
constraint, furthermore, is important for the dynamics of the theory:
since there is no absolute time, there is no Hamiltonian generating
evolution, but only the Hamiltonian constraint. When it is satisfied,
it encodes correlations between the physical fields of gravity and
matter such that evolution in this framework is relational. The
reproduction of a space-time metric in a coordinate dependent way then
requires to choose a gauge and to compute the transformation in gauge
parameters (including the coordinates) generated by the constraints.

It is often useful to describe spatial geometry not by the spatial
metric but by a triad $e^a_i$ which defines three vector fields which
are orthogonal to each other and normalized in each point. This yields
all information about spatial geometry, and indeed the inverse metric
is obtained from the triad by $q^{ab}=e^a_ie^b_i$ where we sum over
the index $i$ counting the triad vector fields. There are differences,
however, between metric and triad formulations. First, the set of
triad vectors can be rotated without changing the metric, which
implies an additional gauge freedom with group SO(3) acting on the
index $i$. Invariance of the theory under those rotations is then
guaranteed by a Gauss constraint in addition to the diffeomorphism and
Hamiltonian constraints.

The second difference will turn out to be more important later on: we
can not only rotate the triad vectors but also reflect them, i.e.\ 
change the orientation of the triad given by $\mathrm{sgn}\det e^a_i$.
This does not change the metric either, and so could be included in
the gauge group as O(3). However, reflections are not connected to the
unit element of O(3) and thus are not generated by a constraint. It
then has to be seen whether or not the theory allows to impose
invariance under reflections, i.e.\ if its solutions are reflection
symmetric. This is not usually an issue in the classical theory since
positive and negative orientations on the space of triads are
separated by degenerate configurations where the determinant of the
metric vanishes. Points on the boundary are usually singularities
where the classical evolution breaks down such that we will never
connect between both sides. However, since there are expectations that
quantum gravity may resolve classical singularities, which indeed are
confirmed in loop quantum cosmology, we will have to keep this issue
in mind and not restrict to only one orientation from the outset.

\subsection{Ashtekar variables}

To quantize a constrained canonical theory one can use Dirac's
prescription \cite{DirQuant} and first represent the classical Poisson
algebra of a suitable complete set of basic variables on phase space
as an operator algebra on a Hilbert space, called
kinematical. This ignores the constraints, which can be written as
operators on the same Hilbert space. At the quantum level the
constraints are then solved by determining their kernel, to be equipped
with an inner product so as to define the physical Hilbert space. If
zero is in the discrete part of the spectrum of a constraint, as e.g.\ 
for the Gauss constraint when the structure group is compact, the
kernel is a subspace of the kinematical Hilbert space to which the
kinematical inner product can be restricted. If, on the other hand,
zero lies in the continuous part of the spectrum, there are no
normalizable eigenstates and one has to construct a new physical
Hilbert space from distributions. This is the case for the
diffeomorphism and Hamiltonian constraints.

To perform the first step we need a Hilbert space of functionals
$\psi[q]$ of spatial metrics. Unfortunately, the space of metrics, or
alternatively extrinsic curvature tensors, is mathematically poorly
understood and not much is known about suitable inner products. At
this point, a new set of variables introduced by Ashtekar
\cite{AshVarLett,AshVar,AshVarReell} becomes essential. This is a
triad formulation, but uses the triad in a densitized form (i.e.\ it
is multiplied with an additional factor of a Jacobian under coordinate
transformations). The densitized triad $E^a_i$ is then related to the
triad by $E^a_i=\left|\det e^b_j\right|^{-1} e^a_i$ but has the same
properties concerning gauge rotations and its orientation (note the
absolute value which is often omitted). The densitized triad is
conjugate to extrinsic curvature coefficients $K_a^i:=K_{ab}e^b_i$:
\begin{equation}
 \{K_a^i(x),E^b_j(y)\}= 8\pi G\delta^b_a\delta^i_j\delta(x,y)
\end{equation}
with the gravitational constant $G$. Extrinsic curvature is then
replaced by the Ashtekar connection
\begin{equation}
 A_a^i=\Gamma_a^i+\gamma K_a^i
\end{equation}
with a positive value for $\gamma$, the Barbero--Immirzi parameter
\cite{AshVarReell,Immirzi}. Classically, this number can be changed
by a canonical transformation of the fields, but it will play a more
important and fundamental role upon quantization. The Ashtekar
connection is defined in such a way that it is conjugate to the triad,
\begin{equation} \label{Symp}
 \{A_a^i(x),E^b_j(y)\}=8\pi\gamma G\delta_a^b\delta_j^i\delta(x,y)
\end{equation}
and obtains its transformation properties as a connection from the
spin connection
\begin{equation} \label{GammaGen}
 \Gamma_a^i = -\epsilon^{ijk}e^b_j (\partial_{[a}e_{b]}^k+
 {\textstyle\frac{1}{2}} e_k^ce_a^l\partial_{[c}e_{b]}^l)\,.
\end{equation}

Spatial geometry is then obtained directly from the densitized triad,
which is related to the spatial metric by
\[
 E^a_iE^b_i=q^{ab}\det q\,.
\]
There is more freedom in a triad since it can be rotated without
changing the metric. The theory is independent of such rotations
provided the Gauss constraint
\begin{equation}
 G[\Lambda]=\frac{1}{8\pi\gamma G}\int_{\Sigma}\mathrm{d}^3x \Lambda^iD_aE^a_i=
 \frac{1}{8\pi\gamma G} \int_{\Sigma}\mathrm{d}^3x\Lambda^i(\partial_aE^a_i+
 \epsilon_{ijk}A^j_aE^a_k)\approx 0
\end{equation}
is satisfied. Independence from any spatial coordinate system or
background is implemented by the diffeomorphism constraint (modulo
Gauss constraint)
\begin{equation}
 D[N^a]= \frac{1}{8\pi\gamma G}\int_{\Sigma}\mathrm{d}^3x
 N^aF_{ab}^iE_i^b \approx 0
\end{equation}
with the curvature $F_{ab}^i$ of the Ashtekar connection. In this
setting, one can then discuss spatial geometry and its quantization.

Space-time geometry, however, is more complicated to deduce since it
requires a good knowledge of the dynamics. In a canonical setting,
dynamics is implemented by the Hamiltonian constraint
\begin{equation}
 H[N]=\frac{1}{16\pi\gamma G} \int_{\Sigma} \mathrm{d}^3x N\left|\det
   E\right|^{-1/2}
 \left(\epsilon_{ijk}F_{ab}^iE^a_jE^b_k -2(1+\gamma^2)
 K^i_{[a}K^j_{b]}E^a_iE^b_j\right)\approx 0
\end{equation}
where extrinsic curvature components have to be understood as
functions of the Ashtekar connection and the densitized triad through
the spin connection.

\subsection{Representation}

The key new aspect is now that we can choose the space of Ashtekar
connections as our configuration space whose structure is much better
understood than that of a space of metrics. Moreover, the formulation
lends itself easily to a background independent quantization. To see
this we need to remember that quantizing field theories requires one
to smear fields, i.e.\ to integrate them over regions in order to
obtain a well-defined algebra without $\delta$-functions as in
(\ref{Symp}). Usually this is done by integrating both configuration
and momentum variables over three-dimensional regions, which requires
an integration measure. This is no problem in ordinary
field theories which are formulated on a background such as Minkowski
or a curved space. However, doing this here for gravity in terms of
Ashtekar variables would immediately spoil any possible background
independence since a background would already occur at this very basic
step.

There is now a different smearing available which does not require a
background metric. Instead of using three-dimensional regions we
integrate the connection along one-dimensional curves $e$ and
exponentiate in a path-ordered manner, resulting in holonomies
\begin{equation}
 h_e(A)={\cal P}\exp\int_e\tau_i A_a^i\dot{e}^a\mathrm{d}t
\end{equation}
with tangent vector $\dot{e}^a$ to the curve $e$ and
$\tau_j=-\frac{1}{2}i\sigma_j$ in terms of Pauli matrices.  The path
ordered exponentiation needs to be done in order to obtain a covariant
object from the non-Abelian connection. The prevalence of holonomies
or, in their most simple gauge invariant form as Wilson loops
${\mathrm{tr}}h_e(A)$ for closed $e$, is the origin of loop quantum
gravity and its name \cite{LoopRep}. Similarly, densitized vector
fields can naturally be integrated over 2-dimensional surfaces,
resulting in fluxes
\begin{equation} \label{Flux}
 F_S(E)=\int_S \tau^i E^a_in_a\mathrm{d}^2y 
\end{equation}
with the co-normal $n_a$ to the surface.

The Poisson algebra of holonomies and fluxes is now well-defined and
one can look for representations on a Hilbert space. We also require
diffeomorphism invariance, i.e.\ there must be a unitary action of the
diffeomorphism group on the representation by moving edges and
surfaces in space. This is required since the diffeomorphism
constraint has to be imposed later. Under this condition, there is
even a unique representation which defines the kinematical Hilbert
space \cite{FluxAlg,Meas,HolFluxRep,SuperSel,WeylRep,LOST}.

We can construct the Hilbert space in the representation where states
are functionals of connections. This can easily be done by using
holonomies as ``creation operators'' starting with a ``ground state''
which does not depend on connections at all.  Multiplying with
holonomies then generates states which do depend on connections but
only along the edges used in the process. These edges can be collected
in a graph appearing as a label of the state. An independent set of
states is given by spin network states \cite{RS:SpinNet} associated
with graphs whose edges are labeled by irreducible representations of
the gauge group SU(2) in which to evaluate the edge holonomies, and
whose vertices are labeled by matrices specifying how holonomies
leaving or entering the vertex are multiplied together. The inner
product on this state space is such that these states, with an
appropriate definition of independent contraction matrices in
vertices, are orthonormal.

Spatial geometry can be obtained from fluxes representing the
densitized triad. Since these are now momenta, they are represented by
derivative operators with respect to values of connections on the flux
surface. States as constructed above depend on the connection only
along edges of graphs such that the flux operator is non-zero only if
there are intersection points between its surface and the graph in the
state it acts on \cite{Holvar}. Moreover, the contribution from each
intersection point can be seen to be analogous to an angular momentum
operator in quantum mechanics which has a discrete spectrum
\cite{Area}.  Thus, when acting on a given state we obtain a finite
sum of discrete contributions and thus a discrete spectrum of flux
operators.  The spectrum depends on the value of the Barbero--Immirzi
parameter, which can accordingly be fixed using implications of the
spectrum such as black hole entropy which gives a value of the order
of but smaller than one \cite{ABCK:LoopEntro,IHEntro,Gamma,Gamma2}.
Moreover, since angular momentum operators do not commute, flux
operators do not commute in general \cite{NonCommFlux}. There is thus
no triad representation which is another reason why using a metric
formulation and trying to build its quantization with functionals on a
metric space is difficult.

There are important basic properties of this representation which we
will use later on. First, as already noted, flux operators have
discrete spectra and, secondly, holonomies of connections are
well-defined operators. It is, however, not possible to obtain
operators for connection components or their integrations directly but
only in the exponentiated form. These are direct consequences of the
background independent quantization and translate to particular
properties of more complicated operators.

\subsection{Function spaces}
\label{s:FuncSp}

A connection 1-form $A_a^i$ can be reconstructed uniquely if all its
holonomies are known \cite{Giles}. It is thus sufficient to
parameterize the configuration space by matrix elements of $h_e$ for
all edges in space. This defines an algebra of functions on the
infinite dimensional space of connections ${\cal A}$, which are
multiplied as ${\mathbb C}$-valued functions. Moreover, there is a
duality operation by complex conjugation, and if the structure group
$G$ is compact a supremum norm exists since matrix elements of
holonomies are then bounded. Thus, matrix elements form an Abelian
$C^*$-algebra with unit as a subalgebra of all continuous functions on
${\cal A}$.

Any Abelian $C^*$-algebra with unit can be represented as the algebra
of {\em all} continuous functions on a compact space $\bar{\cal A}$.
The intuitive idea is that the original space ${\cal A}$, which has
many more continuous functions, is enlarged by adding new points to
it.  This increases the number of continuity conditions and thus
shrinks the set of continuous functions. This is done until only
matrix elements of holonomies survive when continuity is imposed, and
it follows from general results that the enlarged space must be
compact for an Abelian unital $C^*$-algebra.  We thus obtain a
compactification $\bar{\cal A}$, the space of generalized connections
\cite{ALMMT}, which densely contains the space ${\cal A}$.

There is a natural diffeomorphism invariant measure
$\mathrm{d}\mu_{\mathrm{AL}}$ on $\bar{\cal A}$, the
Ashtekar--Lewandowski measure \cite{FuncInt}, which defines the
Hilbert space ${\cal H}=L^2(\bar{{\cal
    A}},\mathrm{d}\mu_{\mathrm{AL}})$ of square integrable functions
on the space of generalized connections. A dense subset $\mathrm{Cyl}$
of functions is given by cylindrical functions
$f(h_{e_1},\ldots,h_{e_n})$ which depend on the connection through a
finite but arbitrary number of holonomies. They are associated with
graphs $g$ formed by the edges $e_1$, \ldots, $e_n$. For functions
cylindrical with respect to two identical graphs the inner product can
be written as
\begin{equation}
 \langle f|g\rangle= \int_{\bar{\cal A}}\mathrm{d}\mu_{AL}(A) f(A)^*g(A) = 
 \int_{\mathrm{SU}(2)^n} \prod_{i=1}^n\mathrm{d}\mu_{\mathrm{H}}(h_i) 
f(h_1,\ldots,h_n)^*g(h_1,\ldots,h_n)
\end{equation}
with the Haar measure $\mathrm{d}\mu_{\mathrm{H}}$ on $G$. The
importance of generalized connections can be seen from the fact that
the space ${\cal A}$ of smooth connections is a subset of measure zero
in $\bar{\cal A}$ \cite{MM}.

With the dense subset $\mathrm{Cyl}$ of ${\cal H}$ we obtain the
Gel'fand triple
\begin{equation}
 \mathrm{Cyl}\subset {\cal H}\subset \mathrm{Cyl}^*
\end{equation}
with the dual $\mathrm{Cyl}^*$ of linear functionals from
$\mathrm{Cyl}$ to the set of complex numbers. Elements of
$\mathrm{Cyl}^*$ are distributions, and there is no inner product on
the full space. However, one can define inner products on certain
subspaces defined by the physical context. Often, those subspaces
appear when constraints with continuous spectra are solved following
the Dirac procedure. Other examples include the definition of
semiclassical or, as we will use in Sec.~\ref{s:Link}, symmetric states.

\subsection{Composite operators}

From the basic operators we can construct more complicated ones which,
with growing degree of complexity, will be more and more ambiguous for
instance from factor ordering choices. Quite simple expressions exist
for the area and volume operator \cite{AreaVol,Area,Vol2} which are
constructed solely from fluxes. Thus, they are less ambiguous since no
factor ordering issues with holonomies arise. This is true because the
area of a surface and volume of a region can be written classically
as functionals of the densitized triad alone,
$A_S=\int_S\sqrt{E^a_in_a E^b_in_b}\mathrm{d}^2y$ and
$V_R=\int_R\sqrt{\left|\det E^a_i\right|}\mathrm{d}^3x$. At the
quantum level, this implies that, just as fluxes, also area and volume
have discrete spectra showing that spatial quantum geometry is
discrete. (For discrete approaches to quantum gravity in general see
\cite{DiscQG}.) All area eigenvalues are known explicitly, but this is
not possible even in principle for the volume operator. Nevertheless,
some closed formulas and numerical techniques exist
\cite{Loll:Simply,RecTh,Recoup,VolNum}.

The length of a curve, on the other hand, requires the co-triad which
is an inverse of the densitized triad and is more problematic. Since
fluxes have discrete spectra containing zero, they do not have densely
defined inverse operators. As we will describe below, it is
possible to quantize those expressions but requires one to use
holonomies. Thus, here we encounter more ambiguities from factor
ordering. Still, one can show that also length operators have discrete
spectra \cite{Len}.

Inverse densitized triad components also arise when we try to
quantize matter Hamiltonians such as
\begin{equation} \label{Hphi}
 H_{\phi}=\int\mathrm{d}^3x\left( \frac{1}{2}
   \frac{p_{\phi}^2+E^a_iE^b_i\partial_a\phi
     \partial_b\phi}{\sqrt{\left|\det E^c_j\right|}}+\sqrt{\left|\det
     E^c_j\right|}V(\phi)\right)
\end{equation}
for a scalar field $\phi$ with momentum $p_{\phi}$ and potential
$V(\phi)$ (not to be confused with volume). The inverse determinant
again cannot be quantized directly by using, e.g., an inverse of the
volume operator which does not exist. This seems, at first, to be a
severe problem not unlike the situation in quantum field theory on a
background where matter Hamiltonians are divergent.  Yet, it turns out
that quantum geometry allows one to quantize these expressions in a
well-defined manner \cite{QSDV}.

To do this we notice that the Poisson bracket of the volume with
connection components,
\begin{equation} \label{ident}
 \{A_a^i,\int\sqrt{\left|\det E\right|}\mathrm{d}^3x\}= 2\pi\gamma G
 \epsilon^{ijk}\epsilon_{abc} \frac{E^b_jE^c_k}{\sqrt{\left|\det
   E\right|}}
\end{equation}
amounts to an inverse of densitized triad components and does allow a
well-defined quantization: we can express the connection component
through holonomies, use the volume operator and turn the Poisson
bracket into a commutator. Since all operators involved have a dense
intersection of their domains of definition, the resulting operator is
densely defined and amounts to a quantization of inverse powers of the
densitized triad.

This also shows that connection components or holonomies are required
in this process, and thus ambiguities can arise even if initially one
starts with an expression such as $\sqrt{\left|\det E\right|}^{-1}$
which only depends on the triad. There are also many different ways to
rewrite expressions as above, which all are equivalent classically but
result in different quantizations. In classical regimes this would not
be relevant, but can have sizeable effects at small scales. In fact,
this particular aspect, which as a general mechanism is a direct
consequence of the background independent quantization with its
discrete fluxes, implies characteristic modifications of the classical
expressions on small scales. We will discuss this and more detailed
examples in the cosmological context in Sec.~\ref{s:Effective}.

\subsection{Hamiltonian constraint}
\label{s:Ham}

Similarly to matter Hamiltonians one can also quantize the Hamiltonian
constraint in a well-defined manner \cite{QSDI}. Again, this requires
to rewrite triad components and to make other regularization
choices. Thus, there is not just one quantization but a class of
different possibilities.

It is more direct to quantize the first part of the constraint
containing only the Ashtekar curvature. (This part agrees with the
constraint in Euclidean signature and Barbero--Immirzi parameter
$\gamma=1$, and so is sometimes called Euclidean part of the
constraint.) Triad components and their inverse determinant are again
expressed as a Poisson bracket using the identity (\ref{ident}), and
curvature components are obtained through a holonomy around a small
loop $\alpha$ of coordinate size $\Delta$ and with tangent vectors
$s_1^a$ and $s_2^a$ at its base point \cite{RS:Ham}:
\begin{equation}
 s_1^as_2^bF_{ab}^i\tau_i= \Delta^{-1}(h_{\alpha}-1) +O(\Delta)\,.
\end{equation}
Putting this together, an expression for the Euclidean part
$H^{\mathrm{E}}[N]$ can then be constructed in the schematic form
\begin{equation} \label{FullH}
 H^{\mathrm{E}}[N]\propto \sum_vN(v)\epsilon^{IJK}\mathrm{tr}\left(
   h_{\alpha_{IJ}}h_{s_K}\{h_{s_K}^{-1},V\}\right)+O(\Delta)
\end{equation}
where one sums over all vertices of a triangulation of space whose
tetrahedra are used to define closed curves $\alpha_{IJ}$ and
transversal edges $s_K$.

An important property of this construction is that coordinate
functions such as $\Delta$ disappear from the leading term, such that
the coordinate size of the discretization is irrelevant. Nevertheless,
there are several choices to be made, such as how a discretization is
chosen in relation to a graph the constructed operator is supposed to
act on, which in later steps will have to be constrained by studying
properties of the quantization. Of particular interest is the holonomy
$h_{\alpha}$ since it creates new edges to a graph, or at least new
spin on existing ones. Its precise behavior is expected to have a
strong influence on the resulting dynamics \cite{S:ClassLim}. In
addition, there are factor ordering choices, i.e.\ whether triad
components appear to the right or left of curvature components. It
turns out that the expression above leads to a well-defined operator
only in the first case, which in particular requires an operator
non-symmetric in the kinematical inner product. Nevertheless, one can
always take that operator and add its adjoint (which in this full
setting does not simply amount to reversing the order of the curvature
and triad expressions) to obtain a symmetric version, such that the
choice still exists. Another choice is the representation chosen to
take the trace, which for the construction is not required to be the
fundamental one \cite{Gaul}.

The second part of the constraint is more complicated since one has to
use the function $\Gamma(E)$ in $K_a^i$. As also developed in
\cite{QSDI}, extrinsic curvature can be obtained through the already
constructed Euclidean part via $K\sim\{H^{\mathrm{E}},V\}$. The
result, however, is rather complicated, and in models one often uses a
more direct way exploiting the fact that $\Gamma$ has a more special
form.  In this way, additional commutators in the general construction
can be avoided, which usually does not have strong effects. Sometimes,
however, these additional commutators can be relevant, which can
always be decided by a direct comparison of different constructions
(see, e.g., \cite{ScalarLorentz}).

\subsection{Open issues}

For an anomaly-free quantization the constraint operators have to
satisfy an algebra mimicking the classical one. There are arguments
that this is the case for the quantization as described above when
each loop $\alpha$ contains exactly one vertex of a given graph
\cite{AnoFree}, but the issue is still open. Moreover, the operators
are quite complicated and it is not easy to see if they have the
correct expectation values in appropriately defined semiclassical
states.

Even if one regards the quantization and semiclassical issues as
satisfactory, one has to face several hurdles in evaluating the
theory. There are interpretational issues of the wave function
obtained as a solution to the constraints, and also the problem of
time or observables emerges \cite{KucharTime}. There is a wild mixture
of conceptual and technical problems at different levels, not the
least because the operators are quite complicated. For instance, as
seen in the rewriting procedure above, the volume operator plays an
important role even if one is not necessarily interested in the volume
of regions.  Since this operator is complicated, without an explicitly
known spectrum, it translates to complicated matrix elements of the
constraints and matter Hamiltonians. Loop quantum gravity should thus
be considered as a framework rather than a uniquely defined theory,
which however has important rigid aspects. This includes the basic
representation of the holonomy-flux algebra and its general
consequences.

All this should not come as a surprise since even classical gravity, at
this level of generality, is complicated enough. Most solutions and
results in general relativity are obtained with approximations or
assumptions, one of the most widely used being symmetry reduction. In
fact, this allows access to the most interesting gravitational
phenomena such as cosmological expansion, black holes and
gravitational waves. Similarly, symmetry reduction is expected to
simplify many problems of full quantum gravity by resulting in simpler
operators and by isolating conceptual problems such that not all of
them need to be considered at once.

\section{Loop cosmology}
\label{s:Effective}

\begin{flushright}
\begin{quote}
  {\em Je abstrakter die Wahrheit ist, die du lehren willst, um so mehr
  mu\ss t du noch die Sinne zu ihr verf\"uhren.}
  
  ({\em The more abstract the truth you want to teach is, the more you have
  to seduce to it the senses.})
\end{quote}

{\sc Friedrich Nietzsche}

Beyond Good and Evil
\end{flushright}

The gravitational field equations, for instance in the case of
cosmology where one can assume homogeneity and isotropy, involve
components of curvature as well as the inverse metric. (Computational
methods to derive information from these equations are described in
\cite{ComputUni}.) Since singularities occur, these components will
become large in certain regimes, but the equations have been tested
only in small curvature regimes. On small length scales such as close
to the big bang, modifications to the classical equations are not
ruled out by observations and can be expected from candidates of
quantum gravity.  Quantum cosmology describes the evolution of a
universe by a constraint equation for a wave function, but some
effects can be included already at the level of effective classical
equations. In loop quantum gravity, the main modification happens
through inverse metric components which, e.g., appear in the kinematic
term of matter Hamiltonians. This one modification is mainly
responsible for all the diverse effects of loop cosmology.

\subsection{Isotropy}

Isotropy reduces the phase space of general relativity to be
2-dimensional since, up to SU(2)-gauge freedom, there is only one
independent component in an isotropic connection and triad,
respectively, which is not already determined by the symmetry. This is
analogous to metric variables, where the scale factor $a$ is the only
free component in the spatial part of an isotropic metric
\begin{equation} \label{FRW}
 \mathrm{d} s^2 = -N(t)^2\mathrm{d} t^2+a(t)^2((1-kr^2)^{-1}\mathrm{d}
 r^2+r^2\mathrm{d}\Omega^2)\,.
\end{equation}
The lapse function $N(t)$ does not play a dynamical role and
correspondingly does not appear in the Friedmann equation
\begin{equation} \label{IsoConstr}
 \left(\frac{\dot{a}}{a}\right)^2+\frac{k}{a^2}= 
 \frac{8\pi G}{3}a^{-3}H_{\mathrm{matter}}(a)
\end{equation}
with the matter Hamiltonian $H_{\mathrm{matter}}$ and the gravitational
constant $G$, and the parameter $k$ taking the discrete values
zero or $\pm1$ depending on the symmetry group or intrinsic spatial
curvature.

\subsubsection{Canonical formulation}

Indeed, $N(t)$ can simply be absorbed into the time coordinate by
defining proper time $\tau$ through $\mathrm{d}\tau=N(t)\mathrm{d} t$.
This is not possible for the scale factor since it depends on time but
multiplies space differentials in the line element. The scale factor
can only be rescaled by an arbitrary constant, which can be normalized
at least in the closed model where $k=1$.

One can understand these different roles of metric components also
from a Hamiltonian analysis of the Einstein--Hilbert action 
\[
 S_{\mathrm{EH}}=\frac{1}{16\pi G}\int\mathrm{d} t\mathrm{d}^3x
\sqrt{-\det g} R[g]
\]
specialized to isotropic metrics (\ref{FRW}) whose Ricci scalar is
\[
 R=6\left(\frac{\ddot{a}}{N^2a}+\frac{\dot{a}^2}{N^2a^2}+
   \frac{k}{a^2}-\frac{\dot{a}}{a}\frac{\dot{N}}{N^3}\right)\,.
\]
The action then becomes
\[
 S=\frac{V_0}{16\pi G}\int\mathrm{d} t Na^3R= \frac{3V_0}{8\pi
   G}\int\mathrm{d} t  N\left(-\frac{a\dot{a}^2}{N^2}+ka\right)
\]
(with the spatial coordinate volume $V_0=\int_{\Sigma}\mathrm{d}^3x$)
after integrating by parts, from which one derives the momenta
\[
 p_a= \frac{\partial L}{\partial\dot{a}}= -\frac{3V_0}{4\pi
   G}\frac{a\dot{a}}{N} \quad,\quad p_N=\frac{\partial
   L}{\partial\dot{N}}=0
\]
illustrating the different roles of $a$ and $N$. Since $p_N$ must
vanish, $N$ is not a degree of freedom but a Lagrange multiplier. It
appears in the canonical action $S=(16\pi G)^{-1}\int\mathrm{d}
t(\dot{a}p_a-NH))$ only as a factor of
\[
 H=-\frac{2\pi G}{3}\frac{p_a^2}{V_0a}-\frac{3}{8\pi G} V_0ak
\]
such that variation with respect to $N$ forces $H$, the Hamiltonian
constraint, to be zero. In the presence of matter, $H$ also contains
the matter Hamiltonian, and its vanishing is equivalent to the
Friedmann equation.

\subsubsection{Connection variables}

Isotropic connections and triads, as discussed in App.~\ref{s:Iso},
are analogously described by single components $\tilde{c}$ and
$\tilde{p}$, respectively, related to the scale factor by
\begin{equation} \label{paiso}
 |\tilde{p}|=\tilde{a}^2=\frac{a^2}{4}
\end{equation}
for the densitized triad component $\tilde{p}$ and
\begin{equation}
 \tilde{c}=\tilde{\Gamma}+\gamma\dot{\tilde{a}}= \frac{1}{2}(k+\gamma\dot{a})
\end{equation}
for the connection component $\tilde{c}$. Both components are canonically
conjugate:
\begin{equation}
 \{\tilde{c},\tilde{p}\}=\frac{8\pi\gamma G}{3}V_0\,.
\end{equation}

It is convenient to absorb factors of $V_0$ into the basic variables,
which is also suggested by the integrations in holonomies and fluxes
on which background independent quantizations are built \cite{Bohr}.
We thus define
\begin{equation}
 p=V_0^{2/3}\tilde{p}\quad, \quad c=V_0^{1/3}\tilde{c}
\end{equation}
together with $\Gamma=V_0^{1/3}\tilde{\Gamma}$. The symplectic
structure is then independent of $V_0$ and so are integrated densities
such as total Hamiltonians. For the Hamiltonian constraint in
isotropic Ashtekar variables we have
\begin{equation}
 H=-\frac{3}{8\pi
   G}(\gamma^{-2}(c-\Gamma)^2+\Gamma^2)\sqrt{|p|}+H_{\mathrm{matter}}(p)=0
\end{equation}
which is exactly the Friedmann equation. (In most earlier papers on
loop quantum cosmology some factors in the basic variables and
classical equations are incorrect due, in part, to the existence of
different and often confusing notations in the loop quantum gravity
literature.\footnote{The author is grateful to Ghanashyam Date and
  Golam Hossain for discussions and correspondence on this issue.})

The part of phase space where we have $p=0$ and thus $a=0$ plays a special
role since this is where isotropic classical singularities are
located. On this subset the evolution equation (\ref{IsoConstr}) with
standard matter choices is singular in the sense that
$H_{\mathrm{matter}}$, e.g.\ 
\begin{equation} \label{HphiIso}
 H_{\phi}(a,\phi,p_{\phi})= \frac{1}{2}|p|^{-3/2}p_{\phi}^2+|p|^{3/2} V(\phi)
\end{equation}
for a scalar $\phi$ with momentum $p_{\phi}$ and potential $V(\phi)$,
diverges and the differential equation does not pose a well-defined
initial value problem there. Thus, once such a point is reached the
further evolution is no longer determined by the theory. Since,
according to singularity theorems \cite{SingTheo,InflSing}, any
classical trajectory must intersect the subset $a=0$ for the matter we
need in our universe, the classical theory is incomplete.

This situation, certainly, is not changed by introducing triad
variables instead of metric variables. However, the situation is
already different since $p=0$ is a submanifold in the classical phase
space of triad variables where $p$ can have both signs (the sign
determining whether the triad is left or right handed, i.e.\ the
orientation). This is in contrast to metric variables where $a=0$ is a
boundary of the classical phase space. There are no implications in
the classical theory since trajectories end there nonetheless, but it
will have important ramifications in the quantum theory (see
Sec.~\ref{s:Dyn}).

\subsubsection{Implications of a loop quantization}

We are now dealing with a simple system with finitely many degrees of
freedom, subject to a constraint. It is well known how to quantize
such a system from quantum mechanics, which has been applied to
cosmology starting with DeWitt \cite{DeWitt}. Here, one chooses a
metric representation for wave functions, i.e.\ $\psi(a)$, on which
the scale factor acts as multiplication operator and its conjugate
$p_a$, related to $\dot{a}$, as a derivative operator. These basic
operators are then used to form the Wheeler--DeWitt operator
quantizing the constraint (\ref{IsoConstr}) once a factor ordering is
chosen.

This prescription is rooted in quantum mechanics which, despite its
formal similarity, is physically very different from cosmology. The
procedure looks innocent, but one should realize that there are
already basic choices involved. Choosing the factor ordering is
harmless, even though results can depend on it \cite{Konto}. More
importantly, one has chosen the Schr\"odinger representation of
the classical Poisson algebra which immediately implies the familiar
properties of operators such as the scale factor with a continuous
spectrum. There are inequivalent representations with different
properties, and it is not clear that this representation which
works well in quantum mechanics is also correct for quantum
cosmology. In fact, quantum mechanics is not very sensitive to the
representation chosen \cite{PolymerParticle} and one can use the most
convenient one.
This is the case because energies and thus oscillation lengths of wave
functions described usually by quantum mechanics span only a limited
range. Results can then be reproduced to arbitrary accuracy in any
representation.  Quantum cosmology, in contrast, has to deal with
potentially infinitely high matter energies, leading to small
oscillation lengths of wave functions, such that the issue of quantum
representations becomes essential.

That the Wheeler--DeWitt representation may not be the right choice is
also indicated by the fact that its scale factor operator has a
continuous spectrum, while quantum geometry which is a well-defined
quantization of the full theory, implies discrete volume spectra.
Indeed, the Wheeler--DeWitt quantization of full gravity exists only
formally, and its application to quantum cosmology simply quantizes
the classically reduced isotropic system. This is much easier, and
also more ambiguous, and leaves open many consistency
considerations. It would be more reliable to start with the full
quantization and introduce the symmetries there, or at least follow
the same constructions of the full theory in a reduced model. If this
is done, it turns out that indeed we obtain a quantum representation
inequivalent to the Wheeler--DeWitt representation, with strong
implications in high energy regimes. In particular, just as the full
theory such a quantization has a volume or $p$ operator with a
discrete spectrum, as derived in Sec.~\ref{s:IsoRep}.

\subsubsection{Effective densities and equations}

The isotropic model is thus quantized in such a way that the operator
$\hat{p}$ has a discrete spectrum containing zero. This immediately
leads to a problem since we need a quantization of $|p|^{-3/2}$ in order
to quantize a matter Hamiltonian such as (\ref{HphiIso}) where not
only the matter fields but also geometry are quantized. However, an
operator with zero in the discrete part of its spectrum does not have
a densely defined inverse and does not allow a direct
quantization of $|p|^{-3/2}$.

This leads us to the first main effect of the loop quantization: It
turns out that despite the non-existence of an inverse operator of
$\hat{p}$ one can quantize the classical $|p|^{-3/2}$ to a well-defined
operator. This is not just possible in the model but also in the full
theory where it even has been defined first \cite{QSDV}. Classically,
one can always write expressions in many equivalent ways, which
usually result in different quantizations. In the case of
$|p|^{-3/2}$, as discussed in Sec.~\ref{s:IsoMatter},
there is a general class of ways to rewrite it in a quantizable manner
\cite{InvScale} which differ in details but all have the same
important properties. This can be parameterized by a function
$d(p)_{j,l}$ \cite{Ambig,ICGC} which replaces the classical $|p|^{-3/2}$
and strongly deviates from it for small $p$ while being very close at
large $p$.  The parameters $j\in \frac{1}{2}{\mathbb N}$ and $0<l<1$
specify quantization ambiguities resulting from different ways of
rewriting.  With the function
\begin{eqnarray}
 p_l(q) &=&\frac{3}{2l}q^{1-l}\left( \frac{1}{l+2}
\left((q+1)^{l+2}-|q-1|^{l+2}\right)\right.\\
 && - \left.\frac{1}{l+1}q
\left((q+1)^{l+1}-\mathrm{sgn}(q-1)|q-1|^{l+1}\right)\right)\nonumber
\end{eqnarray}
we have
\begin{equation} \label{deff}
  d(p)_{j,l}:= |p|^{-3/2} p_l(3|p|/\gamma j\ell_{\mathrm{P}}^2)^{3/(2-2l)}
\end{equation}
which indeed fulfills $d(p)_{j,l}\sim |p|^{-3/2}$ for $|p|\gg
p_*:=\frac{1}{3}j\gamma\ell_{\mathrm{P}}^2$, but is finite with a peak
around $p_*$ and approaches zero at $p=0$ in a manner
\begin{equation} \label{deffsmalla}
 d(p)_{j,l}\sim 3^{3(3-l)/(2-2l)} (l+1)^{-3/(2-2l)} (\gamma
 j)^{-3(2-l)/(2-2l)} \ell_{\mathrm{P}}^{-3(2-l)/(1-l)} |p|^{3/(2-2l)}
\end{equation}
as it follows from $p_l(q)\sim 3q^{2-l}/(1+l)$. Some examples
displaying characteristic properties are shown in Fig.~\ref{Dens} in
Sec.~\ref{s:IsoMatter}.

The matter Hamiltonian obtained in this manner will thus behave
differently at small $p$. At those scales also other quantum effects
such as fluctuations can be important, but it is possible to isolate
the effect implied by the modified density (\ref{deff}). We just need
to choose a rather large value for the ambiguity parameter $j$ such
that modifications become noticeable already in semiclassical regimes.
This is mainly a technical tool to study the behavior of equations, but
can also be used to find constraints on the allowed values of
ambiguity parameters.

We can thus use classical equations of motion, which are corrected for
quantum effects by using the effective matter Hamiltonian
\begin{equation}
 H_{\phi}^{({\mathrm{eff}})}(p,\phi,p_{\phi}):=
\frac{1}{2}d(p)_{j,l}p_{\phi}^2+ |p|^{3/2}
 V(\phi)
\end{equation}
(see Sec.~\ref{s:IsoSemiClass} for details on effective equations).
This matter Hamiltonian changes the classical constraint such that now
\begin{equation} \label{ConsClass}
 H=-\frac{3}{8\pi
   G}(\gamma^{-2}(c-\Gamma)^2+\Gamma^2)\sqrt{|p|}+
 H_{\phi}^{(\mathrm{eff})}(p,\phi,p_{\phi}) =0\,.
\end{equation}
Since the constraint determines all equations of motion, they also
change: we obtain the effective Friedmann equation from $H=0$,
\begin{equation} \label{effFried}
 \left(\frac{\dot{a}}{a}\right)^2+\frac{k}{a^2}=\frac{8\pi
   G}{3}\left(\frac{1}{2}|p|^{-3/2}d(p)_{j,l} p_{\phi}^2+ V(\phi)\right)
\end{equation}
and the effective Raychaudhuri equation from $\dot{c}=\{c,H\}$,
\begin{eqnarray} \label{effRay}
  \frac{\ddot{a}}{a} &=& -\frac{4\pi G}{3|p|^{3/2}}
\left(H_{\mathrm{matter}}(p,\phi,p_{\phi})- 2p\frac{\partial 
H_{\mathrm{matter}}(p,\phi,p_{\phi})}{\partial p}\right)\\
 &=& -\frac{8\pi G}{3}\left( |p|^{-3/2}d(p)_{j,l}^{-1}\dot{\phi}^2 
\left(1-{\textstyle\frac{1}{4}}a\frac{\mathrm{d}
\log(|p|^{3/2}d(p)_{j,l})}{\mathrm{d} a}\right) -V(\phi)\right)\,.
\end{eqnarray}

Matter equations of motion follow similarly as
\begin{eqnarray*}
 \dot{\phi} &=& \{\phi,H\}= d(p)_{j,l}p_{\phi}\\
 \dot{p}_{\phi} &=& \{p_{\phi},H\}= -|p|^{3/2} V'(\phi)
\end{eqnarray*}
which can be combined to the effective Klein--Gordon equation
\begin{equation} \label{effKG}
  \ddot{\phi}=\dot{\phi}\,\dot{a}\frac{\mathrm{d}\log d(p)_{j,l}}{\mathrm{d}
    a}-|p|^{3/2}d(p)_{j,l}V'(\phi)\,.
\end{equation}
Further discussion for different forms of matter can be found in
\cite{Metamorph}.

\subsubsection{Properties and intuitive meaning}
\label{s:Intuitive}

As a consequence of the function $d(p)_{j,l}$ the effective equations
have different qualitative behavior at small versus large scales
$p$. In the effective Friedmann equation (\ref{effFried}) this is most
easily seen by comparing it with a mechanics problem with a standard
Hamiltonian, or energy, of the form
\[
 E=\frac{1}{2}\dot{a}^2 -\frac{2\pi G}{3V_0} a^{-1} d(p)_{j,l}
 p_{\phi}^2-\frac{4\pi G}{3}a^2 V(\phi)=0
\]
restricted to be zero. If we assume a constant scalar potential
$V(\phi)$, there is no $\phi$-dependence and the scalar equations of
motion show that $p_{\phi}$ is constant. Thus, the potential for the
motion of $a$ is essentially determined by the function $d(p)_{j,l}$.

In the classical case, $d(p)=|p|^{-3/2}$ and the potential is negative
and increasing, with a divergence at $p=0$. The scale factor $a$ is
thus driven toward $a=0$ which it will always reach in finite time
where the system breaks down. With the effective density $d(p)_{j,l}$,
however, the potential is bounded from below, and is decreasing from
zero for $a=0$ to the minimum around $p_*$. Thus, the scale factor is
now slowed down before it reaches $a=0$, which depending on the matter
content could avoid the classical singularity altogether.

The behavior of matter is also different as shown by the effective
Klein--Gordon equation (\ref{effKG}). Most importantly, the derivative
in the $\dot{\phi}$-term changes sign at small $a$ since the effective
density is increasing there. Thus, the qualitative behavior of all the
equations changes at small scales, which as we will see gives rise to
many characteristic effects.  Nevertheless, for the analysis of the
equations as well as conceptual considerations it is interesting that
solutions at small and large scales are connected by a duality
transformation \cite{JimDual}, which even exists between effective
solutions for loop cosmology and braneworld cosmology
\cite{DualLoopBrane}.

We have seen that the equations of motion following from an effective
Hamiltonian are expected to display qualitatively different behavior
at small scales. Before discussing specific models in detail, it is
helpful to observe what physical meaning the resulting modifications
have.

Classical gravity is always attractive, which implies that there is
nothing to prevent collapse in black holes or the whole universe. In
the Friedmann equation this is expressed by the fact that the
potential as used before is always decreasing toward $a=0$ where it
diverges. With the effective density, on the other hand, we have seen
that the decrease stops and instead the potential starts to increase
at a certain scale before it reaches zero at $a=0$. This means that at
small scales, where quantum gravity becomes important, the
gravitational attraction turns into repulsion. In contrast to
classical gravity, thus, quantum gravity has a repulsive component
which can potentially prevent collapse. So far this has only been
demonstrated in homogeneous models, but it relies on a general
mechanism which is also present in the full theory.

Not only the attractive nature of gravity changes at small scales, but
also the behavior of matter in a gravitational background.
Classically, matter fields in an expanding universe are slowed down by
a friction term in the Klein--Gordon equation (\ref{effKG}) where
$\dot{a}\mathrm{d}\log a^{-3}/\mathrm{d} a=-3\dot{a}/a$ is negative.
Conversely, in a contracting universe matter fields are excited and
even diverge when the classical singularity is reached. This behavior
turns around at small scales where the derivative $\mathrm{d}\log
d(a)_{j,l}/\mathrm{d} a$ becomes positive. Friction in an expanding
universe then turns into antifriction such that matter fields are
driven away from their potential minima before classical behavior sets
in. In a contracting universe, on the other hand, matter fields are
not excited by antifriction but freeze once the universe becomes small
enough.

These effects do not only have implications for the avoidance of
singularities at $a=0$ but also for the behavior at small but non-zero
scales. Gravitational repulsion can not only prevent collapse of a
contracting universe \cite{BounceClosed} but also, in an expanding
universe, enhance its expansion. The universe then accelerates in an
inflationary manner from quantum gravity effects alone
\cite{Inflation}. Similarly, the modified behavior of matter fields
has implications for inflationary models \cite{Closed}.

\subsubsection{Applications}

There is now one characteristic modification in the matter
Hamiltonian, coming directly from a loop quantization. Its
implications can be interpreted as repulsive behavior on small scales
and the exchange of friction and antifriction for matter, and it leads
to many further consequences.

\begin{figure}[h]
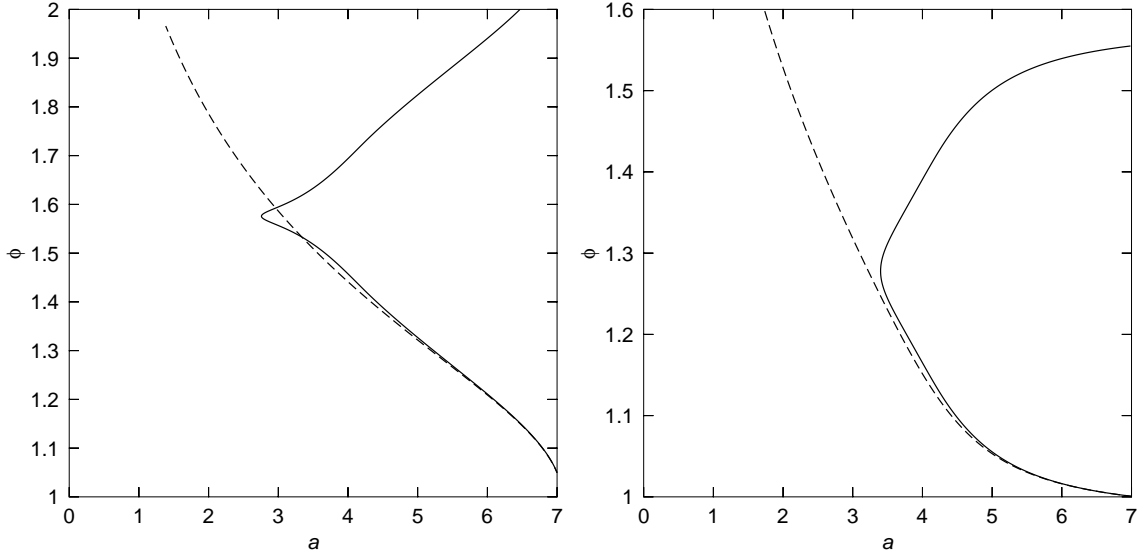

  \centerline{\includegraphics[width=7.5cm,keepaspectratio]{bouncecurv.eps}
 \includegraphics[width=7.5cm,keepaspectratio]{bounceneg.eps}}
  \caption{\it Examples for bouncing solutions with positive curvature 
    (left) or a negative potential (right, negative
    cosmological constant). The solid lines show solutions of
    effective equations with a bounce, while the dashed lines show
    classical solutions running into the singularity at $a=0$ where
    $\phi$ diverges.}
  \label{BounceSol}
\end{figure}

\paragraph{Collapsing phase:} When the universe has collapsed to a
sufficiently small size, repulsion becomes noticeable and bouncing
solutions become possible as illustrated in Fig.~\ref{BounceSol}.
Requirements for a bounce are that the conditions $\dot{a}=0$ and
$\ddot{a}>0$ can be fulfilled at the same time, where the first one
can be evaluated with the Friedmann equation, and the second one with
the Raychaudhuri equation. The first condition can only be fulfilled
if there is a negative contribution to the matter energy, which can
come from a positive curvature term $k=1$ or a negative matter
potential $V(\phi)<0$. In those cases, there are classical solutions
with $\dot{a}=0$, but they generically have $\ddot{a}<0$ corresponding
to a recollapse. This can easily be seen in the flat case with a
negative potential where (\ref{effRay}) is strictly negative with
$\mathrm{d}\log a^3d(a)_{j,l}/\mathrm{d} a\approx 0$ at large scales.

The repulsive nature at small scales now implies a second point where
$\dot{a}=0$ from (\ref{effFried}) at smaller $a$ since the matter
energy now decreases also for $a\to0$. Moreover, the modified
Raychaudhuri equation (\ref{effRay}) has an additional positive term
at small scales such that $\ddot{a}>0$ becomes possible.

Matter also behaves differently through the modified Klein--Gordon
equation (\ref{effKG}). Classically, with $\dot{a}<0$ the scalar
experiences antifriction and $\phi$ diverges close to the
classical singularity. With the modification, antifriction turns into
friction at small scales, damping the motion of $\phi$ such that it
remains finite. In the case of a negative potential \cite{Cyclic} this
allows the kinetic term to cancel the potential term in the Friedmann
equation. With a positive potential and positive curvature, on the
other hand, the scalar is frozen and the potential is canceled by the
curvature term. Since the scalar is almost constant, the behavior
around the turning point is similar to a de Sitter bounce
\cite{BounceClosed,BounceQualitative}. Further, more generic
possibilities for bounces arise from other correction terms
\cite{GenericBounce,KasnerBounce}.

\begin{figure}[h]
  \centerline{\includegraphics[width=15.5cm,keepaspectratio]{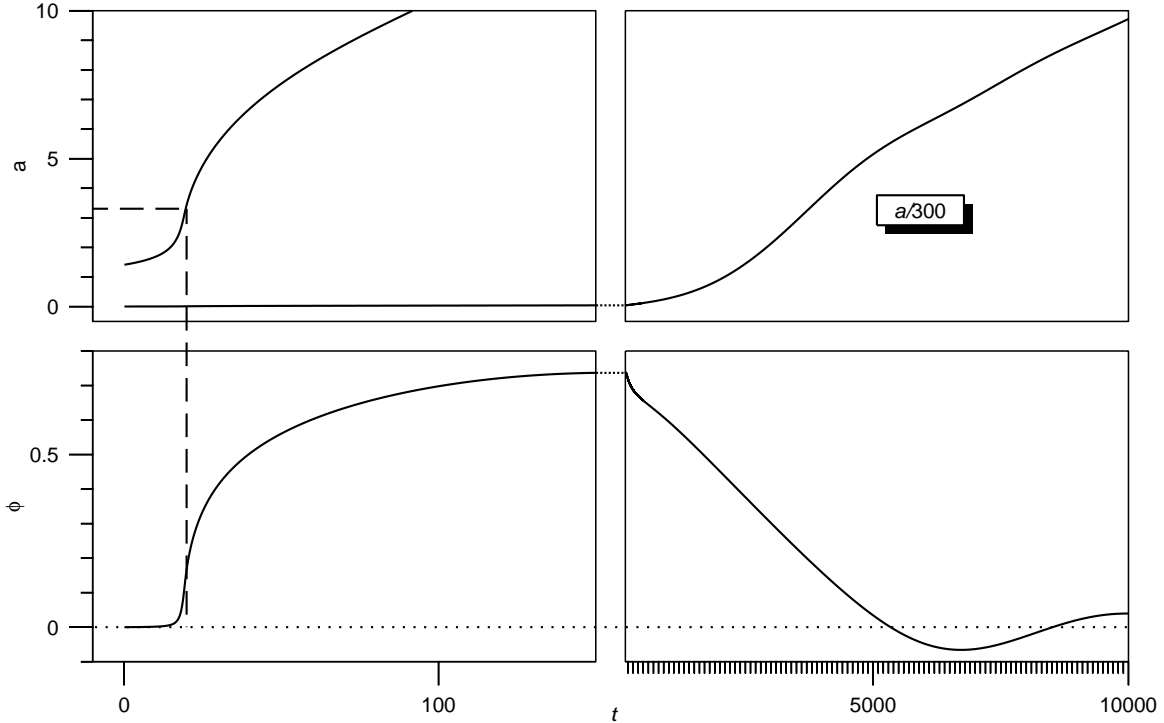}}
  \caption{\it Example for a solution of $a(t)$ and $\phi(t)$ showing early
    loop inflation and later slow-roll inflation driven by a scalar
    which is pushed up its potential by loop effects. The left hand
    side is stretched in time so as to show all details. An idea of
    the duration of different phases can be obtained from
    Fig.~\ref{PushMov}.}
  \label{Push}
\end{figure}

\paragraph{Expansion:} Repulsion can not only prevent collapse but
also accelerates an expanding phase. Indeed, using the behavior
(\ref{deffsmalla}) at small scales in the effective Raychaudhuri
equation (\ref{effRay}) shows that $\ddot{a}$ is generically positive
since the inner bracket is smaller than $-1/2$ for the allowed values
$0<l<1$. Thus, as illustrated by the numerical solution in the upper
left panel of Fig.~\ref{Push}, inflation is realized by quantum
gravity effects for any matter field irrespective of its form,
potential or initial values \cite{Inflation}. The kind of expansion at
early stages is generically super-inflationary, i.e.\ with equation of
state parameter $w<-1$.  For free massless matter fields, $w$ usually
starts very small, depending on the value of $l$, but with a non-zero
potential such as a mass term for matter inflation $w$ is generically
close to exponential: $w_{\mathrm{eff}}\approx -1$. This can be shown
by a simple and elegant argument independently of the precise matter
dynamics \cite{GenericInfl}: The equation of state parameter is
defined as $w=P/\rho$ where $P=-\partial E/\partial V$ is the
pressure, i.e.\ the negative change of energy with respect to volume,
and $\rho=E/V$ energy density. Using the matter Hamiltonian for $E$
and $V=|p|^{3/2}$, we obtain
\[
 P_{\mathrm{eff}}= -{\textstyle \frac{1}{3}}|p|^{-1/2}d'(p)p_{\phi}^2-
 V(\phi)
\]
and thus in the classical case
\[
 w=\frac{\frac{1}{2}|p|^{-3}p_{\phi}^2-V(\phi)}{\frac{1}{2}
   |p|^{-3}p_{\phi}^2+ V(\phi)}
\]
as usually. In the modified case, however, we have
\[
 w_{\mathrm{eff}}= -\frac{\frac{1}{3}|p|^{-1/2}d'(p)p_{\phi}^2+
   V(\phi)}{\frac{1}{2} |p|^{-3/2}d(p)p_{\phi}^2+V(\phi)}\,.
\]

\begin{figure}[h]
 \centerline{\includegraphics[width=15.5cm,keepaspectratio]{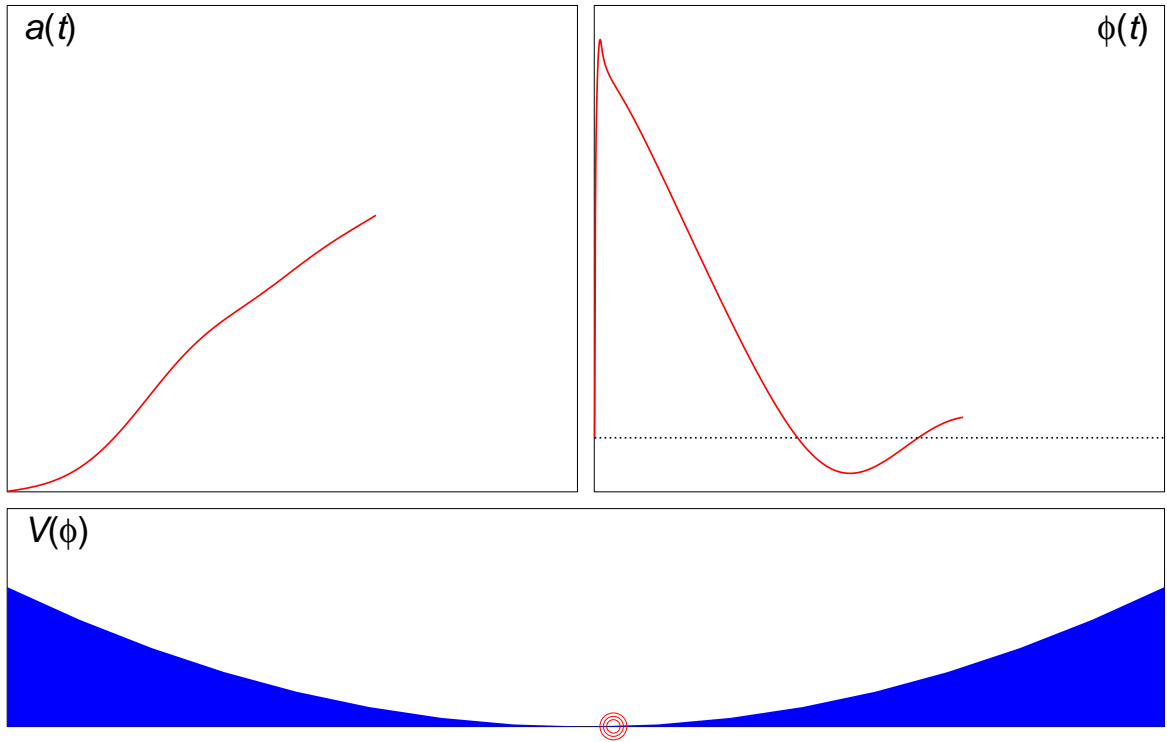}}
  \caption{\it Still of a Movie showing the initial push of a scalar
  $\phi$ up its potential and the ensuing slow-roll phase together
  with the corresponding inflationary phase of $a$. The movie is
  available from the online version \cite{LivRev} of this article at {\tt
  http://relativity.livingreviews.org/Articles/lrr-2005-11/}.}
  \label{PushMov}
\end{figure}

In general, we need to know the matter behavior to know $w$ and
$w_{\mathrm{eff}}$. But we can get generic qualitative information by
treating $p_{\phi}$ and $V(\phi)$ as unknowns determined by $w$ and
$w_{\mathrm{eff}}$. In the generic case, there is no unique solution
for $p_{\phi}^2$ and $V(\phi)$ since, after all, $p_{\phi}$ and $\phi$
change with $t$. They are now subject to two linear equations in terms
of $w$ and $w_{\mathrm{eff}}$, whose determinant must be zero
resulting in
\[
 w_{\mathrm{eff}}=-1+\frac{|p|^{3/2}(w+1)(d(p)- \frac{2}{3}|p|d'(p))}{1-w+
   (w+1)|p|^{3/2}d(p)}\,.
\]
Since for small $p$ the numerator in the fraction approaches zero
faster than the second part of the denominator, $w_{\mathrm{eff}}$
approaches minus one at small volume except for the special case $w=1$
which is realized for $V(\phi)=0$. Note that the argument does not
apply to the case of vanishing potential since then
$p_{\phi}^2=\mathrm{const}$ and $V(\phi)=0$ presents a unique solution
to the linear equations for $w$ and $w_{\mathrm{eff}}$.  In fact, this
case leads in general to a much smaller $w_{\mathrm{eff}}=
-\frac{2}{3}|p|d(p)'/d(p)\approx -1/(1-l)<-1$ \cite{Inflation}.

One can also see from the above formula that $w_{\mathrm{eff}}$,
though close to minus one, is a little smaller than minus one
generically. This is in contrast to single field inflaton models where
the equation of state parameter is a little larger than minus one. As
we will discuss in Sec.~\ref{s:EffResults}, this opens the door to
characteristic signatures distinguishing different models.

Again, also the matter behavior changes, now with classical friction
being replaced by antifriction \cite{Closed}. Matter fields thus move
away from their minima and become excited even if they start close to
a minimum (Fig.~\ref{Push}). Since this does not only apply to the
homogeneous mode, it can provide a mechanism of structure formation as
discussed in Sec.~\ref{s:EffResults}. But also in combination with
chaotic inflation as the mechanism to generate structure does the
modified matter behavior lead to improvements: if we now view the
scalar $\phi$ as an inflaton field, it will be driven to large values
in order to start a second phase of slow-roll inflation which is long
enough. This is satisfied for a large range of the ambiguity
parameters $j$ and $l$ \cite{Robust} and can even leave signatures
\cite{InflationWMAP} in the cosmic microwave spectrum \cite{CMB}: The
earliest moments when the inflaton starts to roll down its potential
are not slow roll, as can also be seen in Figs.~\ref{Push} and
\ref{PushMov} where the initial decrease is steeper. Provided the
resulting structure can be seen today, i.e.\ there are not too many
e-foldings from the second phase, this can lead to visible effects
such as a suppression of power.  Whether or not those effects are to
be expected, i.e.\ which magnitude of the inflaton is generically
reached by the mechanism generating initial conditions, is currently
being investigated at the basic level of loop quantum cosmology
\cite{ASV}. They should be regarded as first suggestions, indicating
the potential of quantum cosmological phenomenology, which have to be
substantiated by detailed calculations including inhomogeneities or at
least anisotropic geometries. In particular the suppression of power
can be obtained by a multitude of other mechanisms.

\paragraph{Model building:} It is already clear that there are
different inflationary scenarios using effects from loop cosmology. A
scenario without inflaton is more attractive since it requires less
choices and provides a fundamental explanation of inflation directly
from quantum gravity. However, it is also more difficult to analyze
structure formation in this context while there are already
well-developed techniques in slow role scenarios.

In these cases where one couples loop cosmology to an inflaton model
one still requires the same conditions for the potential, but
generically gets the required large initial values for the scalar by
antifriction. On the other hand, finer details of the results now
depend on the ambiguity parameters which describe aspects of the
quantization which also arise in the full theory.

It is also possible to combine collapsing and expanding phases in
cyclic or oscillatory models \cite{Oscill}. One then has a history of
many cycles separated by bounces, whose duration depends on details of
the model such as the potential. There can then be many brief cycles
until eventually, if the potential is right, one obtains an
inflationary phase if the scalar has grown high enough. In this way,
one can develop ideas for the pre-history of our universe before the
big bang. There are also possibilities to use a bounce to describe the
structure in the universe. So far, this has only been described in
effective models \cite{Ekpyrotic} using brane scenarios \cite{Roy}
where the classical singularity has been assumed to be absent by yet
to be determined quantum effects. As it turns out, the explicit
mechanism removing singularities in loop cosmology is not compatible
with the assumptions made in those effective pictures. In particular,
the scalar was supposed to turn around during the bounce which is
impossible in loop scenarios unless it encounters a range of positive
potential during its evolution \cite{Cyclic}. Then, however,
generically an inflationary phase commences as in \cite{Oscill} which
is then the relevant regime for structure formation. This shows how
model building in loop cosmology can distinguish scenarios which are
more likely to occur from quantum gravity effects.

Cyclic models can be argued to shift the initial moment of a universe
in the infinite past, but they do not explain how the universe
started. An attempt to explain this is the emergent universe model
\cite{Emergent,Emergent2} where one starts close to a static solution.
This is difficult to achieve classically, however, since the available
fixed points of the equations of motion are not stable and thus a
universe departs too rapidly. Loop cosmology, on the other hand,
implies an additional fixed point of the effective equations which is
stable and allows to start the universe in an initial phase of
oscillations before an inflationary phase is entered
\cite{EmergentLoop,EmergentNat}. This presents a natural realization
of the scenario where the initial scale factor at the fixed point is
automatically small so as to start the universe close to the Planck
phase.

\paragraph{Stability:}

Cosmological equations displaying super-inflation or antifriction are
often unstable in the sense that matter can propagate faster than
light. This has been voiced as a potential danger for loop cosmology,
too \cite{Coule,CouleRev}. An analysis requires inhomogeneous
techniques at least at an effective level, such as those described in
Sec.~\ref{s:InhomIso}. It has been shown that loop cosmology is free
of this problem because the modified behavior for the homogeneous mode
of the metric and matter is not relevant for matter propagation
\cite{Stable}. The whole cosmological picture which follows from the
effective equations is thus consistent.

\subsection{Anisotropies}

Anisotropic models provide a first generalization of isotropic ones to
more realistic situations. They thus can be used to study the
robustness of effects analyzed in isotropic situations and, at the
same time, provide a large class of interesting applications. An
analysis in particular of the singularity issue is important since the
classical approach to a singularity can be very different from the
isotropic one. On the other hand, the anisotropic approach is deemed
to be characteristic even for general inhomogeneous singularities if
the BKL scenario \cite{BKL} is correct.

\subsubsection{Metric variables}

A general homogeneous but anisotropic metric is of the form
\[
\mathrm{d} s^2=-N(t)^2 \mathrm{d} t^2+ \sum_{I,J=1}^3q_{IJ}(t)
\omega^I\otimes \omega^J
\]
with left-invariant 1-forms $\omega^I$ on space $\Sigma$ which, thanks
to homogeneity, can be identified with the simply transitive symmetry
group $S$ as a manifold. The left-invariant 1-forms satisfy the
Maurer--Cartan relations
\[
 \mathrm{d}\omega^I=-\frac{1}{2}C^I_{JK} \omega^J\wedge\omega^K
\]
with the structure constants $C^I_{JK}$ of the symmetry group. In a
matrix parameterization of the symmetry group, one can derive explicit
expressions for $\omega^I$ from the Maurer--Cartan form $\omega^I
T_I=\theta_{MC}=g^{-1}\mathrm{d} g$ with generators $T_I$ of $S$.

The simplest case of a symmetry group is an Abelian one with
$C^I_{JK}=0$, corresponding to the Bianchi I model. In this case, $S$
is given by ${\mathbb R}^3$ or a torus, and left-invariant 1-forms are simply
$\omega^I=\mathrm{d} x^I$ in Cartesian coordinates. Other groups must be
restricted to class A models in this context, satisfying $C^I_{JI}=0$
since otherwise there is no Hamiltonian formulation. The structure
constants can then be parameterized as $C^I_{JK}=\epsilon^I_{JK}
n^{(I)}$.

A common simplification is to assume the metric to be diagonal at all
times, which corresponds to a reduction technically similar to a
symmetry reduction. This amounts to $q_{IJ}=a_{(I)}^2 \delta_{IJ}$ as
well as $K_{IJ}=K_{(I)}\delta_{IJ}$ for the extrinsic curvature with
$K_I=\dot{a}_I$. Depending on the structure constants, there is also
non-zero intrinsic curvature quantified by the spin connection
components
\begin{equation}
 \Gamma_I=\frac{1}{2}\left(\frac{a_J}{a_K}n^J+ \frac{a_K}{a_J}n^K-
   \frac{a_I^2}{a_Ja_K}n^I\right) \quad\mbox{for}\quad
 \epsilon_{IJK}=1\,.
\end{equation}
This influences the evolution as follows from the Hamiltonian
constraint
\begin{eqnarray}
 &&-\frac{1}{8\pi G}\left(
   a_1\dot{a}_2\dot{a}_3+ a_2\dot{a}_1\dot{a}_2+
   a_3\dot{a}_1\dot{a}_2- (\Gamma_2\Gamma_3-n^1\Gamma_1)a_1-
   (\Gamma_1\Gamma_3-n^2\Gamma_2)a_2\right.\nonumber\\
 &&-\left.
   (\Gamma_1\Gamma_2-n^3\Gamma_3)a_3\right)+
 H_{\mathrm{matter}}(a_I)=0\,. \label{BianchiClass}
\end{eqnarray}

In the vacuum Bianchi I case the resulting equations are easy to solve
by $a_I\propto t^{\alpha_I}$ with $\sum_I\alpha_I=\sum_I\alpha_I^2=1$
\cite{Kasner}.  The volume $a_1a_2a_3\propto t$ vanishes for $t=0$
where the classical singularity appears. Since one of the exponents
$\alpha_I$ must be negative, however, only two of the $a_I$ vanish at
the classical singularity while the third one diverges. This already
demonstrates how different the behavior can be from the isotropic one
and that anisotropic models provide a crucial test of any mechanism
for singularity resolution.

\subsubsection{Connection variables}

A densitized triad corresponding to a diagonal homogeneous metric has
real components $p^I$ with $|p^I|=a_Ja_K$ if $\epsilon_{IJK}=1$
\cite{HomCosmo}. Connection components are $c_I=\Gamma_I+\gamma
K_I=\Gamma_I+\gamma\dot{a}_I$ and are conjugate to the
$p_I$, $\{c_I,p^J\}=8\pi\gamma G\delta_I^J$. In terms of triad
variables we now have spin connection components
\begin{equation}
 \Gamma_I = \frac{1}{2}\left(\frac{p^K}{p^J}n^J+ \frac{p^J}{p^K}n^K-
   \frac{p^Jp^K}{(p^I)^2}n^I\right)
\end{equation}
and the Hamiltonian constraint (in the absence of matter)
\begin{eqnarray} \label{H}
 H &=& \frac{1}{8\pi G}\left\{\left[(c_2\Gamma_3+c_3\Gamma_2-\Gamma_2\Gamma_3)
     (1+\gamma^{-2})-
     n^1c_1-\gamma^{-2}c_2c_3\right]
   \sqrt{\left|\frac{p^2p^3}{p^1}\right|} \right.\nonumber\\ 
  &&+\left[(c_1\Gamma_3+c_3\Gamma_1-\Gamma_1\Gamma_3)
     (1+\gamma^{-2})- n^2c_2-\gamma^{-2}c_1c_3\right]
   \sqrt{\left|\frac{p^1p^3}{p^2}\right|} \nonumber\\
  &&\left.+\left[(c_1\Gamma_2+c_2\Gamma_1-\Gamma_1\Gamma_2)
     (1+\gamma^{-2})- n^3c_3-\gamma^{-2}c_1c_2\right]
   \sqrt{\left|\frac{p^1p^2}{p^3}\right|} \right\}\,.
\end{eqnarray}

Unlike in isotropic models, we now have inverse powers of $p^I$ even
in the vacuum case through the spin connection, unless we are in the
Bianchi I model. This is a consequence of the fact that not just
extrinsic curvature, which in the isotropic case is related to the
matter Hamiltonian through the Friedmann equation, leads to
divergences but also intrinsic curvature. These divergences are cut
off by quantum geometry effects as before such that also the dynamical
behavior changes. This can again be dealt with by effective equations
where inverse powers of triad components are replaced by bounded
functions \cite{Spin}. However, even with those modifications,
expressions for curvature are not necessarily bounded unlike in the
isotropic case. This comes from the presence of different classical
scales, $a_I$, such that more complicated expressions as in $\Gamma_I$
are possible, while in the isotropic model there is only one scale
and curvature can only be an inverse power of $p$ which is then
regulated by effective expressions like $d(p)$.

\subsubsection{Applications}
\label{s:HomApp}

\paragraph{Isotropization:} Matter fields are not the only
contributions to the Hamiltonian in cosmology, but also the effect of
anisotropies can be included in this way to an isotropic model. The
late time behavior of this contribution can be shown to behave as
$a^{-6}$ in the shear energy density \cite{Isotropize}, which falls
off faster than any other matter component. Thus, toward later times
the universe becomes more and more isotropic.

In the backward direction, on the other hand, this means that the
shear term diverges most strongly which suggests that this term should
be most relevant for the singularity issue. Even if matter densities
are cut off as discussed before, the presence of bounces would depend
on the fate of the anisotropy term. This simple reasoning is not true,
however, since the behavior of shear is only effective and uses
assumptions about the behavior of matter. It can thus not simply be
extrapolated to early times. Anisotropies are independent degrees of
freedom which affect the evolution of the scale factor. But only in
certain regimes can this contribution be modeled simply by a function
of the scale factor alone; in general one has to use the coupled
system of equations for the scale factor, anisotropies and possible
matter fields.

\begin{figure}[h]
  \centerline{\includegraphics[width=12cm,keepaspectratio]{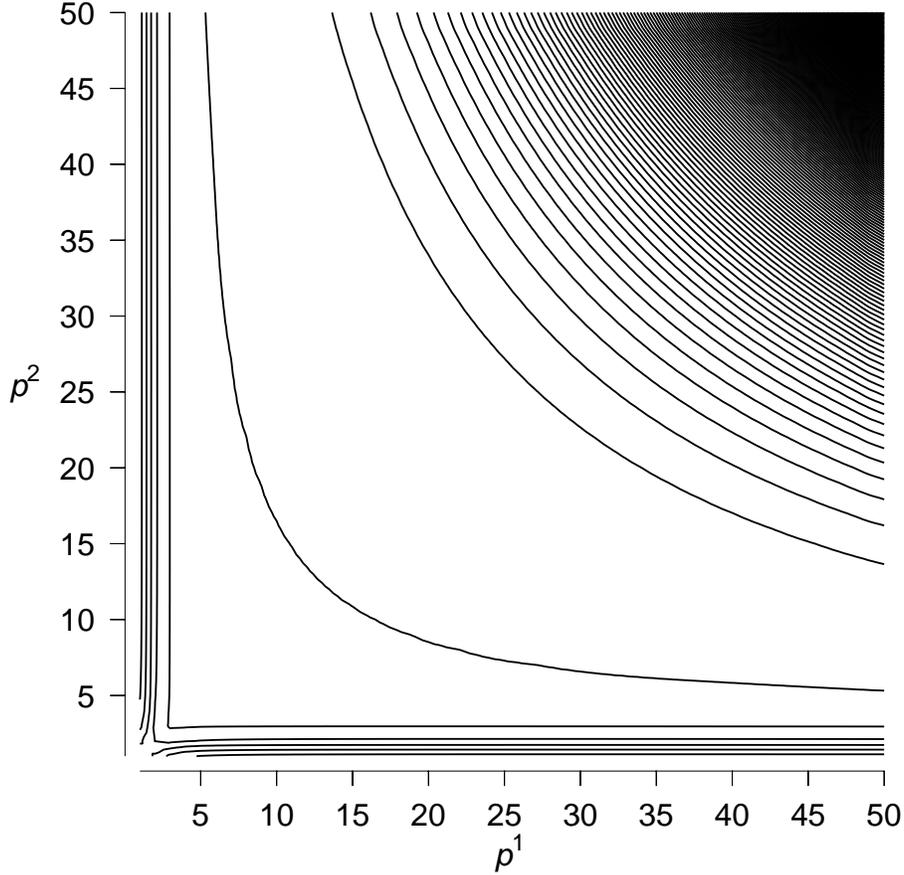}}
  \caption{\it Still of a Movie illustrating the Bianchi IX potential 
  (\ref{BIXPot})
    and the movement of its walls, rising toward zero $p^1$ and $p^2$
    and along the diagonal direction, toward the classical singularity
    with decreasing volume $V=\sqrt{|p^1p^2p^3|}$. The contours are plotted
    for the function $W(p^1,p^2,V^2/(p^1p^2))$. The movie is
  available from the online version \cite{LivRev} of this article at {\tt
  http://relativity.livingreviews.org/Articles/lrr-2005-11/}.}
  \label{BIX}
\end{figure}

\paragraph{Bianchi IX:} Modifications to classical behavior are most
drastic in the Bianchi IX model with symmetry group
$S\cong\mathrm{SU}(2)$ such that $n^I=1$. The classical evolution can
be described by a 3-dimensional mechanics system with a potential
obtained from (\ref{BianchiClass}) such that the kinetic term is
quadratic in derivatives of $a_I$ with respect to a time coordinate
$\tau$ defined by $\mathrm{d}t=a_1a_2a_3\mathrm{d}\tau$. This potential
\begin{eqnarray}
 W(p^I) &=& (\Gamma_2\Gamma_3-n^1\Gamma_1)p^2p^3+
   (\Gamma_1\Gamma_3-n^2\Gamma_2)p^1p^3+
   (\Gamma_1\Gamma_2-n^3\Gamma_3)p^1p^2 \label{BIXPot}\\
 &=& \frac{1}{4} \left(\left(\frac{p^2p^3}{p^1}\right)^2+
   \left(\frac{p^1p^3}{p^2}\right)^2+
   \left(\frac{p^1p^2}{p^3}\right)^2-
   2(p^1)^2-2(p^2)^2-2(p^3)^2\right) \nonumber
\end{eqnarray}
diverges at small $p^I$, in particular (in a direction dependent
manner) at the classical singularity where all $p^I=0$. Fig.~\ref{BIX}
illustrates the walls of the potential which with decreasing volume
push the universe toward the classical singularity.

 As before in
isotropic models, effective equations where the behavior of
eigenvalues of the spin connection components is used do not have this
divergent potential.  Instead, if two $p^I$ are held fixed and the
third approaches zero, the effective quantum potential is cut off and
goes back to zero at small values, which changes the approach to the
classical singularity. Yet, the effective potential is unbounded if
one $p^I$ diverges while another one goes to zero and the situation is
qualitatively different from the isotropic case. Since the effective
potential corresponds to spatial intrinsic curvature, curvature is
{\em not bounded} in anisotropic effective models. However, this is a
statement only about curvature expressions on minisuperspace, and the
more relevant question is what happens to curvature along trajectories
obtained by solving equations of motion. This demonstrates that
dynamical equations must always be considered to draw conclusions for
the singularity issue.

\begin{figure}[h]
  \centerline{\includegraphics[width=12cm,keepaspectratio]{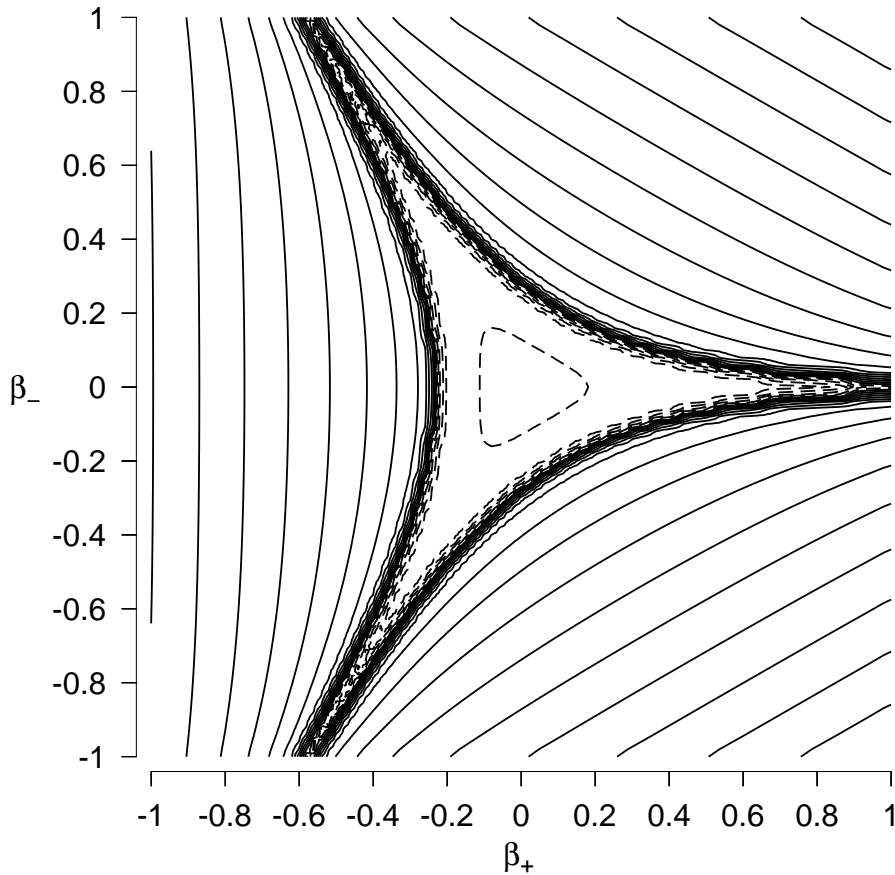}}
  \caption{\it Still of a Movie illustrating the Bianchi IX potential in the
    anisotropy plane and its exponentially rising walls. Positive
    values of the potential are drawn logarithmically with solid
    contour lines and negative values with dashed contour lines. The movie is
  available from the online version \cite{LivRev} of this article at {\tt
  http://relativity.livingreviews.org/Articles/lrr-2005-11/}.}
  \label{BIXLog}
\end{figure}

The approach to the classical singularity is best analyzed in Misner
variables \cite{Mixmaster} consisting of the scale factor
$\Omega:=-\frac{1}{3}\log V$ and two anisotropy parameters
$\beta_{\pm}$ defined such that
\[
 a_1=e^{-\Omega+\beta_++\sqrt{3}\beta_-}\quad,\quad
   a_2=e^{-\Omega+\beta_+-\sqrt{3}\beta_-}\quad,\quad
 a_3=e^{-\Omega-2\beta_+}\,.
\]
The classical potential then takes the form
\[
 W(\Omega,\beta_{\pm})=\frac{1}{2}e^{-4\Omega}\left(e^{-8\beta_+}-
   4e^{-2\beta_+}\cosh(2\sqrt{3}\beta_-)+
   2e^{4\beta_+}(\cosh(4\sqrt{3}\beta_-)-1)\right)
\]
which at fixed $\Omega$ has three exponential walls rising from the
isotropy point $\beta_{\pm}=0$ and enclosing a triangular region
(Fig.~\ref{BIXLog}). 

A cross section of a wall can be obtained by taking $\beta_-=0$ and
$\beta_+$ to be negative, in which case the potential becomes
$W(\Omega,\beta_+,0)\approx\frac{1}{2}e^{-4\Omega-8\beta_+}$.  One
thus obtains the picture of a point moving almost freely until it is
reflected at a wall. In between reflections, the behavior is
approximately given by the Kasner solution described before. This
behavior with infinitely many reflections before the classical
singularity is reached can be shown to be chaotic \cite{NumSing} which
suggests a complicated approach to classical singularities in general.

\begin{figure}[h]
  \centerline{\includegraphics[width=14cm,keepaspectratio]{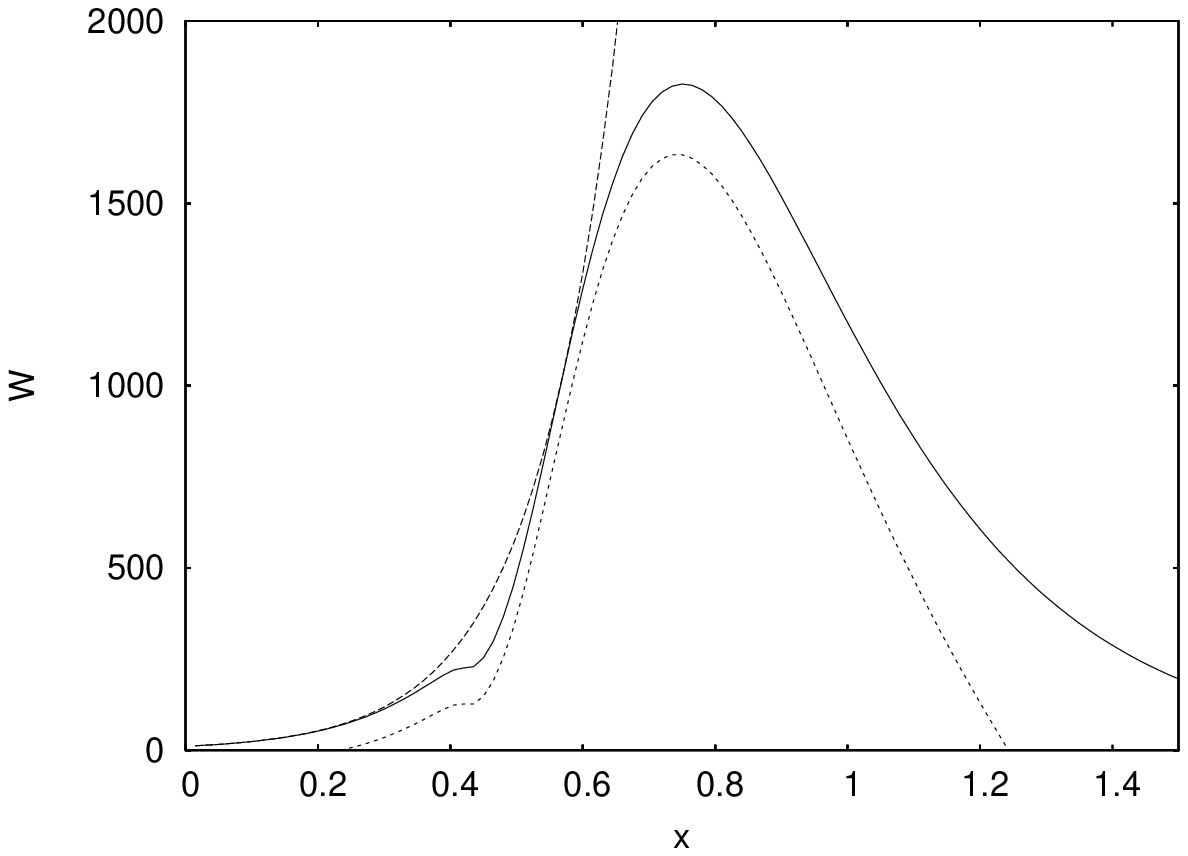}}
  \caption{\it Approximate effective wall of finite height \cite{NonChaos}
  as a function of $x=-\beta_+$, compared to the classical exponential
  wall (upper dashed curve). Also shown is the exact wall
  $W(p^1,p^1,(V/p^1)^2)$ (lower dashed curve), which for $x$ smaller
  than the peak value coincides well with the approximation
  up to a small, nearly constant shift.}
  \label{WallProf}
\end{figure}

\begin{figure}[h]
  \centerline{\includegraphics[width=12cm,keepaspectratio]{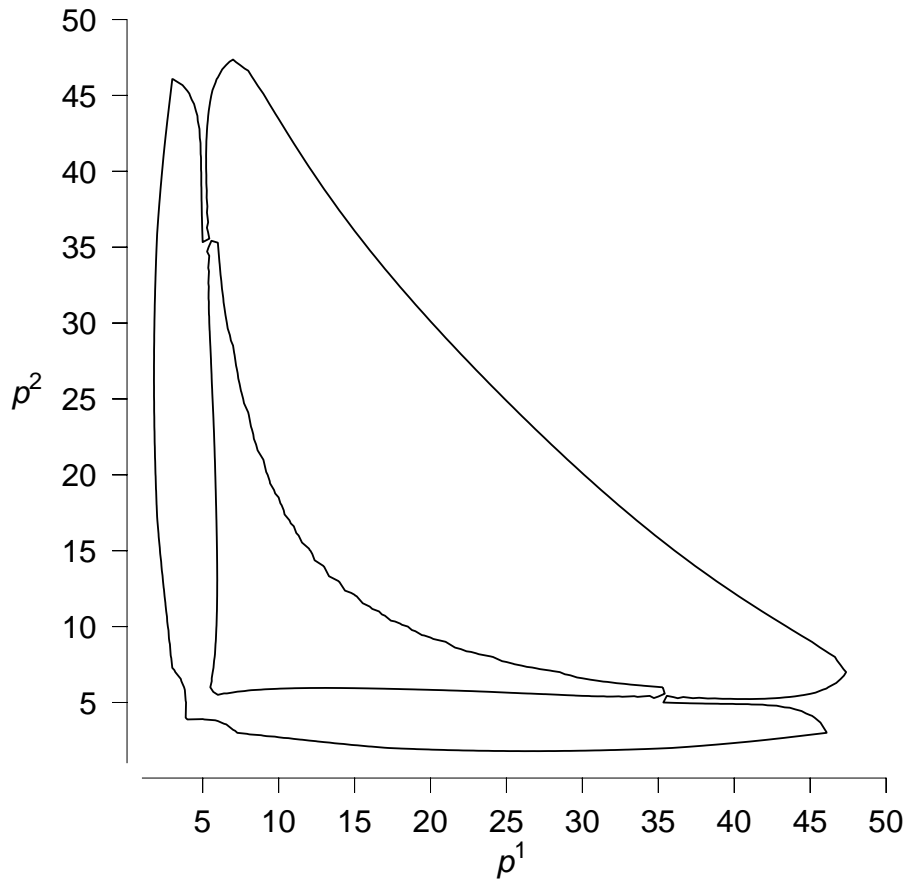}}
  \caption{\it Still of a Movie illustrating the effective Bianchi IX potential
    and the movement and breakdown of its walls. The contours are
    plotted as in Fig.~\ref{BIX}. The movie is
  available from the online version \cite{LivRev} of this article at {\tt
  http://relativity.livingreviews.org/Articles/lrr-2005-11/}.}
  \label{BIXEff}
\end{figure}

\begin{figure}[h]
  \centerline{\includegraphics[width=12cm,keepaspectratio]{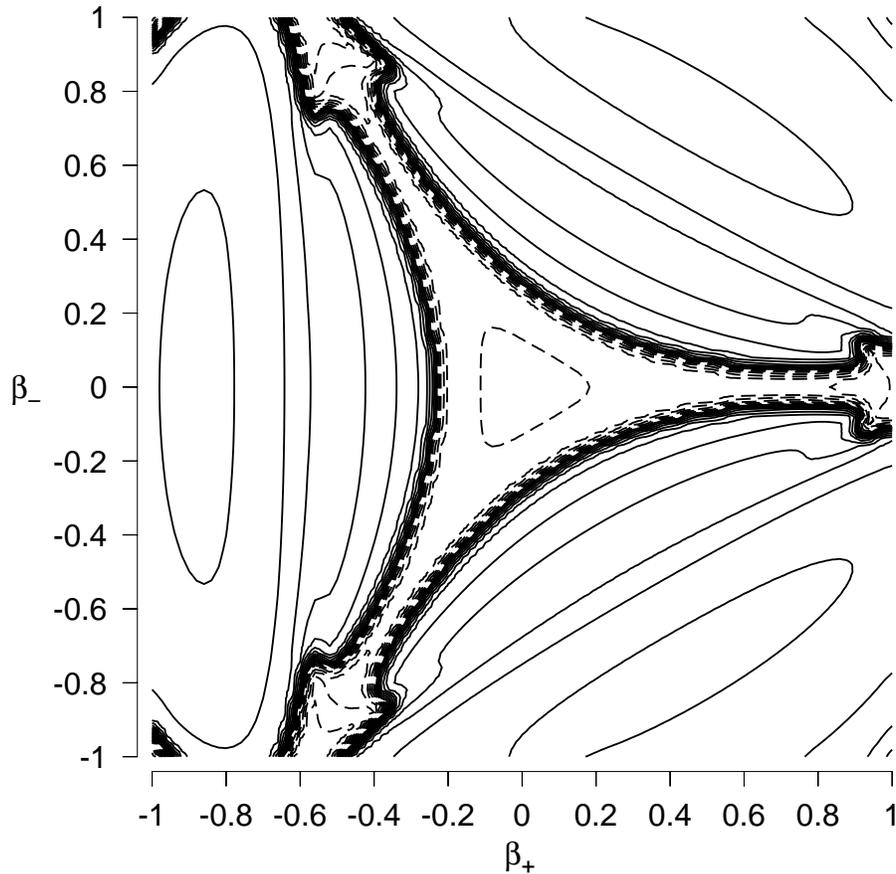}}
  \caption{\it Still of a Movie illustrating the effective Bianchi IX 
 potential in the
    anisotropy plane and its walls of finite height which disappear at
    finite volume. Positive values of the potential are drawn
    logarithmically with solid contour lines and negative values with
    dashed contour lines. The movie is
  available from the online version \cite{LivRev} of this article at {\tt
  http://relativity.livingreviews.org/Articles/lrr-2005-11/}.}
  \label{BIXEffLog}
\end{figure}

With the effective modification, however, the potential for fixed
$\Omega$ does not diverge and the walls, as shown in
Fig.~\ref{WallProf}, break down already at a small but non-zero volume
\cite{NonChaos}. As a function of densitized triad components the
effective potential is illustrated in Fig.~\ref{BIXEff}, and as a
function on the anisotropy plane in Fig.~\ref{BIXEffLog}.  In this
scenario, there are only finitely many reflections which does not lead
to chaotic behavior but instead results in asymptotic Kasner behavior
\cite{ChaosLQC}.

Comparing Fig.~\ref{BIXLog} with Fig.~\ref{BIXEffLog} shows that in
their center they are very close to each other, while strong
deviations occur for large anisotropies. This demonstrates that most
of the classical evolution, which mostly happens in the inner
triangular region, is not strongly modified by the effective
potential. Quantum effects are important only when anisotropies become
too large, for instance when the system moves deep into one of the
three valleys, or the total volume becomes small. In those regimes the
quantum evolution will take over and describe the further behavior of
the system.

\paragraph{Isotropic curvature suppression:}

If we use the potential for time coordinate $t$ rather than $\tau$, it
is replaced by $W/(p^1p^2p^3)$ which in the isotropic reduction
$p^1=p^2=p^3=\frac{1}{4}a^2$ gives the curvature term $ka^{-2}$.
Although the anisotropic effective curvature potential is not bounded
it is, unlike the classical curvature, bounded from above at any fixed
volume. Moreover, it is bounded along the isotropy line and decays
when $a$ approaches zero. Thus, there is a suppression of the
divergence in $ka^{-2}$ when the closed isotropic model is viewed as
embedded in a Bianchi IX model. Similarly to matter Hamiltonians,
intrinsic curvature then approaches zero at zero scale factor.

This is a further illustration for the special nature of isotropic
models compared to anisotropic ones. In the classical reduction, the
$p^I$ in the anisotropic spin connection cancel such that the spin
connection is a constant and no special steps are needed for its
quantization. By viewing isotropic models within anisotropic ones, one
can consistently realize the model and see a suppression of intrinsic
curvature terms. Anisotropic models, on the other hand, do not have,
and do not need, complete suppression since curvature functions can
still be unbounded.

\subsubsection{Implications for inhomogeneities}

Even without implementing inhomogeneous models the previous discussion
allows some tentative conclusions as to the structure of general
singularities. This is based on the BKL picture \cite{BKL} whose basic
idea is to study Einstein's field equations close to a singularity.
One can then argue that spatial derivatives become subdominant
compared to time-like derivatives such that the approach should locally
be described by homogeneous models, in particular the Bianchi IX model
since it has the most freedom in its general solution.

Since spatial derivatives are present, though, they lead to small
corrections and couple the geometries in different spatial points. One
can visualize this by starting with an initial slice which is
approximated by a collection of homogeneous patches. For some time,
each patch evolves independently of the others, but this is not
precisely true since coupling effects have been ignored. Moreover,
each patch geometry evolves in a chaotic manner which means that two
initially nearby geometries depart rapidly from each other. The
approximation can thus be maintained only if the patches are
subdivided during the evolution which goes on without limits in the
approach to the singularity. There is thus more and more inhomogeneous
structure being generated on arbitrarily small scales which leads to a
complicated picture of a general singularity.

This picture can be taken over to the effective behavior of the
Bianchi IX model. Here, the patches do not evolve chaotically even
though at larger volume they follow the classical behavior. The
subdivision thus has to be done also for the initial effective
evolution. At some point, however, when reflections on the potential
walls stop, the evolution simplifies and subdivisions are no longer
necessary. There is thus a lower bound to the scale of structure whose
precise value depends on the initial geometries. Nevertheless, from
the scale at which the potential walls break down one can show that
structure formation stops at the latest when the discreteness scale of
quantum geometry is reached \cite{NonChaos}. This can be seen as a
consistency test of the theory since structure below the discreteness
could not be supported by quantum geometry.

We have thus a glimpse on the inhomogeneous situation with a
complicated but consistent approach to a general classical
singularity. The methods involved, however, are not very robust since
the BKL scenario, which even classically is still at the level of a
conjecture for the general case \cite{NumSing,WS:AR}, would need to be
available as an approximation to quantum geometry. For more reliable
results the methods need to be refined to take into account
inhomogeneities properly.

\subsection{Inhomogeneities}

Allowing for inhomogeneities inevitably means to take a big step from
finitely many degrees of freedom to infinitely many ones. There is no
straightforward way to cut down the number of degrees of freedom to
finitely many ones while being more general than in the
homogeneous context. One possibility would be to introduce a
small-scale cut-off such that only finitely many wave modes arise
(e.g.\ through a lattice as is indeed done in some coherent state
constructions \cite{CohState}). This is in fact expected to happen in
a discrete framework such as quantum geometry, but would at this stage
of defining a model simply be introduced by hand.

\subsubsection{Available approximations}

For the analysis of inhomogeneous situations there are several
different approximation schemes:
\begin{itemize}
\item Use only isotropic quantum geometry and in particular its
  effective description, but couple to inhomogeneous matter fields.
  Problems in this approach are that back-reaction effects are ignored
  (which is also the case in most classical treatments) and that there
  is no direct way how to check modifications used in particular for
  gradient terms of the matter Hamiltonian. So far, this approach has
  led to a few indications of possible effects.
 \item Start with the full constraint operator, write it as the
   homogeneous one plus correction terms from inhomogeneities, and
   derive effective classical equations. This approach is more
   ambitious since contact to the full theory is realized. So far,
   there are not many results since a suitable perturbation scheme has
   to be developed.
 \item There are inhomogeneous symmetric models, such as the
   spherically symmetric one or Einstein--Rosen waves, which have
   infinitely many kinematical degrees of freedom but can be treated
   explicitly. Also here, contact to the full theory is present
   through the symmetry reduction procedure of Sec.~\ref{s:Link}. This
   procedure itself can be tested by studying those models between
   homogeneous ones and the full theory, but results can also be used
   for physical applications involving inhomogeneities. Many issues
   which are of importance in the full theory, such as the anomaly
   problem, also arise here and can thus be studied more explicitly.
\end{itemize}

\subsubsection{Inhomogeneous matter with isotropic quantum geometry}
\label{s:InhomIso}

Inhomogeneous matter fields cannot be introduced directly to isotropic
quantum geometry since after the symmetry reduction there is no space
manifold left for the fields to live on. There are then two different
routes to proceed: One can simply take the classical field Hamiltonian
and introduce effective modifications modeled on what happens to the
isotropic Hamiltonian, or perform a mode decomposition of the matter
fields and just work with the space-independent amplitudes. The latter
is possible since the homogeneous geometry provides a background for
the mode decomposition.

The basic question, for the example of a scalar field, then is how the
metric coefficient $E^a_iE^b_i/\sqrt{\left|\det E\right|}$ in the
gradient term of (\ref{Hphi}) would be replaced effectively. For the
other terms, one can simply use the isotropic modification which is
taken directly from the quantization. For the gradient term, however,
one does not have a quantum expression in this context and a
modification can only be guessed. The problem arises since the
inhomogeneous term involves inverse powers of $E$, while in the
isotropic context the coefficient just reduces to $\sqrt{|p|}$ which
would not be modified at all. There is thus no obvious and unique way
to find a suitable replacement.

A possible route would be to read off the modification from the full
quantum Hamiltonian, or at least from an inhomogeneous model,
which requires a better knowledge of the reduction
procedure. Alternatively, one can take a more phenomenological point
of view and study the effects of possible replacements. If the
robustness of these effects to changes in the replacements is known,
one can get a good picture of possible implications. So far, only
initial steps have been taken and there is no complete program in this
direction.

Another approximation of the inhomogeneous situation has been
developed in \cite{CorrectScalar} by patching isotropic quantum
geometries together to support an inhomogeneous matter field. This can
be used to study modified dispersion relations to the extent that the
result agrees with preliminary calculations performed in the full
theory \cite{GRB,Correct1,Correct2,QFTonCSTI,QFTonCSTII} even at a
quantitative level. There is thus further evidence that symmetric
models and their approximations provide reliable insights into the full
theory.

\subsubsection{Perturbations}

With a symmetric background, a mode decomposition is not only possible
for matter fields but also for geometry. The homogeneous modes can
then be quantized as before, while higher modes are coupled as
perturbations implementing inhomogeneities \cite{Halliwell}. As with
matter Hamiltonians before, one can then also deal with the
gravitational part of the Hamiltonian constraint. In particular, there
are terms with inverse powers of the homogeneous fields which receive
modifications upon quantization. As with gradient terms in matter
Hamiltonians, there are several options for those modifications which
can only be restricted by relating them to the full Hamiltonian. This
would require introducing the mode decomposition, analogously to
symmetry conditions, at the quantum level and writing the full
constraint operator as the homogeneous one plus correction terms.

An additional complication compared to matter fields is that one is
now dealing with infinitely many coupled constraint equations since
the lapse function $N(x)$ is inhomogeneous, too. This function can
itself be decomposed into modes $\sum_nN_n Y_n(x)$, with harmonics
$Y_n(x)$ according to the symmetry, and each amplitude $N_n$ is varied
independently giving rise to a separate constraint. The main
constraint arises from the homogeneous mode, which describes how
inhomogeneities affect the evolution of the homogeneous scale factors.

\subsubsection{Inhomogeneous models}

The full theory is complicated at several different levels of both
conceptual and technical nature. For instance, one has to deal with
infinitely many degrees of freedom, most operators have complicated
actions, and interpreting solutions to all constraints in a
geometrical manner can be difficult. Most of these complications are
avoided in homogeneous models, in particular when effective classical
equations are employed. These equations use approximations of
expectation values of quantum geometrical operators which need to be
known rather explicitly. The question then arises whether one can
still work at this level while relaxing the symmetry conditions and
bringing in more complications of the full theory.

Explicit calculations at a level similar to homogeneous models, at
least for matrix elements of individual operators, are possible in
inhomogeneous models, too. In particular the spherically symmetric
model and cylindrically symmetric Einstein--Rosen waves are of this
class, where the symmetry or other conditions are strong enough to
result in a simple volume operator. In the spherically symmetric model
this simplification comes from the remaining isotropy subgroup
isomorphic to U(1) in generic points, while the Einstein--Rosen model
is simplified by polarization conditions which play a role analogous
to the diagonalization of homogeneous models. With these models one
obtains access to applications for black holes and gravitational waves,
but also to inhomogeneities in cosmology.

In spherical coordinates $x$, $\vartheta$, $\varphi$ a spherically symmetric
spatial metric takes the form
\[
\mathrm{d} s^2=q_{xx}(x,t)\mathrm{d} x^2+
q_{\varphi\varphi}(x,t)\mathrm{d}\Omega^2
\]
with $\mathrm{d}\Omega^2=\mathrm{d}\vartheta^2+
\sin^2\vartheta\mathrm{d}\varphi^2$.  This is related to densitized
triad components by \cite{SphKl1,SphKl2}
\[
 |E^x|=q_{\varphi\varphi}\quad,\quad (E^{\varphi})^2=q_{xx}q_{\varphi\varphi}
\]
which are conjugate to the other basic variables given by the Ashtekar
connection component $A_x$ and the extrinsic curvature component
$K_{\varphi}$:
\[
 \{A_x(x),E^x(y)\}= 8\pi G\gamma\delta(x,y) \quad,\quad \{\gamma
 K_{\varphi}(x),E^{\varphi}(y)\}= 16\pi G\gamma\delta(x,y)\,.
\]
Note that we use the Ashtekar connection for the inhomogeneous
direction $x$ but extrinsic curvature for the homogeneous direction
along symmetry orbits \cite{SphSymmHam}. Connection and extrinsic
curvature components for the $\varphi$-direction are related by
$A_{\varphi}^2=\Gamma_{\varphi}^2+\gamma^2 K_{\varphi}^2$ with the
spin connection component
\begin{equation} \label{Gammaphi}
 \Gamma_{\varphi}= -\frac{E^{x\prime}}{2E^{\varphi}}\,.
\end{equation}
Unlike in the full theory or homogeneous models, $A_{\varphi}$ is not
conjugate to a triad component but to \cite{SphSymm}
\[
 P^{\varphi}=\sqrt{4(E^{\varphi})^2-A_{\varphi}^{-2}(P^{\beta})^2}
\]
with the momentum $P^{\beta}$ conjugate to a U(1)-gauge angle $\beta$.
This is a rather complicated function of both triad and connection
variables such that the volume $V=4\pi\int
\sqrt{|E^x|}E^{\varphi}\mathrm{d} x$ would have a rather complicated
quantization. It would still be possible to compute the full volume
spectrum, but with the disadvantage that volume eigenstates would not
be given by triad eigenstates such that computations of many operators
would be complicated \cite{SphSymmVol}. This can be avoided by using
extrinsic curvature which is conjugate to the triad component
\cite{SphSymmHam}.  Moreover, this is also in accordance with a
general scheme to construct Hamiltonian constraint operators for the
full theory as well as symmetric models
\cite{QSDI,CosmoIII,BlackHoles}.

The constraint operator in spherical symmetry is given by
\begin{equation} \label{HSphSymm}
 H[N] = -(2G)^{-1}\int_B\mathrm{d} x N(x) |E^x|^{-1/2}\left(
 (K_{\varphi}^2 E^{\varphi}+2
 K_{\varphi}K_x E^x)+(1-\Gamma_{\varphi}^2)E^{\varphi}+ 
2\Gamma_{\varphi}' E^x \right)
\end{equation}
accompanied by the diffeomorphism constraint
\begin{equation} \label{DSphSymm}
 D[N^x]=(2G)^{-1} \int_B N^x(x)
 (2E^{\varphi}K_{\varphi}'-K_xE^{x\prime})\,.
\end{equation}
We have expressed this in terms of $K_x$ for simplicity, keeping in
mind that as the basic variable for quantization we will later use the
connection component $A_x$.

Since the Hamiltonian constraint contains the spin connection
component $\Gamma_{\varphi}$ given by (\ref{Gammaphi}), which contains
inverse powers of densitized triad components, one can expect
effective classical equations with modifications similar to the
Bianchi IX model. However, the situation is now much more complicated
since we have a system constrained by many constraints with a
non-Abelian algebra. Simply replacing the inverse of $E^{\varphi}$ by a
bounded function as before will change the constraint algebra and thus
most likely lead to anomalies. It is currently open if a more refined
replacement can be done where not only the spin connection but also
the extrinsic curvature terms are modified. This issue has the
potential to shed light on many questions related to the anomaly
issue. It is one of the cases where models between homogeneous ones,
where the anomaly problem trivializes, and the full theory are most
helpful.

\subsubsection{Results}
\label{s:EffResults}

There are some results obtained for inhomogeneous systems. We have
already discussed glimpses from the BKL picture, which used loop
results only for anisotropic models. Methods described in this section
have led to some preliminary insights into possible cosmological
scenarios.

\paragraph{Matter gradient terms and small-$a$ effects:} 

When an inhomogeneous matter Hamiltonian is available it is possible
to study its implications on the cosmic microwave background with
standard techniques. With modifications of densities there are then
different regimes since the part of the inflationary era responsible
for the formation of currently visible structure can be in the
small-$a$ or large-$a$ region of the effective density.

The small-$a$ regime below the peak of effective densities has more
dramatic effects since inflation can here be provided by quantum
geometry effects alone and the matter behavior changes to be
anti-frictional \cite{Inflation,Closed}. Mode evolution in this regime
has been investigated for a particular choice of gradient term and
using a power-law approximation for the effective density at small
$a$, with the result that there are characteristic signatures
\cite{PowerLoop}. As in standard inflation models the spectrum is
nearly scale invariant, but its spectral index is slightly larger than
one (blue tilt) as compared to slightly smaller than one (red tilt)
for single-field inflaton models. Since small scale factors at early
stages of inflation generate structure which today appears on the
largest scales, this implies that low multipoles of the power spectrum
should have a blue tilt. The running of the spectral index in this
regime can also be computed but depends only weakly on ambiguity
parameters.

The main parameter then is the duration of loop inflation. In the
simplest scenario one can assume only one inflationary phase which
would require huge values for the ambiguity parameter $j$. This is
unnatural and would imply that the spectrum is blue on almost all
scales which is in conflict with present observations. Thus, not only
conceptual arguments but also cosmological observations point to smaller
values for $j$, which is quite remarkable.

In order to have sufficient inflation to make the universe big enough
one then needs additional stages provided by the behavior of matter
fields. One still does not need an inflaton since now the details of
the expansion after the structure generating phase are less
important. Any matter field being driven away from its potential
minimum during loop inflation and rolling down its potential
thereafter suffices. Depending on the complexity of the model there
can be several such phases.

\paragraph{Matter gradient terms and large-$a$ effects:} 

At larger scale factors above the peak of effective densities there
are only perturbative corrections from loop effects. This has been
investigated with the aim of finding trans-Planckian corrections to
the microwave background, also here with a particular gradient
term. In this model, cancellations have been observed which imply that
corrections appear only at higher orders of the perturbation series
and are too weak to be observable \cite{PowerPert}.

A common problem of both analyses is that the robustness of the
observed effects has not yet been studied. This is in particular a
pressing problem since one modification of the gradient term has been
chosen without further motivation. Moreover, the modifications in both
examples were different. Without a more direct derivation of the
modifications from inhomogeneous models or the full theory one can
only rely on a robustness analysis to show that the effects can be
trusted. In particular the cancellation in the second example must be
shown to be realized for a larger class of modifications.

\paragraph{Non-inflationary structure formation:}

Given a modification of the gradient term one obtains effective
equations for the matter field which for a scalar results in a
modified Klein--Gordon equation. After a mode decomposition one can
then easily see that all the modes behave differently at small scales
with the classical friction replaced by anti-friction as in
Sec.~\ref{s:Intuitive}. Thus, not only the average value of the field
is driven away from its potential minimum but also higher modes are
being excited. The coupled dynamics of all the modes thus provides a
scenario for structure formation which does not rely on inflation but
on the anti-friction effect of loop cosmology.

Even though all modes experience this effect, they do not all see it
in the same way. The gradient term implies an additive contribution to
the potential proportional to $k^2$ for a mode of wave number $k$,
which also depends on the metric in a way determined by the gradient
term modification. For larger scales, the additional term is not
essential and their amplitudes will be pushed to similar magnitudes,
suggesting scale invariance for them.  The potential relevant for
higher modes, however, becomes steeper and steeper such that they are
less excited by anti-friction and retain a small initial amplitude. In
this way, the structure formation scenario provides a dynamical
mechanism for a small-scale cut-off, possibly realizing older
expectations \cite{InflCutoff,InflCutOff2}.

\paragraph{Stability:}

As already noted, inhomogeneous matter Hamiltonians can be used to
study the stability of cosmological equations in the sense that matter
does not propagate faster than light. The modified behavior of
homogeneous modes has led to the suspicion that loop cosmology is not
stable \cite{Coule,CouleRev} since other cosmological models
displaying super-inflation have this problem. A detailed analysis of
the loop equations, however, shows that the equations as they arise
from modifications are automatically stable. While the
homogeneous modes display super-inflationary and anti-frictional
behavior, they are not relevant for matter propagation. Modes relevant
for propagation, on the other hand, are modified differently in such a
manner that the total behavior is stable \cite{Stable}. Most
importantly, this is an example where an inhomogeneous matter
Hamiltonian with its modifications must be used and the qualitative
result of stability can be shown to be robust under possible changes
of the effective modification. This shows that reliable conclusions
can be drawn for important issues without a precise definition of the
effective inhomogeneous behavior.

\subsection{Summary}

Loop cosmology is an effective description of quantum effects in
cosmology, obtained in a framework of a background independent and
non-perturbative quantization. There is mainly one change compared to
classical equations coming from modified densities in matter
Hamiltonians or also anisotropy potentials. These modifications are
non-perturbative as they contain inverse powers of the Planck length
and thus the gravitational constant, but also perturbative corrections
arise from curvature terms, which are now being studied.

The non-perturbative modification alone is responsible for a
surprising variety of phenomena which all improve the behavior in
classical cosmology. Nevertheless, the modification had not been
motivated by phenomenology but derived through the background
independent quantization. Details of its derivation in cosmological
models and its technical origin will now be reviewed in
Sec.~\ref{s:Analog}, before we come to a discussion of the link to the
full theory in Sec.~\ref{s:Link}.

\section{Loop quantization of symmetric models}
\label{s:Analog}

\begin{flushright}
\begin{quote}
{\em Analogies prove nothing, but they can make one feel more at home.}
\end{quote}

{\sc Sigmund Freud}

Introductory Lectures on Psychoanalysis
\end{flushright}

In full loop quantum gravity, the quantum representation is crucial
for the foundation of the theory. The guiding theme there is
background independence which requires one to smear the basic fields
in a particular manner to holonomies and fluxes. In this section we
will see what implications this has for composite operators and the
physical effects they entail.  We will base this analysis on symmetric
models in order to be able to perform explicit calculations.

Symmetries are usually introduced in order to simplify calculations or
make them possible in the first place. However, symmetries can
sometimes also lead to complications in conceptual questions if the
additional structure they provide is not fully taken into account. In
the present context, it is important to realize that the action of a
symmetry group on a space manifold provides a partial background such
that the situation is always slightly different from the full theory.
If the symmetry is strong, such as in homogeneous models, other
representations such as the Wheeler--DeWitt representation can be
possible even though the fact that a background has been used may not
be obvious. While large scale physics is not very sensitive to the
representation used, it becomes very important on the smallest scales
which we have to take into account when the singularity issue is
considered.

Instead of looking only at one symmetric model where one may have
different possibilities to choose the basic representation, one should
thus keep the full view on different models as well as the full
theory. In fact, in loop quantum gravity it is possible to relate
models and the full theory such that symmetric states and basic
operators, and thus the representation, can be derived from the unique
background independent representation of the full theory. We will
describe this in detail in Sec.~\ref{s:Link}, after having
discussed the construction of quantum models in the present section.
Without making use of the relation to the full theory, one can
construct models by {\em analogy}. This means that quantization steps
are modeled on those which are known to be crucial in the full theory,
which starts with the basic representation and continues to the
Hamiltonian constraint operator. One can then disentangle places where
additional input as compared to the full theory is needed and which
implications it has.

\subsection{Symmetries and backgrounds}

It is impossible to introduce symmetries in a completely background
independent manner. The mathematical action of a symmetry group is
defined by a mapping between abstract points which do not exist in a
diffeomorphism invariant setting (if one, for instance, considers only
equivalence classes up to arbitrary diffeomorphisms).

More precisely, while the full theory has as background only a
differentiable or analytic manifold $\Sigma$, a symmetric model has as
background a symmetric manifold $(\Sigma,S)$ consisting of a
differentiable or analytic manifold $\Sigma$ together with an action
of a symmetry group $S\colon\Sigma\to\Sigma$. How strong the
additional structure is depends on the symmetry used. The strongest
symmetry in gravitational models is realized with spatial isotropy
which implies a unique spatial metric up to a scale factor. The
background is thus equivalent to a conformal space.

All constructions in a given model must take its symmetry into account
since otherwise its particular dynamics, for instance, could not be
captured. The structure of models thus depends on the different types
of background realized for different symmetry groups. This can not
only lead to simplifications but also to conceptual differences, and
it is always instructive to keep the complete view on different models as
well as the full theory. Since the loop formalism is general enough to
encompass all relevant models, there are many ways to compare and
relate different systems. It is thus possible to observe
characteristic features of (metric) background independence even in
cases where more structure is available.

\subsection{Isotropy}

Isotropic models are described purely in terms of the scale factor
$a(t)$ such that there is only a single kinematical degree of
freedom. In connection variables, this is parameterized by the triad
component $p$ conjugate to the connection component $c$.

\subsubsection{Representation}
\label{s:IsoRep}

If we restrict ourselves to invariant connections of a given form, it
suffices to probe them with only special holonomies. For an isotropic
connection $A_a^i=\tilde{c}\Lambda^i_I\omega_a^I$ (see
App.~\ref{s:Iso}) we can choose holonomies along one integral curve of
a symmetry generator $X_I$.  They are of the form
\begin{equation} \label{isohol}
 h_I=\exp\int A_a^iX_I^a\tau_i= \cos{\textstyle \frac{1}{2}}\mu
 c+2\Lambda_I^i\tau_i \sin{\textstyle\frac{1}{2}}\mu c
\end{equation}
where $\mu$ depends on the parameter length of the curve and can be
any real number (thanks to homogeneity, path ordering is not
necessary). Since knowing the values $\cos\frac{1}{2}\mu$ and
$\sin\frac{1}{2}\mu c$ for all $\mu$ uniquely determines the value of
$c$, which is the only gauge invariant information contained in the
connection, these holonomies describe the configuration space of
connections completely.

This illustrates how symmetric configurations allow one to simplify
the constructions behind the full theory. But it also shows which
effects the presence of a partial background can have on the formalism
\cite{Bohr}. In the present case the background enters through the
left-invariant 1-forms $\omega^I$ defined on the spatial manifold
whose influence is contained in the parameter $\mu$. All information
about the edge used to compute the holonomy is contained in this
single parameter which leads to degeneracies compared to the full
theory. Most importantly, one cannot distinguish between the parameter
length and the spin label of an edge: Taking a power of the holonomy
in a non-fundamental representation simply rescales $\mu$, which could
just as well come from a longer parameter length. That this is related
to the presence of a background can be seen by looking at the roles of
edges and spin labels in the full theory. There, both concepts are
independent and appear very differently. While the embedding of an
edge, including its parameter length, is removed by diffeomorphism
invariance, the spin label remains well-defined and is important for
ambiguities of operators. In the model, however, the full
diffeomorphism invariance is not available such that some information
about edges remains in the theory and merges with the spin label.
Issues like that have to be taken into account when constructing
operators in a model and comparing with the full theory.

\paragraph{Compactification:}

The functions appearing in holonomies for isotropic connections define
the algebra of functions on the classical configuration space which,
together with fluxes, is to be represented on a Hilbert space. This
algebra does not contain arbitrary continuous functions of $c$ but
only almost periodic ones of the form \cite{Bohr}
\begin{equation}
 f(c)=\sum_{\mu}f_{\mu}\exp(i\mu c/2)
\end{equation}
where the sum is over a countable subset of ${\mathbb R}$. This is
analogous to the full situation, reviewed in Sec.~\ref{s:FuncSp},
where matrix elements of holonomies define a special algebra of
continuous functions of connections. As in this case, the algebra can
be represented as the set of {\em all} continuous functions on a
compact space, called its spectrum. This compactification can be
imagined as being obtained from enlarging the classical configuration
space ${\mathbb R}$ by adding points, and thus more continuity
conditions, until only functions of the given algebra survive as
continuous ones. A well-known example is the one point
compactification which is the spectrum of the algebra of continuous
functions $f$ for which $\lim_{x\to-\infty}f(x)=\lim_{x\to\infty}f(x)$
exists. In this case, one just needs to add a single point at
infinity.

In the present case, the procedure is more complicated and leads to
the Bohr compactification $\bar{\mathbb R}_{\mathrm{Bohr}}$ which
contains ${\mathbb R}$ densely. It is very different from the one
point compactification, as can be seen from the fact that the only
function which is continuous on both spaces is the zero function. In
contrast to the one point compactification, the Bohr compactification
is an Abelian group, just like ${\mathbb R}$ itself. Moreover, there
is a one-to-one correspondence between irreducible representations of
${\mathbb R}$ and irreducible representations of $\bar{\mathbb
  R}_{\mathrm{Bohr}}$, which can also be used as the definition of the
Bohr compactification. Representations of $\bar{\mathbb R}_{\mathrm{Bohr}}$
are thus labeled by real numbers and given by $\rho_{\mu}\colon
\bar{\mathbb R}_{\mathrm{Bohr}}\to{\mathbb C}, c\mapsto e^{i\mu c}$. As
with any compact group, there is a unique normalized Haar measure
$\mathrm{d}\mu(c)$ given by
\begin{equation}
 \int_{\bar{\mathbb R}_{\mathrm{Bohr}}} f(c)\mathrm{d}\mu(c)=
 \lim_{T\to\infty}\frac{1}{2T} \int_{-T}^T f(c)\mathrm{d} c
\end{equation}
where on the right hand side the Lebesgue measure on ${\mathbb R}$ is
used.

The Haar measure defines the inner product for the Hilbert space
$L^2(\bar{\mathbb R}_{\mathrm{Bohr}},\mathrm{d}\mu(c))$ of square
integrable functions on the quantum configuration space. As one can
easily check, exponentials of the form $\langle c|\mu\rangle=e^{i\mu
  c/2}$ are normalized and orthogonal to each other for different $\mu$,
\begin{equation}
 \langle \mu_1|\mu_2\rangle= \delta_{\mu_1,\mu_2}
\end{equation}
which demonstrates that the Hilbert space is not separable.

Similarly to holonomies, one needs to consider fluxes only for special
surfaces, and all information is contained in the single number
$p$. Since it is conjugate to $c$, it is quantized to a derivative
operator
\begin{equation}
 \hat{p}=-\frac{1}{3}i\gamma\ell_{\mathrm{P}}^2 
\frac{\mathrm{d}}{\mathrm{d} c}
\end{equation}
whose action 
\begin{equation} \label{TriadOp}
 \hat{p}|\mu\rangle = \frac{1}{6}\gamma\ell_{\mathrm{P}}^2\mu
 |\mu\rangle =: p_{\mu}|\mu\rangle
\end{equation}
on basis states $|\mu\rangle$ can easily be
determined. In fact, the basis states are eigenstates of the flux
operator which demonstrates that the flux spectrum is discrete (all
eigenstates are normalizable).

This property is analogous to the full theory with its discrete flux
spectra, and similarly it implies discrete quantum geometry. We thus
see that the discreteness survives the symmetry reduction in this
framework \cite{CosmoII}. Similarly, the fact that only holonomies are
represented in the full theory but not connection components is
realized in the model, too. In fact, we have so far represented only
exponentials of $c$, and one can see that these operators are not
continuous in the parameter $\mu$. Thus, an operator quantizing $c$
directly does not exist on the Hilbert space. These properties are
analogous to the full theory, but very different from the
Wheeler--DeWitt quantization. In fact, the resulting representations
in isotropic models are inequivalent. While the representation is not
of crucial importance when only small energies or large scales are
involved \cite{PolymerParticle}, it becomes essential at small scales
which are in particular realized in cosmology.

\subsubsection{Matter Hamiltonian}
\label{s:IsoMatter}

We now know how the basic quantities $p$ and $c$ are quantized, and
can use the operators to construct more complicated ones. Of
particular importance, also for cosmology, are matter Hamiltonians
where now not only the matter field but also geometry is
quantized. For an isotropic geometry and a scalar, this requires us to
quantize $|p|^{-3/2}$ for the kinetic term and $|p|^{3/2}$ for the potential
term. The latter can be defined readily as $|\hat{p}|^{3/2}$, but for
the former we need an inverse power of $p$. Since $\hat{p}$ has a
discrete spectrum containing zero, a densely defined inverse does not
exist.

At this point, one has to find an alternative route to the
quantization of $d(p)=|p|^{-3/2}$, or else one could only conclude that
there is no well-defined quantization of matter Hamiltonians as a
manifestation of the classical divergence. In the case of loop quantum
cosmology it turns out, following a general scheme of the full theory
\cite{QSDV}, that one can reformulate the classical expression in an
equivalent way such that quantization becomes possible. One
possibility is to write, similarly to (\ref{ident})
\[
 d(p)=\left(\frac{1}{3\pi \gamma G}\sum_{I=1}^3\mathrm{tr} 
\left(\tau_Ih_I\{h_I^{-1},\sqrt{V}\}\right)\right)^6 
\]
where we use holonomies of isotropic connections and the volume
$V=|p|^{3/2}$. In this expression we can insert holonomies as
multiplication operators and the volume operator, and turn the Poisson
bracket into a commutator. The result
\begin{equation}
 \widehat{d(p)} = \left(8i\gamma^{-1}\ell_{\mathrm{P}}^{-2} 
(\sin{\textstyle\frac{1}{2}}c
   \sqrt{\hat{V}}
 \cos{\textstyle\frac{1}{2}}c - \cos{\textstyle\frac{1}{2}}c
 \sqrt{\hat{V}} \sin{\textstyle\frac{1}{2}}c)\right)^6
\end{equation}
is not only a densely defined operator but even bounded, which one can
easily read off from the eigenvalues \cite{InvScale}
\begin{equation}
 \widehat{d(p)}|\mu\rangle = \left(4\gamma^{-1}\ell_{\mathrm{P}}^{-2}
 (\sqrt{V_{\mu+1}}-\sqrt{V_{\mu-1}}\,)\right)^6 |\mu\rangle
\end{equation}
with $V_{\mu}=|p_{\mu}|^{3/2}$ and $p_{\mu}$ from (\ref{TriadOp}).

\begin{figure}[h]
  \centerline{\includegraphics[width=7.5cm,keepaspectratio]{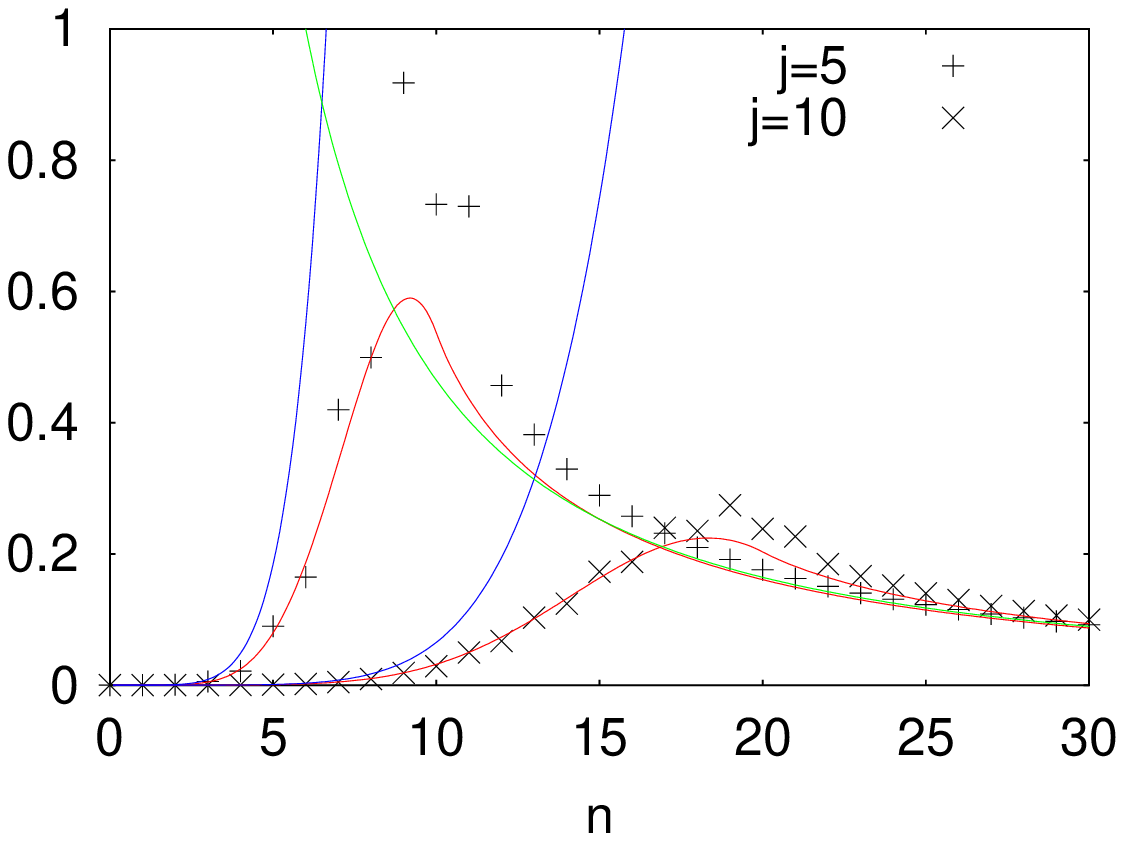} \includegraphics[width=7.5cm,keepaspectratio]{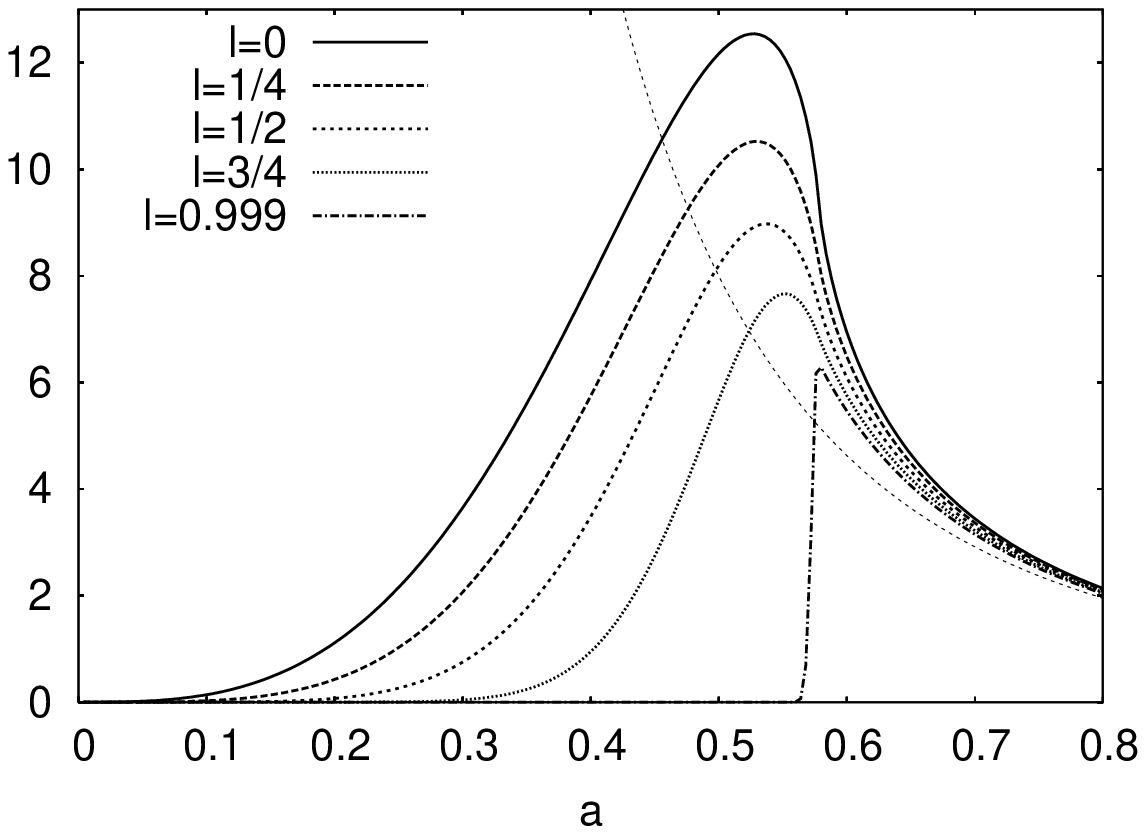}}
  \caption{\it Discrete subset of eigenvalues of $\widehat{d(p)}$
    (left) for two choices of $j$ (and $l=\frac{3}{4}$), together with
    the approximation $d(p)_{j,l}$ from (\ref{deff}) and small-$p$
    power laws.  The classical divergence at small $p$, where the
    behavior is strongly modified, is cut off. The right panel shows
    the dependence of the initial increase on $l$.}
\label{Dens}
\end{figure}

Rewriting a classical expression in such a manner can always be done in
many equivalent ways, which in general all lead to different
operators. In the case of $|p|^{-3/2}$, we highlight the choice of the
representation in which to take the trace (understood as the
fundamental representation above) and the power of $|p|$ in the
Poisson bracket ($\sqrt{V}=|p|^{3/4}$ above). This freedom can be
parameterized by two ambiguity parameters $j\in\frac{1}{2}{\mathbb N}$
for the representation and $0<l<1$ for the power such that
\[
 d(p)= \left(\frac{3}{8\pi\gamma Glj(j+1)(2
 j+1)}\sum_{I=1}^3\mathrm{tr}_j(\tau_I
 h_I\{h_I^{-1},|p|^l\})\right)^{3/(2-2l)}\,.
\]
Following the same procedure as above, we obtain eigenvalues
\cite{Ambig,ICGC}
\[  
\widehat{d(p)}^{(\mu)}_{j,l} =
\left(\frac{9}{\gamma\ell_{\mathrm{P}}^2lj(j+1)(2j+1)} \sum_{k=-j}^j
k|p_{\mu+2k}|^l\right)^{3/(2-2l)}
\]
which, for larger $j$, can be approximated by (\ref{deff}), see also
Fig.~\ref{Dens}. This provides the basis for loop cosmology as
described in Sec.~\ref{s:Effective}.

Notice that operators for the scale factor, volume or their inverse
powers do not refer to observable quantities. It can thus be
dangerous, though suggestive, to view their properties as possible
bounds on curvature. The importance of operators for inverse volume
comes from the fact that this appears in matter Hamiltonians, and thus
the Hamiltonian constraint of gravity. Properties of those operators
such as their boundedness or unboundedness can then determine the
dynamical behavior (see, e.g., \cite{DegFull}).

\subsubsection{Hamiltonian constraint}
\label{s:IsoHam}

Dynamics is controlled by the Hamiltonian constraint, which
classically gives the Friedmann equation. Since the classical
expression (\ref{ConsClass}) contains the connection component $c$, we
have to use holonomy operators. In the quantum algebra we only have
almost periodic functions at our disposal, which does not include
polynomials such as $c^2$. Quantum expressions can therefore only
coincide with the classical one in appropriate limits, which in
isotropic cosmology is realized for small extrinsic curvature, i.e.\ 
small $c$ in the flat case. We thus need an almost periodic function
of $c$ which for small $c$ approaches $c^2$. This can easily be found,
e.g., the function $\sin^2 c$. Again, the procedure is not unique
since there are many such possibilities, e.g.\ 
$\delta^{-2}\sin^2\delta c$, and more quantization ambiguities ensue.
In contrast to the density $|p|^{-3/2}$, where we also used holonomies
in the reformulation, the expressions are not equivalent to each other
classically but only in the small curvature regime. As we will discuss
shortly, the resulting new terms have the interpretation of higher
order corrections to the classical Hamiltonian.

One can restrict the ambiguities to some degree by modeling the
expression on that of the full theory. This means that one does not
simply replace $c^2$ by an almost periodic function, but uses
holonomies tracing out closed loops formed by symmetry generators
\cite{CosmoIII,IsoCosmo}. Moreover, the procedure can be embedded in a
general scheme which encompasses different models and the full theory
\cite{QSDI,CosmoIII,BlackHoles}, further reducing ambiguities. In
particular models with non-zero intrinsic curvature on their symmetry
orbits, such as the closed isotropic model, can then be included in
the construction.
One issue to keep in mind is the fact that ``holonomies'' are treated
differently in models and the full theory. In the latter case, they
are ordinary holonomies along edges, which can be shrunk and then
approximate connection components. In models, on the other hand, one
sometimes uses direct exponentials of connection components without
integration. In such a case, connection components are approximated
only when they are small; if they are not, the corresponding objects
such as the Hamiltonian constraint receive infinitely many correction
terms of higher powers in curvature (similarly to effective actions).
The difference between both ways of dealing with holonomies can be
understood in inhomogeneous models, where they are both realized for
different connection components.

In the flat case the construction is easiest, related to the Abelian
nature of the symmetry group. One can directly use the exponentials
$h_I$ in (\ref{isohol}), viewed as 3-dimensional holonomies along
integral curves, and mimic the full constraint where one follows a
loop to get curvature components of the connection $A_a^i$. Respecting
the symmetry, this can be done in the model with a square loop in two
independent directions $I$ and $J$. This yields the product
$h_Ih_Jh_I^{-1}h_J^{-1}$, which appears in a trace, as in
(\ref{FullH}), together with a commutator $h_K[h_K^{-1},\hat{V}]$
using the remaining direction $K$.  The latter, following the general
scheme of the full theory reviewed in Sec.~\ref{s:Ham}, quantizes the
contribution $\sqrt{|p|}$ to the constraint, instead of directly using
the simpler $\sqrt{|\hat{p}|}$.

Taking the trace one obtains a diagonal operator
\[
 \sin({\textstyle\frac{1}{2}}\delta c)\hat{V}
\cos({\textstyle\frac{1}{2}}\delta c)-
\cos({\textstyle\frac{1}{2}}\delta c)\hat{V}
\sin({\textstyle\frac{1}{2}}\delta c)
\]
in terms of the volume operator, as well as the multiplication
operator 
\[
 \sin^2({\textstyle\frac{1}{2}}\delta c)
\cos^2({\textstyle\frac{1}{2}}\delta c)=
\sin^2(\delta c)\,.
\]
In the triad representation where instead of
working with functions $\langle c|\psi\rangle=\psi(c)$ one works with
the coefficients $\psi_{\mu}$ in an expansion
$|\psi\rangle=\sum_{\mu}\psi_{\mu}|\mu\rangle$, this operator is the
square of a difference operator. The constraint equation thus takes
the form of a difference equation \cite{IsoCosmo,Closed,Bohr}
\begin{eqnarray} \label{DiffIso}
 (V_{\mu+5\delta}-V_{\mu+3\delta})e^{ik}\psi_{\mu+4\delta}(\phi)- 
(2+k^2\gamma^2\delta^2)
 (V_{\mu+\delta}-V_{\mu-\delta})\psi_{\mu}(\phi)&&\nonumber\\
+(V_{\mu-3\delta}-V_{\mu-5\delta})e^{-ik}\psi_{\mu-4\delta}(\phi)=
-\frac{16\pi}{3} G\gamma^3\delta^3\ell_{\mathrm{P}}^2 
\hat{H}_{\mathrm{matter}}(\mu)\psi_{\mu}(\phi)
\end{eqnarray}
for the wave function $\psi_{\mu}$ which can be viewed as an evolution
equation in internal time $\mu$. (Note that this equation is not valid
for $k=-1$ since the derivation via a Hamiltonian formulation is not
available in this case.) Thus, discrete spatial geometry implies a
discrete internal time \cite{CosmoIV}. The equation above results in
the most direct way from a non-symmetric constraint operator with
gravitational part acting as
\[
 \hat{H}|\mu\rangle=
 \frac{3}{16\pi G\gamma^3\delta^3\ell_{\mathrm{P}}^2}
 (V_{\mu+\delta}-V_{\mu-\delta}) 
 (e^{-ik}|\mu+4\delta\rangle-(2+k^2\gamma^2\delta^2)|\mu\rangle+
 e^{ik}|\mu-4\delta\rangle)\,.
\]
One can symmetrize this operator and obtain a difference equation with
different coefficients, which we do here after multiplying the
operator with $\widehat{\mathrm{sgn} p}$ for reasons that will be
discussed in the context of singularities in Sec.~\ref{s:Sing}. The
resulting difference equation is
\begin{eqnarray} \label{DiffIsoSymm}
 (|\Delta_{\delta} V|(\mu+4\delta)+|\Delta_{\delta} V|(\mu))
 e^{ik}\psi_{\mu+4\delta}(\phi)
 - 2(2+k^2\gamma^2\delta^2) 
 |\Delta_{\delta}V|(\mu)\psi_{\mu}(\phi)&&\nonumber\\
+(|\Delta_{\delta} V|(\mu-4\delta)+|\Delta_{\delta} V|(\mu))
e^{-ik}\psi_{\mu-4\delta}(\phi)
=
-\frac{32\pi}{3} G\gamma^3\delta^3\ell_{\mathrm{P}}^2 
\hat{H}_{\mathrm{matter}}(\mu)\psi_{\mu}(\phi)
\end{eqnarray}
where $|\Delta_{\delta} V|(\mu):= \mathrm{sgn}(\mu) (V_{\mu+\delta}-
V_{\mu-\delta})= |V_{\mu+\delta}- V_{\mu-\delta}|$.

Since $\sin c|\mu\rangle=-\frac{1}{2}i(|\mu+2\rangle-|\mu-2\rangle)$,
the difference equation is of higher order, even formulated on an
uncountable set, and thus has many independent solutions. Most of
them, however, oscillate on small scales, i.e.\ between $\mu$ and
$\mu+m\delta$ with small integer $m$. Others oscillate only on larger
scales and can be viewed as approximating continuum solutions. The
behavior of all the solutions leads to possibilities for selection
criteria of different versions of the constraint since there are
quantization choices. Most importantly, one chooses the routing of
edges to construct the square holonomy, again the spin of a
representation to take the trace \cite{Gaul,AmbigConstr}, and factor
ordering choices between quantizations of $c^2$ and $\sqrt{|p|}$. All
these choices also appear in the full theory such that one can draw
conclusions for preferred cases there.

When the symmetry group is not Abelian and there is non-zero intrinsic
curvature, the construction is more complicated. For non-Abelian
symmetry groups integral curves as before do not form a closed loop
and one needs a correction term related to intrinsic curvature
components \cite{IsoCosmo,Spin}. Moreover, the classical regime is not
as straightforward to specify since connection components are not
necessarily small when there is intrinsic curvature. A general scheme
encompassing intrinsic curvature, other symmetric models and the full
theory will be discussed in Sec.~\ref{s:GenCons}.

\subsubsection{Semiclassical limit and correction terms}
\label{s:IsoSemiClass}

When replacing $c^2$ by holonomies we have modified the constraint as
a function on the classical phase space. This is necessary since
otherwise the function cannot be quantized, but is different from the
quantization of densities because now the replacements are not
equivalent to the original constraint classically. Also the limit
$\lim_{\delta\to0} \delta^{-2}\sin^2\delta c$, which would give the
classical result, does not exist at the operator level.

This situation is different from the full theory, again related to the
presence of a partial background \cite{Bohr}. There, the parameter
length of edges used to construct appropriate loops is irrelevant and
thus can shrink to zero. In the model, however, changing the edge
length with respect to the background does change the operator and the
limit does not exist. Intuitively, this can be understood as follows:
The full constraint operator (\ref{FullH}) is a vertex sum obtained after
introducing a discretization of space used to choose loops
$\alpha_{IJ}$. This classical regularization sums over all tetrahedra
in the discretization, whose number diverges in the limit where the
discretization size shrinks to zero. In the quantization, however,
almost all these contributions vanish since a tetrahedron must contain
a vertex of a state in order to contribute non-trivially. The result
is independent of the discretization size once it is fine enough, and
the limit can thus be taken trivially.

In a homogeneous model, on the other hand, contributions from
different tetrahedra of the triangulation must be identical owing to
homogeneity. The coordinate size of tetrahedra drops out of the
construction in the full background independent quantization, as
emphasized in Sec.~\ref{s:Ham}, which is part of the reason for the
discretization independence. In a homogeneous configuration the number
of contributions thus increases in the limit, but their size does not
change. This results in an ill-defined limit as we have already seen
within the model itself.

The difference between models and the full theory is thus only a
consequence of the symmetry and not of different approaches. This will
also become clear later in inhomogeneous models where one obtains a
mixture between the two situations. Moreover, in the full theory one
has a situation similar to symmetric models if one does not only look
at the operator limit when the regularization is removed but also
checks the classical limit on semiclassical states. In homogeneous
models, the expression in terms of holonomies implies corrections to
the classical constraint when curvature becomes larger. This is in
analogy to other quantum field theories where effective actions
generally have higher curvature terms. In the full theory, those
correction terms can be seen when one computes expectation values of
the Hamiltonian constraint in semiclassical states peaked at classical
configurations for the connection and triad. When this classical
configuration becomes small in volume or large in curvature,
correction terms to the classical constraint arise. In this case, the
semiclassical state provides the background with respect to which
these corrections appear. In a homogeneous model, the symmetry already
provides a partial background such that correction terms can be
noticed already for the constraint operator itself.

\paragraph{WKB approximation:}

There are different procedures to make contact between the difference
equation and classical constraints. The most straightforward way is to
expand the difference operators in a Taylor series, assuming that the
wave function is sufficiently smooth. On large scales, this indeed
results in the Wheeler--DeWitt equation as a continuum limit in a
particular ordering \cite{SemiClass}. From then on, one can use the
WKB approximation or Wigner functions as usually. 

That this is possible may be surprising because as just discussed the
continuum limit $\delta\to0$ does not exist for the constraint
operator. And indeed, the limit of the constraint equation, i.e.\ the
operator applied to a wave function, does not exist in general. Even
for a wave function the limit $\delta\to0$ does not exist in general
since some solutions are sensitive to the discreteness and do not have
a continuum limit at all. When performing the Taylor expansion we
already assumed certain properties of the wave function such that the
continuum limit does exist. This then reduces the number of
independent wave functions to that present in the Wheeler--DeWitt
framework, subject to the Wheeler--DeWitt equation. That this is
possible demonstrates that the constraint in terms of holonomies does
not have problems with the classical limit.

The Wheeler--DeWitt equation results at leading order, and in addition
higher order terms arise in an expansion of difference operators in
terms of $\delta$ or $\gamma$. Similarly, after the WKB or other
semiclassical approximation there are correction terms to the
classical constraint in terms of $\gamma$ as well as $\hbar$
\cite{EffHam}.

This procedure is intuitive, but it is not suitable for inhomogeneous
models where the Wheeler--DeWitt representation becomes ill-defined.
One can evade this by performing the continuum and semiclassical limit
together. This again leads to corrections in terms of $\gamma$ as well
as $\hbar$ which are mainly of the following form \cite{DiscCorr}:
matter Hamiltonians receive corrections through the modified density
$d(p)$, and there are similar terms in the gravitational part
containing $\sqrt{|p|}$. These are purely from triad coefficients, and
similarly connection components lead to higher order corrections as
well as additional contributions summarized in a quantum geometry
potential. A possible interpretation of this potential in analogy to
the Casimir effect has been put forward in \cite{CasimirGrav}. A
related procedure to extract semiclassical properties from the
difference operator, based on the Bohmian interpretation of quantum
mechanics, has been discussed in \cite{CausalLQGCosmo}.

\paragraph{Effective formulation:}

In general, one does not only expect higher order corrections for a
gravitational action but also higher derivative terms. The situation
is then qualitatively different since not only correction terms to a
given equation arise, but also new degrees of freedom coming from
higher derivatives being independent of lower ones. In a WKB
approximation, this could be introduced by parameterizing the
amplitude of the wave function in a suitable way, but it has not been
worked out yet. An alternative approach makes use of a geometrical
formulation of quantum mechanics \cite{Schilling} which not only
provides a geometrical picture of the classical limit but also a
clear-cut procedure for computing effective Hamiltonians in analogy to
effective actions \cite{EffAc}.

Instead of using linear operators on a Hilbert space one can formulate
quantum mechanics on an infinite-dimensional phase space. This space
is directly obtained from the Hilbert space where the inner product
defines a metric as well as symplectic form on its linear vector space
(which in this way even becomes K\"ahler). This formulation brings
quantum mechanics conceptually much closer to classical physics which
also facilitates a comparison in a semiclassical analysis.

We thus obtain a quantum phase space with infinitely many degrees of
freedom, together with a flow defined by the Schr\"odinger
equation. Operators become functions on this phase space through
expectation values. Coordinates can be chosen by suitable
parameterizations of a general wave function, in particular using the
expectation values $q=\langle\hat{q}\rangle$ and
$p=\langle\hat{p}\rangle$ together with uncertainties and higher
moments. The projection $\pi\colon {\cal H}\to{\mathbb R}^2,
\psi\mapsto
(\langle\psi|\hat{q}|\psi\rangle,\langle\psi|\hat{p}|\psi\rangle)$
defines the quantum phase space as a fiber bundle over the classical
phase space with infinite-dimensional fibers. Sections of this bundle
can be defined by embedding the classical phase space into the quantum
phase space by means of suitable semiclassical states.

For a harmonic oscillator this embedding can be done by coherent
states which are preserved by the quantum evolution. This means that
the quantum flow is tangential to the embedding of the classical phase
space such that it agrees with the classical flow. The harmonic
oscillator thus does not receive quantum corrections as is well known
from effective actions for free field theories. Other systems,
however, behave in a more complicated manner where in general states
spread. This means that additional coordinates of the quantum phase
space are dynamical and may become excited. If this is the case, the
quantum flow differs from the classical flow and an effective
Hamiltonian arises with correction terms which can be computed
systematically. This effective Hamiltonian is given by the expectation
value $\langle\hat{H}\rangle$ in approximate coherent states
\cite{Perturb,Josh,SemiClassEmerge}. In these calculations one can
include higher degrees of freedom along the fibers which, through the
effective equations of motion, can be related to higher derivatives or
higher curvature in the case of gravity.

For a constrained system, such as gravity, one has to compute the
expectation value of the Hamiltonian constraint, i.e.\ first go to the
classical picture and then solve equations of motion. Otherwise, there
would simply be no effective equations left after the constraints
would already have been solved. This is the same procedure as in
standard effective actions, which one can also formulate in a
constrained manner if one chooses to parameterize time.
Indeed, also for non-constrained systems agreement between the
geometrical way to derive effective equations and standard path
integral methods has been shown for perturbations around a harmonic
oscillator \cite{EffAc}.

\subsection{Homogeneity}

A Hamiltonian formulation is available for all homogeneous models of
Bianchi class A \cite{MacCallum}, which have structure constants
$C^I_{JK}$ fulfilling $C^I_{JI}=0$. The structure constants also
determine left-invariant 1-forms $\omega^I$ in terms of which one can
write a homogeneous connection as $A_a^i=\tilde{\phi}_I^i\omega^I_a$
(see App.~\ref{s:Hom}) where all freedom is contained in the
$x$-independent $\tilde{\phi}_I^i$. A homogeneous densitized triad can
be written in a dual form with coefficients $\tilde{p}^I_i$ conjugate
to $\tilde{\phi}_I^i$. As in isotropic models, one absorbs powers of
the coordinate volume to obtain variables $\phi_I^i$ and $p^I_i$.

The kinematics is the same for all class A models, except possibly for
slight differences in the diffeomorphism constraint
\cite{AshSam,CosmoI}.  Connection components define a distinguished
triple of su(2) elements $\tilde{\phi}_I^i\tau_i$, one for each
independent direction of space.  Holonomies in those directions are
then obtained as $h_I^{(\mu_I)}=
\exp(\mu_I\phi_I^i\tau_i)\in\mathrm{SU}(2)$ with parameters $\mu_I$
for the edge lengths. Cylindrical functions depend on those
holonomies, i.e.\ are countable superpositions of terms
$f(h_1^{(\mu_1)},h_2^{(\mu_2)},h_3^{(\mu_3)})$. A basis can be written
down as spin network states
\[
 f(h_1^{(\mu_1)},h_2^{(\mu_2)},h_3^{(\mu_3)})=
 \rho_{j_1}(h_1^{(\mu_1)})^{A_1}_{B_1}
 \rho_{j_2}(h_2^{(\mu_2)})^{A_2}_{B_2}
 \rho_{j_3}(h_3^{(\mu_3)})^{A_3}_{B_3} K_{A_1A_2A_3}^{B_1B_2B_3}
\]
where the matrix $K$ specifies how the representation matrices are
contracted to a gauge invariant function of $\phi_I^i$. There are
uncountably many such states for different $\mu_I$ and thus the
Hilbert space is non-separable.  In contrast to isotropic models, the
general homogeneous theory is genuinely SU(2) and therefore not much
simpler than the full theory for individual calculations.

As a consequence of homogeneity we observe the same degeneracy as in
isotropic models where both spin and edge length appear similarly as
parameters. Spins are important to specify the contraction $K$ and
thus appear, e.g., in the volume spectrum. For this one needs to know
the spins, and it is not sufficient to consider only products
$j_I\delta_I$. On the other hand, there is still a degeneracy of spin
and edge length and keeping both $j_I$ and $\delta_I$ independent
leaves too many parameters. It is therefore more difficult to
determine what the analog of the Bohr compactification is in this
case.

\subsubsection{Diagonalization}
\label{s:Diag}

The situation simplifies if one considers diagonal models, which is
usually also done in classical considerations since it does not lead
to much loss of information. In a metric formulation, one requires the
metric and its time derivative to be diagonal, which is equivalent to
a homogeneous densitized triad
$p^I_i=p^{(I)}\Lambda^I_i$ and connection
$\phi_I^i=c_{(I)}\Lambda^i_I$ with real numbers $c_I$
and $p^I$ (where coordinate volume has been absorbed as described in
App.~\ref{s:Hom}) which are conjugate to each other,
$\{c_I,p^J\}=8\pi\gamma G\delta^J_I$, and internal directions
$\Lambda_I^i$ as in isotropic models \cite{HomCosmo}. In fact, the
kinematics becomes similar to isotropic models, except that there are
now three independent copies.  The reason for the simplification is
that we are able to separate off the gauge degrees of freedom in
$\Lambda_I^i$ from gauge invariant variables $c_I$ and $p^I$ (except
for remaining discrete gauge transformations changing the signs of two
of the $p^I$ and $c_I$ together). In a general homogenous connection,
gauge-dependent and gauge-invariant parameters are mixed together in
$\phi_I^i$ which both react differently to a change in $\mu_I$.
This makes it more difficult to discuss the structure of relevant
function spaces without assuming diagonalization.

As mentioned, the variables $p^I$ and $c_I$ are not completely gauge
invariant since a gauge transformation can flip the sign of two
components $p^I$ and $c_I$ while keeping the third fixed. There is
thus a discrete gauge group left, and only the total sign
$\mathrm{sgn}(p^1p^2p^3)$ is gauge invariant in addition to the absolute
values.

Quantization can now proceed simply by using as Hilbert space the
triple product of the isotropic Hilbert space, given by square
integrable functions on the Bohr compactification of the real
line. This results in states
$|\psi\rangle=\sum_{\mu_1,\mu_2,\mu_3}\psi_{\mu_1,\mu_2,\mu_3}
|\mu_1,\mu_2,\mu_3\rangle$ expanded in an orthonormal basis
\[
\langle c_1,c_2,c_3|\mu_1,\mu_2,\mu_3\rangle=
e^{i(\mu_1c_1+\mu_2c_2+\mu_3c_3)/2}\,.
\]
Gauge invariance under discrete gauge transformations requires
$\psi_{\mu_1,\mu_2,\mu_3}$ to be symmetric under a flip of two signs
in $\mu_I$. Without loss of generality one can thus assume that $\psi$
is defined for all real $\mu_3$ but only non-negative $\mu_1$ and $\mu_2$.

Densitized triad components are quantized by
\[
 \hat{p}^I|\mu_1,\mu_2,\mu_3\rangle= \frac{1}{2}\mu_I\gamma\ell_{\mathrm{P}}^2
 |\mu_1,\mu_2,\mu_3\rangle
\]
which directly give the volume operator
$\hat{V}=\sqrt{|\hat{p}^1\hat{p}^2\hat{p}^3|}$ with spectrum
\[
 V_{\mu_1,\mu_2,\mu_3}=
 ({\textstyle\frac{1}{2}}\gamma\ell_{\mathrm{P}}^2)^{3/2}
   \sqrt{|\mu_1\mu_2\mu_3|} \,.
\]
Moreover, after dividing out the remaining discrete gauge freedom the
only independent sign in triad components is given by the orientation
$\mathrm{sgn}(\hat{p}^1\hat{p}^2\hat{p}^3)$ which again leads to a
doubling of the metric minisuperspace with a degenerate subset in
the interior, where one of the $p^I$ vanishes.

\subsubsection{Dynamics}

The Hamiltonian constraint can be constructed in the standard manner
and its matrix elements can be computed explicitly thanks to the
simple volume spectrum. There are holonomy operators for all three
directions, and so in the triad representation the constraint equation
becomes a partial difference equation for $\psi_{\mu_1,\mu_2,\mu_3}$
in three independent variables. Its (lengthy) form can be found in
\cite{HomCosmo} for the Bianchi I model and in \cite{Spin} for all
other class A models.

Simpler cases arise in so-called locally rotationally symmetric (LRS)
models, where a non-trivial isotropy subgroup is assumed.
Here, only two independent parameters $\mu$ and $\nu$ remain, where
only one, e.g.\ $\nu$ can take both signs if discrete gauge freedom is
fixed, and the vacuum difference equation is, e.g.\ for Bianchi I,
\begin{eqnarray} \label{DiffLRS}
&& 2\delta\sqrt{|\nu+2\delta|} (\psi_{\mu+2\delta,\nu+2\delta}-
\psi_{\mu-2\delta,\nu+2\delta})\nonumber\\
&& +{\textstyle\frac{1}{2}} (\sqrt{|\nu+\delta|}-\sqrt{|\nu-\delta|})
\left((\mu+4\delta)\psi_{\mu+4\delta,\nu}-
  2\mu\psi_{\mu,\nu}+ 
(\mu-4\delta)\psi_{\mu-4\delta,\nu}\right)\nonumber\\
&&-2\delta\sqrt{|\nu-2\delta|}(\psi_{\mu+2\delta,\nu-2\delta}-
\psi_{\mu-2\delta,\nu-2\delta})\nonumber\\ 
&=& 0
\end{eqnarray}
from the non-symmetric constraint and
\begin{eqnarray} \label{DiffLRSSymm}
&& 2\delta(\sqrt{|\nu+2\delta|}+\sqrt{|\nu|}) (\psi_{\mu+2\delta,\nu+2\delta}-
\psi_{\mu-2\delta,\nu+2\delta})\nonumber\\
&& +(\sqrt{|\nu+\delta|}-\sqrt{|\nu-\delta|})
\left((\mu+2\delta)\psi_{\mu+2\delta,\nu}-
  \mu\psi_{\mu,\nu}+ 
(\mu-2\delta)\psi_{\mu-2\delta,\nu}\right)\nonumber\\
&&-2\delta(\sqrt{|\nu-2\delta|}+\sqrt{|\nu|})(\psi_{\mu+2\delta,\nu-2\delta}-
\psi_{\mu-2\delta,\nu-2\delta})\nonumber\\ 
&=& 0
\end{eqnarray}
from the symmetric version (see also \cite{BHInt}).  This leads to a
reduction between fully anisotropic and isotropic models with only two
independent variables, and provides a class of interesting systems to
analyze effects of anisotropies.

\subsection{Inhomogeneous models}

Homogeneous models provide a rich generalization of isotropic ones,
but inhomogeneities lead to stronger qualitative differences. To start
with, at least at the kinematical level one has infinitely many
degrees of freedom and is thus always dealing with field theories.
Studying field theoretical implications does not require going
immediately to the full theory since there are many inhomogeneous
models of physical interest.

We will describe some 1-dimensional models with one inhomogeneous
coordinate $x$ and two others parameterizing symmetry orbits. A
general connection is then of the form (with coordinate differentials
$\omega_y$ and $\omega_z$ depending on the symmetry)
\begin{equation} \label{Agen}
 A=A_x(x) \Lambda_x(x) \mathrm{d} x+ A_y(x) \Lambda_y(x) \omega_y+ A_z(x)
\Lambda_z(x) \omega_z+ \mbox{field independent terms}
\end{equation}
with three real functions $A_I(x)$ and three internal directions
$\Lambda_I(x)$ normalized to $\mathrm{tr}(\Lambda_I^2)=-\frac{1}{2}$
which in general are independent of each other. The situation in a
given point $x$ is thus similar to general homogeneous models with
nine free parameters.  Correspondingly, there are not many
simplifications from this general form, and one needs analogs of the
diagonalization employed for homogeneous models. What is required
mathematically for simplifications to occur is a connection with
internally perpendicular components, i.e.\ 
$\mathrm{tr}(\Lambda_I\Lambda_J)=-\frac{1}{2}\delta_{IJ}$ in each
point. This arises in different physical situations.

\subsubsection{Einstein--Rosen waves}
\label{s:ER}

One class of 1-dimensional models is given by cylindrically symmetric
gravitational waves, with connections and triads
\begin{eqnarray}
 A &=& A_x(x)\tau_3\mathrm{d} x+ (A_1(x)\tau_1+A_2(x)\tau_2)\mathrm{d} z+
(A_3(x)\tau_1+A_4(x)\tau_2)\mathrm{d}\varphi \\
 E &=& E^x(x)\tau_3\frac{\partial}{\partial x}+
(E^1(x)\tau_1+E^2(x)\tau_2)\frac{\partial}{\partial z}+
(E^3(x)\tau_1+E^4(x)\tau_2)\frac{\partial}{\partial\varphi}
\end{eqnarray}
in cylindrical coordinates. This form is more special than
(\ref{Agen}), but still not simple enough for arbitrary $A_1$, $A_2$,
$A_3$ and $A_4$. Einstein--Rosen waves
\cite{EinsteinRosen,BicakSchmidt} are a special example of cylindrical
waves subject to the polarization condition $A_2A_4+A_1A_3=0$, and
analogously for triad components.  This is just what is needed to
restrict the model to internally perpendicular connection components
and is thus analogous to diagonalization in a homogeneous model.

\paragraph{Canonical variables:}

A difference to homogeneous models, however, is that the internal
directions of a connection and a triad do not need to be identical,
which in homogeneous models with internal directions $\Lambda_I^i$ is
the case as a consequence of the Gauss constraint
$\epsilon^{ijk}\phi_I^jp^I_k=0$. With inhomogeneous fields, now, the
Gauss constraint reads
\begin{equation}
 E^{x\prime}+A_1E^2-A_2E^1+A_3E^4-A_4E^3=0
\end{equation}
or, after splitting off norms and internal directions 
\begin{eqnarray}
 A_z:=\sqrt{A_1^2+A_2^2} \quad &,&\quad A_{\varphi}:=\sqrt{A_3^2+A_4^2}\\
 \Lambda_z^A:=\frac{A_1\tau_1+A_2\tau_2}{A_z} \quad &,& \quad
\Lambda_{\varphi}^A:=\frac{A_3\tau_1+A_4\tau_2}{A_{\varphi}}
\end{eqnarray}
and analogously $E^z$, $E^{\varphi}$, $\Lambda_E^z$ and
$\Lambda_E^{\varphi}$,
\begin{equation}
 E^{x\prime}+(A_zE^z+A_{\varphi}E^{\varphi})\sin\alpha=0
\end{equation}
with $\sin\alpha:=-2\mathrm{tr}(\Lambda^A_z\Lambda_E^z\tau_3)$. If $E^x$ is
not constant, $\alpha$ cannot be zero and thus connections and triads
have different internal directions.

As a consequence, $E^z$ is not conjugate to $A_z$, anymore, and
instead the momentum of $A_z$ is $E^z\cos\alpha$ \cite{SphSymm}. This
seems to make a quantization more complicated since the momenta will
be quantized to simple flux operators, but do not directly determine
the geometry such as the volume $V=4\pi\int\mathrm{d}
x\sqrt{|E^xE^zE^{\varphi}|}$. For this one would need to know the
angle $\alpha$ which depends on both connections and triads. Moreover,
it would not be obvious how to obtain a discrete volume spectrum since
then volume does not depend only on fluxes.

It turns out that there is a simple canonical transformation which
allows one to work with canonical variables $E^z$ and $E^{\varphi}$
playing the role of momenta of $A_z\cos\alpha$ and
$A_{\varphi}\cos\alpha$ \cite{SphSymmHam}. This seems to be
undesirable, too, since now the connection variables are modified
which play an important role for holonomies. That these canonical
variables are very natural, however, follows after one considers the
structure of spin connections and extrinsic curvature tensors in this
model. The new canonical variables are then simply given by
$A_z\cos\alpha=\gamma K_z$, $A_{\varphi}\cos\alpha=\gamma
K_{\varphi}$, i.e.\ proportional to extrinsic curvature components.
Thus, in the inhomogeneous model we simply replace connection
components with extrinsic curvature in homogeneous directions (note
that $A_x$ remains unchanged) while momenta remain elementary triad
components. This is part of a broader scheme which is also important
for the Hamiltonian constraint operator (Sec.~\ref{s:GenCons}).

\paragraph{Representation:}

With the polarization condition the kinematics of the quantum theory
simplifies. Relevant holonomies are given by
$ h_e(A)=\exp (\frac{1}{2}i\smallint_e A_x(x)\mathrm{d} x)$
along edges in the 1-dimensional manifold and
\[
h_v^z(A)=\exp (i\gamma\nu_vK_z(v))\quad,\quad h_v^{\varphi}(A)=\exp
(i\gamma\mu_vK_{\varphi}(v))
\]
in vertices $v$ with real $\mu_v,\nu_v\geq0$. Cylindrical functions
depend on finitely many of those holonomies, whose edges and vertices
form a graph in the 1-dimensional manifold. Flux operators, i.e.\ 
quantized triad components, act simply by
\begin{eqnarray}
 \hat{E}^x(x) T_{g,k,\mu} &=& \frac{\gamma\ell_{\mathrm{P}}^2}{8\pi}
\frac{k_{e^+(x)}+k_{e^-(x)}}{2} T_{g,k,\mu} \label{Exspec}\\
 \int_{\cal I}\hat{E}^zT_{g,k,\mu} &=& \frac{\gamma\ell_{\mathrm{P}}^2}{4\pi}
\sum_{v\in{\cal I}} \nu_v T_{g,k,\mu}\label{Pzspec}\\ 
\int_{\cal I}\hat{E}^{\varphi}T_{g,k,\mu} &=&
\frac{\gamma\ell_{\mathrm{P}}^2}{4\pi}
\sum_{v\in{\cal I}} \mu_v T_{g,k,\mu}\label{Ppspec}\\
\end{eqnarray}
on a spin network state
\begin{eqnarray} \label{SpinNetwork}
 T_{g,k,\mu}(A) &=& \prod_{e\in g} \rho_{k_e}(h_e)\prod_{v\in
 V(g)}\rho_{\mu_v}(\gamma K_{\varphi}(v)) \rho_{\nu_v}(\gamma K_z(v))
\rho_{k_v}(\beta(v))\nonumber\\ &=& \prod_{e\in g}
 \exp\left({\textstyle\frac{1}{2}}ik_e\smallint_e A_x(x)\mathrm{d} x\right)
 \prod_{v\in 
 V(g)} e^{i\gamma\mu_v K_{\varphi}(v)} e^{i\gamma\nu_v
K_z(v)}e^{ik_v\beta(v)}
\end{eqnarray}
which also depend on the gauge angle $\beta$ determining the internal
direction of $\Lambda^E_z$. If we solve the Gauss constraint at the
quantum level, the labels $k_v$ will be such that a gauge invariant
spin network only depends on the gauge invariant combination $A_x+\beta'$.

Since triad components have simple quantizations, one can directly
combine them to get the volume operator and its spectrum
\begin{equation}\label{VolSpecInhom}
 V_{k,\mu,\nu}=\frac{\gamma^{3/2}\ell_{\mathrm{P}}^3}{4\sqrt{\pi}}
 \sum_v\sqrt{\mu_v\nu_v|k_{e^+(v)}+k_{e^-(v)}|}\,.
\end{equation}
The labels $\mu_v$ and $\nu_v$ are always non-negative, and the local
orientation is given through the sign of edge labels $k_e$.

Commutators between holonomies and the volume operator will
technically be similar to homogeneous models, except that there are
more possibilities to combine different edges. Accordingly, one can
easily compute all matrix elements of composite operators such as the
Hamiltonian constraint. The result is only more cumbersome because
there are more terms to keep track of. Again as in diagonal
homogeneous cases, the triad representation exists and one can
formulate the constraint equation there. Now, however, one has
infinitely many coupled difference equations for the wave function
since the lapse function is inhomogeneous (one obtains one difference
equation for each vertex). 

There are obvious differences to cases considered previously owing to
inhomogeneity. For instance, each edge label can take positive or
negative values, or go through zero during evolution corresponding to
the fact that a spatial slice does not need to intersect the classical
singularity everywhere. Also the structure of coefficients of the
difference equations, though qualitatively similar to homogeneous
models, is changed crucially in inhomogeneous models, mainly due to
the volume eigenvalues (\ref{VolSpecInhom}). Now, $k_{e^+}$, say, and
thus $E^x$ can be zero without volume eigenvalues in neighboring
vertices having zero volume.

\subsubsection{Spherical symmetry}

For spherically symmetric models, a connection has the form
(App.~\ref{s:SphSymm})
\begin{equation} \label{Asphsymm}
 A=A_x(x)\tau_3\mathrm{d} r+ (A_1(x)\tau_1+A_2(x)\tau_2)\mathrm{d}
 \vartheta+ (A_1(x)\tau_2-A_2(x)\tau_1)
 \sin\vartheta\mathrm{d}\varphi+ \tau_3\cos\vartheta\mathrm{d}\varphi
\end{equation}
whose field-dependent terms automatically have perpendicular internal
directions. In this case, it is not diagonalization as in the
polarization condition for Einstein--Rosen waves but a non-trivial
isotropy subgroup which leads to this property. The kinematical
quantization is then simplified as discussed before, with the only
difference being that there is only one type of vertex holonomy
\[
 h_v(A)=\exp(i\gamma\mu_vK_{\varphi}(v))
\]
as a consequence of a non-trivial isotropy subgroup.  The Hamiltonian
constraint can again be computed explicitly \cite{SphSymmHam}.

\subsubsection{Tolman--Bondi reduction}

Spherically symmetric models are usually used for applications to
non-rotating black holes, but they can also be useful for cosmological
purposes. They are particularly interesting as models for the
evolution of inhomogeneities as perturbations, which can be applied to
gravitational collapse but also cosmology. In such a context one often
reduces the spherically symmetric configuration even further by
requiring a spatial metric
\[
 \mathrm{d} s^2= q_{xx}(x,t)\mathrm{d}
 x^2+q_{\varphi\varphi}(x,t)\mathrm{d}\Omega^2 
\]
where $q_{xx}$ is related to $q_{\varphi\varphi}$ by
$\partial_x\sqrt{q_{xx}}=\sqrt{q_{\varphi\varphi}}$. One example for such a
metric is the spatial part of a flat Friedmann--Robertson--Walker
space-time, where $q_{\varphi\varphi}(x,t)=x^2a(t)^2$. This allows one to
study perturbations around a homogeneous space-time, which can also be
done at the quantum level.

\subsection{Loop inspired quantum cosmology}

The constructions described so far in this section follow all the
steps in the full theory as closely as possible. Most
importantly, one obtains quantum representations inequivalent to those
used in a Wheeler--DeWitt quantization, which results in many further
implications. This has inspired investigations where not all the steps
of loop quantum gravity are followed, but only the same type of
representation, i.e.\ the Bohr Hilbert space in an isotropic model, is
used. Other constructions, based on ADM rather than Ashtekar
variables, are then done in the most straightforward way rather than a
way suggested by the full theory \cite{BohrADM}.

In isotropic models the results are similar, but already here one can
see conceptual differences. Since the model is based on ADM variables,
in particular using the metric and not triads, it is not clear what
the additional sign factor $\mathrm{sgn}(\mu)$, which is then
introduced by hand, means geometrically. In loop quantum cosmology it
arose naturally as orientation of triads, even before its role in
removing the classical singularity, to be discussed in
Sec.~\ref{s:Sing}, had been noticed. (The necessity of having both
signs available is also reinforced independently by kinematical
consistency considerations in the full theory \cite{Flux}.)  In
homogeneous models the situation is even more complicated since sign
factors are still introduced by hand, but not all of them are removed
by discrete gauge transformations as in Sec.~\ref{s:Diag} (see
\cite{Modesto} as opposed to \cite{BHInt}).  Those models are useful
to illuminate possible effects, but they also demonstrate how new
ambiguities, even with conceptual implications, arise if guidance from
a full theory is lost.

In particular the internal time dynamics is more ambiguous in those
models and thus not usually considered. There are then only arguments
that the singularity could be avoided through boundedness of relevant
operators, but those statements are not generic in anisotropic models
\cite{Spin} or even the full theory \cite{BoundFull}. Moreover, even
if all curvature quantities could be shown to be bounded, the
evolution could still stop (as happens classically where not any
singularity is also a curvature singularity).

\subsection{Dynamics}
\label{s:Dyn}

\begin{flushright}
\begin{quote}
  {\em Because irrational numbers are always the result of calculations,
  never the result of direct measurement, might it not be possible in
  physics to abandon irrational numbers altogether and work only with
  the rational numbers? That is certainly possible, but it would be a
  revolutionary change. \ldots
  
  At some future time, when much more is known about space and time
  and the other magnitudes of physics, we may find that all of them
  are discrete.}
\end{quote}

{\sc Rudolf Carnap}

 An Introduction to the Philosophy of Science
\end{flushright}

So far we have mainly described the kinematical construction of
symmetric models in loop quantum gravity up to the point where the
Hamiltonian constraint appears. Since many dynamical issues in
different models appear in a similar fashion, we discuss them in this
section with a common background. The main feature is that dynamics is
formulated by a difference equation which by itself, compared to the
usual appearance of differential equations, implies new properties of
evolution. Depending on the model there are different classes, which
even within a given model are subject to quantization choices. Yet,
since there is a common construction procedure many characteristic
features are very general.

\subsubsection{Curvature}

Classically, curvature encodes the dynamics of geometry and does so in
quantum gravity, too. On the other hand, quantum geometry is most
intuitively understood in eigenstates of geometry, e.g.\ a triad
representation if it exists, in which curvature is unsharp. Anyway,
only solutions to the Hamiltonian constraint are relevant which in
general are peaked neither on spatial geometry nor on extrinsic
curvature. The role of curvature thus has a different, less direct
meaning in quantum gravity. Still, it is instructive to quantize
classical expressions for curvature in special situations, such as
$a^{-2}$ in isotropy. Since the resulting operator is bounded, it has
played an influential role on the development of statements regarding
the fate of classical singularities.

However, one has to keep in mind that isotropy is a very special case,
as emphasized before, and already anisotropic models shed quite a
different light on curvature quantities. Isotropy is special because
there is only one classical spatial length scale given by the scale
factor $a$, such that intrinsic curvature can only be a negative power
such as $a^{-2}$ just for dimensional reasons. That the modification
is not obvious by quantization in the model is illustrated by
comparing the intrinsic curvature term $ka^{-2}$, which remains
unmodified and thus unbounded in the purely isotropic quantization,
with the term coming from a matter Hamiltonian where the classical
divergence of $a^{-3}$ is cut off.

In an anisotropic model we do have different classical scales and thus
dimensionally also terms like $a_1a_2^{-3}$ are possible. It is then
not automatic that the quantization is bounded even if $a_2^{-3}$ were
to be bounded. As an example for such quantities consider the spatial
curvature scalar given by $W(p^1,p^2,p^3)/(p^1p^2p^3)$ with $W$ in
(\ref{BIXPot}) through the spin connection components. When quantized
and then reduced to isotropy, one does obtain a cut-off to the
intrinsic curvature term $ka^{-2}$ as mentioned in
Sec.~\ref{s:HomApp}, but the anisotropic expression remains unbounded
on minisuperspace.  The limit to vanishing triad components is
direction dependent and the isotropic case picks out a vanishing
limit. However, in general this is not the limit taken by the
dynamical trajectories. Similarly, in the full theory one can show
that inverse volume operators are not bounded even, in contrast to
anisotropic models, on states where the volume eigenvalue vanishes
\cite{BoundFull}. However, this is difficult to interpret since
nothing is known about its relevance for dynamics, and even the
geometrical role of spin labels, and thus of the configurations
considered, is unclear.

It is then quantum dynamics which is necessary to see what properties
are relevant and how degenerate configurations are approached. This
should allow one to check if the classical boundary a finite
distance away is removed in quantum gravity. This can only happen if
quantum gravity provides candidates for a region beyond the classical
singularity, and means to probe how to evolve there. The most crucial
aim is to prevent incompleteness of space-time solutions or their
quantum replacements. Even if curvature would be finite, by itself it
would not be enough since one could not tell if the singularity
persists as incompleteness. Only a demonstration of continuing
evolution can ultimately show that singularities are absent.

\subsubsection{General construction}
\label{s:GenCons}

Not all steps in the construction of the full constraint can be taken
over immediately to a model since symmetry requirements have to be
respected. It is thus important to have a more general construction
scheme which shows how generic different steps are, and whether or not
crucial input in a given symmetric situation is needed. 

We have already observed one such issue, which is the appearance of
holonomies but also simple exponentials of connection components
without integration. This is a consequence of different transformation
properties of different connection components in a reduced context.
Components along remaining inhomogeneous directions, such as $A_x$ for
Einstein--Rosen waves, play the role of connection components in the
model, giving rise to ordinary holonomies. Other components, such as
$A_z$ and $A_{\varphi}$ in Einstein--Rosen waves or all components in
homogeneous models, transform as scalars and thus only appear in
exponentials without integration. In the overall picture, we have the
full theory with only holonomies, homogeneous models with only
exponentials, and inhomogeneous models in between where both
holonomies and exponentials appear.

Another crucial issue is that of intrinsic curvature encoded in the
spin connection. In the full theory, the spin connection does not have
any covariant meaning and in fact can locally be made to vanish. In
symmetric models, however, some spin connection components can become
covariantly well-defined since not all coordinate transformations are
allowed within a model. In isotropic models, for instance, the spin
connection is simply given by a constant proportional to the curvature
parameter. Of particular importance is the spin connection when one
considers semiclassical regimes because intrinsic curvature does not
need to become small there in contrast to extrinsic curvature. Since
the Ashtekar connection mixes the spin connection and extrinsic
curvature, its semiclassical properties can be rather complicated in
symmetric models.

The full constraint is based on holonomies around closed loops in
order to approximate Ashtekar curvature components when the loop
becomes small in a continuum limit. For homogeneous directions,
however, one cannot shrink the loop and instead works with
exponentials of the components. One thus approximates the classical
components only when arguments of the exponential are small. If
these arguments were always connection components, one would not
obtain the right semiclassical properties because those components can
remain large. In models one thus has to base the construction for
homogeneous directions on extrinsic curvature components, i.e.\
subtract off the spin connection from the Ashtekar connection. For
inhomogeneous directions, on the other hand, this is not possible
since one needs a connection in order to define a holonomy.

At first sight this procedure seems rather ad hoc and even goes half a
step back to ADM variables since extrinsic curvature components are
used. However, there are several places where this procedure turns out
to be necessary for a variety of independent reasons. We have already
seen in Sec.~\ref{s:ER} that inhomogeneous models can lead to a
complicated volume operator when one insists on using all Ashtekar
connection components. When one allows for extrinsic curvature
components in the way just described, on the other hand, the volume
operator becomes straightforward. This appeared after performing a
canonical transformation which rests non-trivially on the form of
inhomogeneous spin connections and extrinsic curvature tensors.

Moreover, in addition to the semiclassical limit used above as
justification one also has to discuss local stability of the resulting
evolution equation \cite{FundamentalDisc}: Since higher order
difference equations have additional solutions, one must ensure that
they do not become dominant in order not to spoil the continuum limit.
This is satisfied with the above construction, while it is generically
violated if one were to use only connection components.

There is thus a common construction scheme available based on
holonomies and exponentials. As already discussed, this is responsible
for correction terms in a continuum limit, but also gives rise to the
constraint equation being a difference equation in a triad
representation, whenever it exists. In homogeneous models the
structure of the resulting difference equation is clear, but there are
different open possibilities in inhomogeneous models. This is
intimately related to the issue of anomalies, which also appears only
in inhomogeneous models.

With a fixed choice, one has to solve a set of coupled difference
equations for a wave function on superspace. The basic question then
always is what kind of initial or boundary value problem has to be
used in order to ensure the existence of solutions with suitable
properties, e.g.\ in a semiclassical regime. Once this is specified
one can already discuss the singularity problem since one needs to
find out if initial conditions in one semiclassical regime together
with boundary conditions away from classical singularities suffice for
a unique solution on all of superspace. A secondary question is how
this equation can be interpreted as evolution equation for the wave
function in an internal time. This is not strictly necessary and can
be complicated owing to the problem of time in general. Nevertheless,
when available, an evolution interpretation can be helpful for
interpretations.

\subsubsection{Singularities}
\label{s:Sing}

\begin{flushright}
\begin{quote}
 {\em  Il n'est rien de plus pr\'ecieux que le temps, puisque c'est le prix
  de l'\'eternit\'e.}

 ({\em There is nothing more precious than time, for
  it is the price of eternity.})
\end{quote}

{\sc Louis Bourdaloue}

Sermon sur la perte de temps

\end{flushright}

In the classical situation, we always have trajectories on superspace
running into singular submanifolds where some or all densitized triad
components vanish. In semiclassical regimes one can think of physical
solutions as wave packets following these trajectories in internal
time, but at smaller triad components spreading and deformations from
a Gaussian become stronger. Moreover, discreteness becomes essential
and properties of difference equations need to be taken into account
in order to see what is happening at the singular submanifolds.

The simplest situation is given by isotropic models where superspace
is one dimensional with coordinate $p$. Minisuperspace is thus
disconnected classically with two sides separated by the classical
singularity at $p=0$. At this point, classical energy densities
diverge and there is no well-defined initial value problem to evolve
further. (Sometimes, formal extensions of solutions beyond a classical
singularity exist \cite{HawkingEllis}, but they are never unique and
unrelated to the solution preceding the singularity. This shows that a
resolution of singularities has not only to provide a new region, but
also an evolution there uniquely from initial values at one side.) A
Wheeler--DeWitt quantization would similarly lead to diverging matter
Hamiltonian operators and the initial value problem for the wave
function generically breaks down. In isotropic loop quantum cosmology
we have already seen that the matter Hamiltonian does not have
diverging contributions from inverse metric components even at the
classical singularity. Nevertheless, the evolution could break down if
highest order coefficients in the difference equation become zero.
This indeed happens with the non-symmetric constraint (\ref{DiffIso})
or (\ref{DiffLRS}), but in these cases can be seen not to lead to any
problems: some coefficients can become zero such that the wave
function at $\mu=0$ remains undetermined by initial conditions, but
the wave function at the other side of the classical singularity is
still determined uniquely. There is no breakdown of evolution, and
thus no singularity \cite{Sing}. As one can see, this relies on
crucial properties of the loop representation with well-defined
inverse metric components and a difference rather than differential
equation \cite{CosmoIV}.

Also the structure of difference equations is important, depending on
some choices. Most important is the factor ordering or symmetrization
chosen. As just discussed, the ordering used earlier leads to
non-singular evolution but with the wave function at the classical
singularity itself remaining undetermined. In anisotropic models one
can symmetrize the constraint and obtain a difference equation, such
as (\ref{DiffLRSSymm}), whose leading order coefficients never vanish.
Evolution then never stops and even the wave function at the classical
singularity is determined. In the isotropic case, direct
symmetrization would lead to a break-down of evolution, which thus
provides an example for singular quantum evolution and demonstrates
the non-triviality of continuing evolution: The leading order
coefficient would then be
$V_{\mu-3\delta}-V_{\mu-5\delta}+V_{\mu+\delta}-V_{\mu-\delta}$ which
vanishes if and only if $\mu=2\delta$. Thus, in the backward evolution
$\psi_{-2\delta}$ remains undetermined, just as $\psi_0$ is
undetermined in the non-symmetric ordering. However, now
$\psi_{-2\delta}$ would be needed to evolve further. Since it is not
determined by initial data, one would need to prescribe this value, or
else the evolution stops. There is thus a new region at negative
$\mu$, but evolution does not continue uniquely between the two sides.
In such a case, even though curvature is bounded, the quantum system
would be singular. Similar behavior happens in other orderings such as
when triads are ordered to the left. Note that also in the full theory
one cannot order triads to the left since otherwise the constraint
would not be densely defined \cite{QSDI}.

The breakdown of the symmetric ordering in isotropic models is special
and related to the fact that all directions degenerate. The breakdown
does not happen for a symmetric ordering in anisotropic or even
inhomogeneous systems. One can avert it in isotropic cases by
multiplying the constraint with $\mathrm{sgn}\hat{p}$ before
symmetrizing, so that the additional factor of $\mathrm{sgn}\mu$ leads
to non-zero coefficients as in (\ref{DiffIsoSymm}).

This is the general scheme which also applies in more complicated
cases. The prime example for the general homogeneous behavior is given
by the Kasner evolution of the Bianchi I model. Here, the approach to
the singularity is not isotropic but given in such a way that two of
the three diagonal metric components become zero while the third one
diverges. This would lead to a different picture than described before
since the classical singularity then lies at the infinite boundary of
metric or co-triad minisuperspace. Also unlike in the isotropic case,
densities or curvature potentials are not necessarily bounded in
general as functions on minisuperspace, and the classical dynamical
approach is important.  In densitized triad variables, however, we
have a situation as before since here all components approach zero,
although at different rates. Now the classical singularity is in the
interior of minisuperspace and one can study the evolution as before,
again right through the classical singularity. Note that densitized
triad variables were required for a background independent
quantization, and now independently for non-singular evolution.

Other homogeneous models are more complicated since for them Kasner
motion takes place with a potential given by curvature components.
Approximate Kasner epochs arise when the potential is negligible,
intermitted by reflections at the potential walls where the direction
of Kasner motion in the anisotropy plane changes. Still, since in each
Kasner epoch the densitized triad components decrease, the classical
singularity remains in the interior and is penetrated by the discrete
quantum evolution.

One can use this for indications as to the general inhomogeneous
behavior by making use of the BKL scenario. If this can be justified,
in each spatial point the evolution of geometry is given by a
homogeneous model. For the quantum formulation this indicates that
also here classical singularities are removed. However, it is by no
means clear whether the BKL scenario applies at the quantum level since
even classically it is not generally established. If the scenario is
not realized (or if some matter systems can change the local
behavior), diverging $p$ are possible and the behavior would
qualitatively be very different. One thus has to study the
inhomogeneous quantum evolution directly as done before for
homogeneous cases.

In the 1-dimensional models described here classical singularities
arise when $E^x$ becomes zero. Since this is now a field, it depends
on the point $x$ on the spatial manifold where the slice hits the
classical singularity. At each such place, midisuperspace opens up to
a new region not reached by the classical evolution, where the sign of
$E^x(x)$ changes and thus the local orientation of the triad. Again,
the kinematics automatically provides us with these new regions just
as needed, and quantum evolution continues. Also, the scheme is
realized much more non-trivially, and now even the non-symmetric
ordering is not allowed. This is a consequence of the fact that
$k_e=0$ for a single edge label does not imply that neighboring volume
eigenvalues vanish. There is thus no obvious decoupling in a
non-singular manner, and it shows how less symmetric situations put
more stringent restrictions on the allowed dynamics.  Still, the
availability of other possibilities, maybe with leading coefficients
which can vanish and result in decoupling, needs to be analyzed. Most
importantly, the symmetric version still leads to non-singular
evolution even in those inhomogeneous cases which have local
gravitational degrees of freedom \cite{SphSymmSing}.

There is thus a general scheme for the removal of singularities: in
the classical situation one has singular boundaries of superspace
which cannot be penetrated. Densitized triad variables then lead to
new regions, given by a change in the orientation factor
$\mathrm{sgn}\det E$ which, however, does not help classically since
singularities remain as interior boundaries. For the quantum situation
one has to look at the constraint equation and see whether or not it
uniquely allows to continue a wave function to the other side (which
does not require time parameters even though they may be helpful if
available). This usually depends on factor ordering and other choices
which arise in the construction of constraint operators and play a
role also for the anomaly issue. One can thus fix ambiguities by
selecting a non-singular constraint if possible.  However, the
existence of non-singular versions, as realized in a natural fashion
in homogeneous models, is a highly non-trivial and by no means
automatic property of the theory showing its overall consistency.

In inhomogeneous models the issue is more complicated. We thus have a
situation where the theory, which so far is well-defined, can be
tested by trying to extend results to more general cases. It should
also be noted that different models should not require different
quantization choices unless symmetry itself is clearly responsible (as
happens with the orientation factor in the symmetric ordering for an
isotropic model, or when non-zero spin connection components receive
covariant meaning in models), but that there should rather be a common
scheme leading to non-singular behavior. This puts further strong
conditions on the construction, and is possible only if one knows how
models and the full theory are related.

\subsubsection{Initial/boundary value problems}

In isotropic models the gravitational part of the constraint
corresponds to an ordinary difference operator which can be
interpreted as generating evolution in internal time. One thus needs
to specify only initial conditions to solve the equation. The number
of conditions is large since, first, the procedure to construct the
constraint operator usually results in higher order equations and,
second, this equation relates values of a wave function $\psi_{\mu}$
defined on an uncountable set. In general, one thus has to choose a
function on a real interval unless further conditions are used.

This can be achieved, for instance, by using observables which can
reduce the kinematical framework back to wave functions defined on a
countable discrete lattice \cite{Velhinho}. Similar restrictions can
come from semiclassical properties or the physical inner product
\cite{IsoSpinFoam}, all of which has not yet been studied in
generality.

The situation in homogeneous models is similar, but now one has
several gravitational degrees of freedom only one of which is
interpreted as internal time. One has a partial difference equation
for a wave function on a minisuperspace with boundary, and initial as
well as boundary conditions are required \cite{HomCosmo}. Boundary
conditions are imposed only at non-singular parts of mini-superspace
such as $\mu=0$ in LRS models (\ref{DiffLRS}). They must not be
imposed at places of classical singularities, of course, where instead
the evolution must continue just as at any regular part.

In inhomogeneous models, then, there are not only many independent
kinematical variables but also many difference equations for only one
wave function on midisuperspace. These difference equations are of a
similar type as in homogeneous models, but they are coupled in
complicated ways. Since one has several choices in the general
construction of the constraint, there are different possibilities for
the way how difference equations arise and are coupled. Not all of
them are expected to be consistent, i.e.\ in many cases some of the
difference equations will not be compatible such that there would be
no non-zero solution at all. This is related to the anomaly
issue since the commutation behavior of difference operators is
important for properties and the existence of common solutions.

So far, the evolution operator in inhomogeneous models has not been
studied in detail, and solutions in this case remain poorly
understood. The difficulty of this issue can be illustrated by the
expectations in spherical symmetry where there is only one classical
physical degree of freedom. If this is to be reproduced for
semiclassical solutions of the quantum constraint, there must be a
subtle elimination of infinitely many kinematical degrees of freedom
such that in the end only one physical degree of freedom remains.
Thus, from the many parameters needed in general to specify a solution
to a set of difference equations, only one can remain when
compatibility relations between the coupled difference equations and
semiclassicality conditions are taken into account.

How much this cancellation depends on semiclassicality and asymptotic
infinity conditions remains to be seen. Some influence is to be
expected since classical behavior should have a bearing on the correct
reproduction of classical degrees of freedom.  However, it may also
turn out that the number of solutions to the quantum constraint is
more sensitive to quantum effects. It is already known from isotropic
models that the constraint equation can imply additional conditions
for solutions beyond the higher order difference equation, as we will
discuss in Sec.~\ref{s:DynIn}. This usually arises at the place of
classical singularities where the order of the difference equation can
change. Since the quantum behavior at classical singularities is
important here, the number of solutions can be different from the
classically expected freedom, even when combined with possible
semiclassical requirements far away from the singularity. We will now
first discuss these requirements in semiclassical regimes, followed by
more information on possibly arising additional conditions for
solutions.

\subsubsection{Pre-classicality and boundedness}
\label{s:PreClass}

The high order of difference equations implies that there are in
general many independent solutions, most of which are oscillating on
small scales, i.e.\ when the labels change only slightly. One
possibility to restrict the number of solutions then is to require
suppressed or even absent oscillations on small scales \cite{DynIn}.
Intuitively, this seems to be a pre-requisite for semiclassical
behavior and has thus been called pre-classicality. It can be
motivated by the fact that a semiclassical solution should not be
sensitive to small changes of, e.g., the volume by amounts of Planck
size. However, even though the criterion sounds intuitively
reasonable, there is so far no justification through more physical
arguments involving observables or measurement processes to extract
information from wave functions. The status of pre-classicality as a
selection criterion is thus not final.

Moreover, pre-classicality is not always consistent in all disjoint
classical regimes or with other conditions. For instance, as discussed
in the following section, there can be additional conditions on wave
functions arising from the constraint equation at the classical
singularity. Such conditions do not arise in classical regimes, but
they nevertheless have implications for the behavior of wave functions
there through the evolution equation \cite{GenFunc,GenFuncBI}.
Pre-classicality also may not be possible to impose in all
disconnected classical regimes. If the evolution equation is locally
stable, which is a basic criterion for constructing the constraint,
choosing initial values in classical regimes which do not have
small-scale oscillations guarantees that oscillations do not build up
through evolution in a classical regime \cite{FundamentalDisc}.
However, when the solution is extended through the quantum regime
around a classical singularity, oscillations do arise and do not in
general decay after a new supposedly classical regime beyond the
singularity is entered. It is thus not obvious that indeed a new
semiclassical region forms even if the quantum evolution for the wave
function is non-singular. On the other hand, evolution does continue
to large volume and macroscopic regions, which is different from other
scenarios such as \cite{StringInHom} where inhomogeneities have been
quantized on a background.

A similar issue is the boundedness of solutions, which also is
motivated intuitively by referring to the common probability
interpretation of quantum mechanics \cite{ClosedExp} but must be
supported by an analysis of physical inner products. The issue arises
in particular in classically forbidden regions where one expects
exponentially growing and decaying solutions. If a classically
forbidden region extends to infinite volume, as happens for models of
recollapsing universes, the probability interpretation would require
that only the exponentially decaying solution is realized. As before,
such a condition at large volume is in general not consistent in all
asymptotic regions or with other conditions arising in quantum
regimes.

Both issues, pre-classicality and boundedness, seem to be reasonable,
but their physical significance has to be founded on properties of the
physical inner product. They are rather straightforward to analyze in
isotropic models without matter fields, where one is dealing with
ordinary difference equations. However, other cases can be much more
complicated such that conclusions drawn from isotropic models alone
can be misleading. Moreover, numerical investigations have to be taken
with care since in particular for boundedness an exponentially
increasing contribution can easily arise from numerical errors and
dominate the exact, potentially bounded solution.

One thus needs analytical or at least semi-analytical techniques to
deal with these issues. For pre-classicality one can advantageously
use generating function techniques \cite{GenFunc} if the difference
equation is of a suitable form, e.g.\ has only coefficients with
integer powers of the discrete parameter. The generating function
$G(x):=\sum_n\psi_nx^n$ for a solution $\psi_n$ on an equidistant
lattice then solves a differential equation equivalent to the
difference equation for $\psi_n$. If $G(x)$ is known, one can use its
pole structure to get hints for the degree of oscillations in
$\psi_n$. In particular the behavior around $x=-1$ is of interest to
rule out alternating behavior where $\psi_n$ is of the form
$\psi_n=(-1)^n\xi_n$ with $\xi_n>0$ for all $n$ (or at least all $n$
larger than a certain value). At $x=-1$ we then have $G(-1)=\sum_n
\xi_n$, which is less convergent than the value for a non-alternating
solution $\psi_n=\xi_n$ resulting in $G(-1)=\sum_n(-1)^n\xi_n$.  One
can similarly find conditions for the pole structure to guarantee
boundedness of $\psi_n$, but the power of the method depends on the
form of the difference equation. More general techniques are available
for the boundedness issue, and also for alternating behavior, by
mapping the difference equation to a continued fraction which can be
evaluated analytically or numerically \cite{ContFrac}.  One can then
systematically find initial values for solutions which are guaranteed
to be bounded.

\subsubsection{Dynamical initial conditions}
\label{s:DynIn}

\begin{flushright}
\begin{quote}
  {\em I am Aton when I am alone in the Nun, but I am Re when it appears,
  in the moment when it starts to govern what it has created.}
\end{quote}

 Book of the Dead
\end{flushright}

The traditional subject of quantum cosmology is the imposition of
initial conditions for the wave function of a universe in order to
guarantee its uniqueness. In the Wheeler--DeWitt framework this is
done at the singularity $a=0$, sometimes combined with final
conditions in the classical regime. One usually uses intuitive
pictures as guidance, akin to Lemaitre's primitive atom whose decay is
supposed to have created the world, Tryon's and Vilenkin's tunneling
event from nothing, or the closure of space-time into a Euclidean
domain by Hartle and Hawking.

In the latter approaches, which have been formulated as initial
conditions for solutions of the Wheeler--DeWitt equation
\cite{tunneling,nobound}, the singularity is still present at $a=0$,
but re-interpreted as a meaningful physical event through the
conditions. In particular, the wave function is still supported at the
classical singularity, i.e.\ $\psi(0)\not=0$, in contrast to DeWitt's
original idea of requiring $\psi(0)=0$ as a means to argue for the
absence of singularities in quantum gravity \cite{DeWitt}. DeWitt's
initial condition is in fact, though most appealing conceptually, not
feasible in general since it does not lead to a well-posed initial
value formulation in more complicated models: the only solution would
then vanish identically. Zeh tried to circumvent this problem, for
instance by proposing an ad hoc Planck potential which is noticeable
only at the Planck scale and makes the initial problem well-defined
\cite{SIC}. However, the problem remains that in general there is no
satisfying origin of initial values.

In all these ideas the usual picture in physics has been taken that
there are dynamical laws describing the general behavior of a physical
system, and independently initial or boundary conditions to select a
particular situation. This is reasonable since usually one can prepare
a system, corresponding to choosing initial and boundary values, and
then study its behavior as determined by the dynamical laws. For
cosmology, however, this is not appropriate since there is no way to
prepare the universe.

At this point, there is a new possibility opened up by loop quantum
cosmology where the dynamical law and initial conditions can be part
of the same entity \cite{DynIn,BoundProp,Essay}. This is a specialty
of difference equations whose order can change locally, in contrast to
differential equations. Mathematically, such a difference equation
would be called singular since its leading order coefficients can
become zero. However, physically we have already seen that the
behavior is non-singular since the evolution does not break down.

The difference equation follows from the constraint equation, which is
the primary object in canonical quantum gravity. As discussed before,
it is usually of high order in classical regimes, where the number of
solutions can be restricted, e.g., by pre-classicality. But this at
most brings us to the number of solutions which we have for the
Wheeler--DeWitt equation such that one needs additional conditions as
in this approach. The new aspect now is that this can follow from the
constraint equation itself: since the order of the difference equation
can become smaller at the classical singularity, there are less
solutions than expected from the semiclassical behavior. In the
simplest models this is just enough to result in a unique solution up
to norm, as appropriate for a wave function describing a universe. In
those cases, the dynamical initial conditions are comparable to
DeWitt's initial condition, albeit in a manner which is well-posed
even in some cases where DeWitt's condition is not \cite{Scalar}.

In general, the issue is not clear but should be seen as a new option
presented by the discrete formulation of loop quantum cosmology. Since
there can be many conditions to be imposed on wave functions in
different regimes, one has to see in each model whether or not
suitable non-zero solutions remain at all. In fact, some first
investigations indicate that different requirements taken together can
be very restrictive \cite{GenFuncBI}, which seems to relate well with
the non-separability of the kinematical Hilbert space
\cite{PreClassBI}.  So far, only homogeneous models have been
investigated in detail, but the mechanism of decoupling is known not
to be realized in an identical manner in inhomogeneous models.

Inhomogeneous models can add qualitatively new ingredients also to the
issue of initial conditions due to the fact that there are many
coupled difference equations. There can then be consistency conditions
for solutions to the combined system which can strongly restrict the
number of independent solutions. This may be welcome, e.g.\ in
spherical symmetry where a single physical parameter remains, but the
restriction can easily become too strong even below the classically
expected one. Since the consistency between difference equations is
related to the anomaly issue, there may be an important role played by
quantum anomalies. While classically anomalies should be absent, the
quantum situation can be different since it also takes the behavior at
the classical singularity into account and is supposed to describe the
whole universe. Anomalies can then be precisely what one needs in
order to have a unique wave function of a universe even in
inhomogeneous cases where initially there is much more freedom. This
does not mean that anomalies are simply ignored or taken lightly since
it is difficult to arange having the right balance between many
solutions and no non-zero solution at all. However, quantum cosmology
suggests that it is worthwhile to have a less restricted,
unconventional view on the anomaly issue.

\subsection{Summary}

There is a general construction of a loop representation in the full
theory and its models, which is characterized by compactified
connection spaces and discrete triad operators. Strong simplifications
of some technical and conceptual steps occur in diverse models. Such a
general construction allows a view not only on the simplest case,
isotropy, but on essentially all representative systems for gravity.

Most important is the dynamics, which in the models discussed here can
be formulated by a difference equation on superspace. A general scheme
for a unique extension of wave functions through classical
singularities is realized, such that the quantum theory is
non-singular. This general argument, which has been verified in many
models, is quite powerful since it does not require detailed knowledge
of or assumptions about matter. It is independent of the availability
of a global internal time, and so the problem of time does not present
an obstacle. Moreover, a complicated discussion of quantum observables
can be avoided since once it is known that a wave function can be
continued uniquely one can extract relational information at both
sides of the classical singularity. (If observables would distinguish
both sides with their opposite oriantations, they would strongly break
parity even on large scales in contradiction with classical gravity.)
Similarly, information on the physical inner product is not required
since there is a general statement for all solutions of the constraint
equation. The uniqueness of an extension through the classical
singularity thus remains even if some solutions have to be excluded
for the physical Hilbert space or factored out if they have zero norm.

This is far from saying that observables or the physical inner product
are irrelevant for an understanding of dynamical processes. Such
constructions can, fortunately, be avoided for a general statement of
non-singular evolution in a wide class of models. For details of the
transition and to get information of the precise form of space-time at
the other side of classical singularities, however, all those objects
are necessary and conceptual problems in their context have to be
understood. 

So far, the transition has often been visualized by intuitive pictures
such as a collapsing universe turning its inside out when orientation
is reversed. An hourglass presents a picture for the importance of
discrete quantum geometry close to the classical singularity and the
emergence of continuous geometry on large scales: away from the
bottleneck of the hourglass, its sand seems to be sinking down almost
continuously. Directly at the bottleneck with its small circumference,
however, one can see that time measured by the hourglass proceeds in
discrete steps --- one grain at a time.

The main remaining issue for the mechanism to remove singularities
then is the question how the models, where it has been demonstrated,
are related to the full theory and to what extent they are
characteristic for full quantum geometry.

\section{Models within the full theory}
\label{s:Link}

\begin{flushright}
\begin{quote}
 {\em If he uses a model at all, he is always aware that it pictures only
  certain aspects of the situation and leaves out other aspects. The
  total system of physics is no longer required to be such that all
  parts of its structure can be clearly visualized. \ldots
  
  A physicist must always guard against taking a visual model as more
  than a pedagogical device or makeshift help. At the same time, he
  must also be alert to the possibility that a visual model can, and
  sometimes does, turn out to be literally accurate. Nature sometimes
  springs such surprises.}
\end{quote}

{\sc Rudolf Carnap}

 An Introduction to the Philosophy of Science
\end{flushright}

In the preceding section, the link between models and the full theory
was given through the same basic variables and kind of representation
used, as well as a general construction scheme for the Hamiltonian
constraint operator. The desired simplifications were realized thanks
to the symmetry conditions, but not too strongly since basic features
of the full theory are still recognizable in models. For instance,
even though possible in many ways and often made use of, we did not
employ special gauges or coordinate or field dependent transformations
obscuring the relation. The models are thus as close to the full
theory as possible while making full use of simplifications in order
to have explicit applications.

Still, there are always some differences not all of which are easy to
disentangle. For instance, we have discussed possible degeneracies
between spin labels and edge lengths of holonomies which can arise in
the presence of a partial background and lead to new ambiguity
parameters not present in the full theory. The question thus arises
what the precise relation between models and the full theory is, or
even how and to what extent a model for a given symmetry type can be
derived from the full theory.

This is possible for the basic representation: The symmetry and the
partial background it provides can be used to define natural
subalgebras of the full holonomy/flux algebra by using holonomies and
fluxes along symmetry generators and averaging in a suitable manner.
Since the full representation is unique and cyclic, it induces
uniquely a representation of models which is taken directly from the
full theory. This will now be described independently for states and
basic operators to provide the idea and to demonstrate the role of the
extra structure involved.  See also \cite{SphSymmSing} and
\cite{AnisoPert} for illustrations in the context of spherical
symmetry and anisotropy, respectively.

\subsection{Symmetric states}

One can imagine to construct states which are invariant under a given
action of a symmetry group on space by starting with a general
state and naively summing over all its possible translates by
elements of the symmetry group. For instance on spin network states,
the symmetry group acts by moving the graph underlying the spin
network, keeping the labels fixed. Since states with different graphs
are orthogonal to each other, the sum over uncountably many different
translates cannot be normalizable. In simple cases, such as for graphs
with a single edge along a symmetry generator, one can easily make
sense of the sum as a distribution. But this is not clear for
arbitrary states, in particular for states whose graphs have vertices,
which on the other hand would be needed for sufficient generality. A
further problem is that any such action of a symmetry group is a
subgroup of the diffeomorphism group. At least on compact space
manifolds where there are no asymptotic conditions for diffeomorphisms
in the gauge group, it then seems that any group averaged
diffeomorphism invariant state would already be symmetric with respect
to arbitrary symmetries, which is obviously not sensible.

In fact, symmetries and (gauge) diffeomorphisms are conceptually very
different, even though mathematically they are both expressed by group
actions on a space manifold. Gauge diffeomorphisms are generated by
first class constraints of the theory which in canonical quantum
gravity are imposed in the Dirac manner \cite{DirQuant} or following
refined algebraic quantization \cite{AlgQuant}, conveniently done by
group averaging \cite{Refined}. Symmetries, however, are additional
conditions imposed on a given theory to extract a particular sector of
special interest. They can also be formulated as constraints added to
the theory, but these constraints must be second class for a well
defined framework: one obtains a consistent reduced theory, e.g.\ with
a non-degenerate symplectic structure, only if configuration and
momentum variables are required to be symmetric in the same (or dual)
way.

In the case of gravity in Ashtekar variables, the symmetry type
determines, along the lines of App.~\ref{s:InvConn} the form of
invariant connections and densitized triads defining the phase space
of the reduced model. At the quantum level, however, one cannot keep
connections and triads on the same footing since a polarization is
required. One usually uses the connection representation in loop
quantum gravity such that states are functionals on the space of
connections. In a minisuperspace quantization of the classically
reduced model states would then be functionals only of invariant
connections for the given symmetry type.  This suggests to define
symmetric states in the full theory to be those states whose support
contains invariant connections as a dense subset \cite{SymmRed,PhD}
(one requires only a dense subset because possible generalized
connections must be allowed for).  As such, they must necessarily be
distributional, as already expected from the naive attempt at
construction. Symmetric states thus form a subset of the
distributional space $\mathrm{Cyl}^*$. In this manner, only the
reduced degrees of freedom are relevant, i.e.\ the reduction is
complete, and all of them are indeed realized, i.e.\ the reduction is
not too strong. Moreover, an ``averaging'' map from a non-symmetric
state to a symmetric one can easily be defined by restricting the
non-symmetric state to the space of invariant connections and
requiring it to vanish everywhere else.

This procedure defines states as functionals, but since there is no
inner product on the full $\mathrm{Cyl}^*$ this does not automatically
result in a Hilbert space. Appropriately defined subspaces of
$\mathrm{Cyl}^*$ nevertheless often carry natural inner products which
is also the case here. In fact, since the reduced space of invariant
connections can be treated by the same mathematical techniques as the
full space, it carries an analog of the full Ashtekar--Lewandowski
measure and this is indeed induced from the unique representation of
the full theory. The only difference is that in general an invariant
connection is not only determined by a reduced connection but also by
scalar fields (see App.~\ref{s:InvConn}). As in the full theory, this
space ${\cal A}_{\mathrm{inv}}$ of reduced connections and scalars is
compactified to the space $\bar{\cal A}_{\mathrm{inv}}$ of generalized
invariant connections on which the reduced Hilbert space is defined.
One thus arrives at the same Hilbert space for the subset of symmetric
states in $\mathrm{Cyl}^*$ as used before for reduced models, e.g.\ 
using the Bohr compactification in isotropic models. The new
ingredient now is that these states have meaning in the full theory as
distributions, whose evaluation on normalizable states depends on the
symmetry type and partial background structure used.

That the symmetric Hilbert space obtained in this manner is identical
to the reduced loop quantization of Sec.~\ref{s:Analog} does not
happen by definition but is a result of the procedure. The support of
a distribution is by definition a closed subset of the configuration
space, and would thus be larger than just the set of generalized
invariant connections if $\bar{\cal A}_{\mathrm{inv}}$ would not be a
closed subset in $\bar{\cal A}$. In such a case, the reduction at the
quantum level would give rise to more degrees of freedom than a loop
quantization of the classically reduced model. As shown in
\cite{SymmRed}, however, the set of invariant connections is a closed
subset of the full space of connections such that loop quantum
cosmology can be interpreted as a minisuperspace quantization.

\subsection{Basic operators}

In the classical reduction, symmetry conditions are imposed on both
connections and triads, but so far at the level of states only
connections have been taken into account. Configuration and momentum
variables play different roles in any quantum theory since a
polarization is necessary. As we based the construction on the
connection representation, symmetric triads have to be implemented at
the operator level. (There cannot be additional reduction steps at the
state level since, as we already observed, states just implement the
right number of reduced degrees of freedom.)

Classically, the reduction of phase space functions is simply done by
pull back to the reduced phase space. The flow generated by the
reduced functions then necessarily stays in the reduced phase space
and defines canonical transformations for the model. An analog
statement in the corresponding quantum theory would mean that the
reduced state space would be fixed by full operators such that their
action (or dual action on distributions) could directly be used in the
model without further work. This, however, is not the case with the
reduction performed so far. We have considered only connections in the
reduction of states, and also classically a reduction to a subspace
${\cal A}_{\mathrm{inv}}\times {\cal E}$ where connections are
invariant but not triads would be incomplete. First, this would not
define a phase space of its own with a non-degenerate symplectic
structure. More important in this context is the fact that this
subspace would not be preserved by the flow of reduced functions.

As an example (see also \cite{SphSymm} for a different discussion in
the spherically symmetric model) we consider a diagonal homogeneous
model, such as Bianchi I for simplicity, with connections of the form
$A_a^i\mathrm{d} x^a= \tilde{c}_{(I)}\Lambda^i_I\omega^I$ and look at
the flow generated by the full volume
$V=\int\mathrm{d}^3x\sqrt{\left|\det E\right|}$. It is straightforward
to evaluate the Poisson bracket
\[
 \{A_a^i(x),V\}= 2\pi\gamma G\epsilon_{abc}\epsilon^{ijk}
 E^b_jE^c_k/\sqrt{\left|\det E\right|}
\]
already used in (\ref{ident}).  A point on ${\cal
  A}_{\mathrm{inv}}\times {\cal E}$ characterized by
$\tilde{c}_{(I)}\Lambda^i_I$ and an arbitrary triad thus changes
infinitesimally by
\[
 \delta(\tilde{c}_{(I)}\Lambda^i_I)= 2\pi\gamma G\epsilon_{Ibc}\epsilon^{ijk}
 E^b_jE^c_k/\sqrt{\left|\det E\right|}
\]
which does not preserve the invariant form: First, on the right hand
side we have arbitrary fields $E$ such that
$\delta(\tilde{c}_{(I)}\Lambda^i_I)$ is not homogeneous. Second, even
if we would restrict ourselves to homogeneous $E$,
$\delta(\tilde{c}_{(I)}\Lambda^i_I)$ would not be of the original
diagonal form. This is the case only if
$\delta(\tilde{c}_{(I)}\Lambda^i_I)=\Lambda_I^i\delta(\tilde{c}_{(I)})$
since only the $\tilde{c}_I$ are canonical variables. The latter
condition is satisfied only if
\[
 \epsilon^{ijk}\Lambda_I^j\delta(\tilde{c}_{(I)}\Lambda^i_I)=
 4\pi\gamma G\epsilon_{Ibc} \Lambda_{(I)}^j
 E^b_iE^c_j/\sqrt{\left|\det E\right|}
\]
vanishes, which is not the case in general. This condition is true
only if $E^a_i\propto \Lambda^a_i$, i.e.\ if we restrict the triads to
be of diagonal homogeneous form just as the connections.

A reduction of only one part of the canonical variables is thus
incomplete and leads to a situation where most phase space functions
generate a flow which does not stay in the reduced space. Analogously,
the dual action of full operators on symmetric distributional states
does not in general map this space to itself. Thus, an arbitrary full
operator maps a symmetric state to a non-symmetric one and cannot be
used to define the reduced operator. In general, one needs a second
reduction step which implements invariant triads at the level of
operators by an appropriate projection of its action back to the
symmetric space. This can be quite complicated, and fortunately there
are special full operators adapted to the symmetry for which this step
is not necessary.

From the above example it is clear that those operators must be linear
in the momenta $E^a_i$ for otherwise one would have a triad remaining
after evaluating the Poisson bracket which on ${\cal
  A}_{\mathrm{inv}}\times {\cal E}$ would not be symmetric everywhere.
Fluxes are linear in the momenta, so we can try $p^K(z_0):=
\int_{S_{z_0}}\mathrm{d}^2y \Lambda_{(K)}^kE^a_k\omega_a^K$ where
$S_{z_0}$ is a surface in the $IJ$-plane at position $z=z_0$ in the
$K$-direction.  By choosing a surface along symmetry generators $X_I$
and $X_J$ this expression is adapted to the symmetry, even though it
is not fully symmetric yet since the position $z_0$ has to be chosen.
Again, we compute the Poisson bracket
\[
 \{A_a^i(x),p^K(z_0)\}=8\pi\gamma
 G\Lambda^i_{(K)}\int_{S_{z_0}}\delta(x,y)\omega^K_a(y)\mathrm{d}^2 y
\]
resulting in
\[
 \delta(\tilde{c}_{(I)}\Lambda_I^i)=8\pi\gamma G \Lambda^i_I
 \delta(z,z_0)\,.
\]
Also here the right hand side is not homogeneous, but we have
$\epsilon^{ijk}\Lambda_I^j\delta(\tilde{c}_{(I)}\Lambda_I^k)=0$ such
that the diagonal form is preserved. The violation of homogeneity is
expected since the flux is not homogeneous. This can easily be
remedied by ``averaging'' the flux in the $K$-direction to
\[
 p^K:=\lim_{N\to\infty} N^{-1} \sum_{\alpha=1}^N p^K(\alpha
 N^{-1}L_0)
\]
where $L_0$ is the coordinate length of the $K$-direction if it is
compact. For any finite $N$ the expression is well-defined and can
directly be quantized, and the limit can be performed in a
well-defined manner at the quantum level of the full theory. 

Most importantly, the resulting operator preserves the form of
symmetric states for the diagonal homogeneous model in its dual
action, corresponding to the flux operator of the reduced model as
used before. In averaging the full operator the partial background
provided by the group action has been used, which is responsible for
the degeneracy between edge length and spin in one reduced flux label.
Similarly, one can obtain holonomy operators along the $I$-direction
which preserve the form of symmetric states after averaging them along
the $J$ and $K$ directions (in such a way that the edge length is
variable in the averaging limit). Thus, the dual action of full
operators is sufficient to derive all basic operators of the model
from the full theory. The representation of states and basic
operators, which was seen to be responsible for most effects in loop
quantum cosmology, is thus directly linked to the full theory. This,
then, defines the cosmological sector of loop quantum gravity.

\subsection{Quantization before reduction}

When quantizing a model after a classical reduction there is much
freedom even in choosing the basic representation. For instance, in
homogeneous models one can use the Wheeler--DeWitt formulation based
on the Schr\"odinger representation of quantum mechanics. In other
models one could choose different smearings, e.g.\ treating triad
components by holonomies and connection components by fluxes, since
transformation properties can change from the reduced point of view
(see, e.g., \cite{SphSymm}).  There is thus no analog of the
uniqueness theorem of the full theory, and models constructed in this
manner would have much inherent freedom even at a basic level. With
the link to the full theory, however, properties of the unique
representation there are transferred directly to models, resulting in
analogous properties such as discrete fluxes and an action only of
exponentiated connection components. This is sufficient for a
construction by analogy of composite operators, such as the
Hamiltonian constraint according to the general scheme.

If the basic representation is taken from the full quantization, one
makes sure that many consistency conditions of quantum gravity are
already observed. This can never be guaranteed when classically
reduced models are quantized since then many consistency conditions
trivialize as a consequence of simplifications in the model. In
particular, background independence requires special properties, as
emphasized before. A symmetric model, however, always incorporates a
partial background and within a model alone one cannot determine which
structures are required for background independence. In loop quantum
cosmology, on the other hand, this is realized thanks to the link to
the full theory. Even though a model in loop quantum cosmology can
also be seen as obtained by a particular minisuperspace quantization,
it is distinguished by the fact that its representation is derived by
quantizing before performing the reduction.

In general, symmetry conditions take the form of second class
constraints since they are imposed for both connections and triads. It
is often said that second class constraints always have to be solved
classically before the quantization because of quantum uncertainty
relations. This seems to make impossible the above statement that
symmetry conditions can be imposed after quantizing. It is certainly
true that there is no state in a quantum system satisfying all second
class constraints of a given reduction. Also using distributional
states, as required for first class constraints with zero in the
continuous spectrum, does not help. The reduction described above thus
does not simply proceed in this way by finding states, normalizable or
distributional, in the full quantization. Instead, the reduction is
done at the operator algebra level, or alternatively the selection of
symmetric states is accompanied by a reduction of operators which at
least for basic ones can be performed explicitly. In general terms,
one does not look for a sub-representation of the full quantum
representation, but for a representation of a suitable subalgebra of
operators related to the symmetry. This gives a well-defined map from
the full basic representation to a new basic representation for the
model. In this map, non-symmetric degrees of freedom are removed
irrespective of the uncertainty relations from the full point of view.

Since the basic representations of the full theory and the model are
related, it is clear that similar ambiguities arise in the
construction of composite operators. Some of them are inherited
directly, such as the representation label $j$ one can choose when
connection components are represented through holonomies \cite{Gaul}.
Other ambiguities are reduced in models since many choices can result
in the same form or are restricted by adaptations to the symmetry.
This is for instance the case for positions of new vertices created by
the Hamiltonian constraint. However, also new ambiguities can arise
from degeneracies such as that between spin labels and edge lengths
resulting in the parameter $\delta$ in Sec.~\ref{s:IsoHam}. Also
factor ordering can appear more ambiguously in a model and lead to
less unique operators than in the full theory. As a simple example we
can consider a system with two degrees of freedom $(q_1,p_1;q_2,p_2)$
constrained to be equal to each other: $C_1=q_1-q_2$, $C_2=p_1-p_2$.
In the unconstrained plane $(q_1,q_2)$, angular momentum is given by
$J=q_1p_2-q_2p_1$ with an unambiguous quantization. Classically, $J$
vanishes on the constraint surface $C_1=0=C_2$, but in the quantum
system ambiguities arise: $q_1$ and $p_2$ commute before but not after
reduction. There is thus a factor ordering ambiguity in the reduction
which is absent in the unconstrained system. Since angular momentum
operators formally appear in the volume operator of loop quantum
gravity, it is not surprising that models have additional factor
ordering ambiguities in their volume operators. Fortunately, they are
harmless and result, e.g., in differences as an isotropic volume
spectrum $|\mu|^{3/2}$ compared to $\sqrt{(|\mu|-1)|\mu|(|\mu|+1)}$
where the second form \cite{CosmoII} is closer to SU(2) as compared to
U(1) expressions.

\subsection{Minisuperspace approximation}

Most physical applications in quantum gravity are obtained in mini- or
midisuperspace approximations by focusing only on degrees of freedom
relevant for a given situation of interest. Other degrees of freedom
and their interactions with the remaining ones are ignored so as to
simplify the complicated full dynamics. Their role in particular for
the evolution, however, is not always clear, and so one should check
what happens if they are gradually tuned in.

There are examples, in the spirit of \cite{MiniValid}, where
minisuperspace results are markedly different from less symmetric
ones. In those analyses, however, already the classical reduction is
unstable, or back reaction is important, and thus solutions which
start almost symmetric move away rapidly from the symmetric
submanifold of the full phase space. The failure of a minisuperspace
quantization in those cases can thus already be decided classically
and is not a quantum gravity issue. Even a violation of uncertainty
relations, which occurs in any reduction at the quantum level, is not
automatically dangerous but only if corresponding classical models are
unstable.

As for the general approach to a classical singularity, the
anisotropic behavior and not so much inhomogeneities is considered to
be essential. Isotropy can indeed be misleading, but the anisotropic
behavior is more characteristic. In fact, relevant features of full
calculations on a single vertex \cite{BoundFull} agree with the
anisotropic \cite{HomCosmo,Spin}, but not the isotropic behavior
\cite{IsoCosmo}. Also patching of homogeneous models to form an
inhomogeneous space reproduces some full results even at a
quantitative level \cite{CorrectScalar}. The main differences and
simplifications of models can be traced back to an effective
Abelianization of the full SU(2)-gauge transformations, which is not
introduced by hand in this case but a consequence of symmetries. It is
also one of the reasons why geometrical configurations in models are
usually easier to interpret than in the full theory. Most importantly,
it implies strong conceptual simplifications since it allows a triad
representation in which the dynamics can be understood more
intuitively than in a connection representation. Explicit results in
models have thus been facilitated by this property of basic variables,
and therefore a comparison with analogous situations in the full
theory is most interesting in this context, and most important as a
test of models.

If one is using a quantization of a classically reduced system, it can
only be considered a model for full quantum gravity. Relations between
different models and the full theory are important in order to specify
to what degree such models approximate the full situation, and where
additional correction terms by the ignored degrees of freedom have to
be taken into account. This is under systematic investigation in loop
quantum cosmology.

\subsection{Quantum geometry from models to the full theory}

By now, many models are available explicitly and can be compared with
each other and the full theory. Original investigations were done in
isotropic models which in many respects are special, but important
aspects of the loop quantization are now known to be realized in all
models and sometimes the full theory without contradictions so
far. There is thus a consistent picture of singularity-free dynamical
behavior together with candidates for characteristic phenomenology.

There are certainly differences between models, which can be observed
already for geometrical spectra such as area or volume. Akin to level
splitting in atoms or molecules, spectra become more complicated when
symmetry is reduced \cite{AreaOp,SphSymmVol,MG9}. Also the behavior of
densities or curvatures on arbitrary geometrical configurations can be
different in different models. In isotropic models, densities are
bounded which is a kinematical statement but in this case important
for a singularity free evolution. It is important here since
minisuperspace is just one-dimensional and so dynamical
trajectories could not pass regions of unbounded curvature should they
exist. Anisotropic models are more characteristic for the approach to
classical singularities, and here curvature expressions in general
remain unbounded if all of minisuperspace is considered. Again, this
is only kinematical, and here the dynamics tells us that evolution
does not proceed along directions of unbounded curvature. This is
similar in inhomogeneous models studied so far.

In the full theory the situation becomes again more complicated since
here densities can be unbounded even on degenerate configurations of
vanishing volume eigenvalue \cite{BoundFull}. In this case, however,
it is not known what the significance for evolution is, or even the
geometrical meaning of the degenerate configurations.

As an analogy one can, as before, take spectroscopy of atoms and level
splitting. Essential properties, such as the stability of the hydrogen
atom in quantum mechanics as opposed to the classical theory, are
unchanged if complicated interactions are taken into account. It is
important to take into account, in this context, that stability can
and does change if arbitrary interactions would be considered, rather
than realistic ones which one already fixed from other observations.
Hydrogen then remains stable under those realistic interactions, but
its properties would change drastically if any possible interaction
term would be considered. Similarly, it is not helpful to consider the
behavior of densities on arbitrary geometries unless it is known which
configurations are important for dynamics or at least their
geometrical role is clear. Dynamics in the canonical picture is
encoded in the Hamiltonian constraint, and including it (or suitable
observables) in the analysis is analogous, in the picture of atomic
spectra, to making use of realistic gravitational interaction terms.
In the full theory, such an analysis is currently beyond reach, but it
has been extensively studied in loop quantum cosmology. Since the
non-singular behavior of models, whether or not curvature is bounded,
is a consequence of basic effects and the representation derived from
the full theory, it can be taken as reliable information on the
behavior in quantum geometry.

\section{Philosophical ramifications}

In the context of loop quantum cosmology or loop quantum gravity in
general some wider issues arise which have already been touched
briefly. This has to be seen in the general context of what one should
expect from quantum theories of gravity for which there are several
quite different approaches. These issues deal with questions about the
uniqueness of theories or solutions and what information is accessible
in one universe. Also the role of time plays a more general role, and
the related question of unitarity or determinism.

\subsection{Unique theories, unique solutions}

{
\begin{flushright}
\begin{quote}
 {\em  It is often the case that, before quantitative concepts can be
  introduced into a field of science, they are preceded by comparative
  concepts that are much more effective tools for describing,
  predicting, and explaining than the cruder classificatory concepts.}
\end{quote}

{\sc Rudolf Carnap}

 An Introduction to the Philosophy of Science
\end{flushright}
}

The rise of loop quantum gravity presents an unprecedented situation
in physics where full gravity is tackled in a background independent
and non-perturbative manner. Not surprisingly, the result is often
viewed skeptically since it is very different from other well-studied
quantum field theories. Usually, intuition in quantum field theory
comes either from models which are so special that they are completely
integrable, or from perturbative expansions around free field
theories. Since no relevant ambiguities arise in this context,
ambiguities in other frameworks are usually viewed with
suspicion. A similar treatment is not possible for gravity because a
complete formulation as a perturbation series around a free theory is
unavailable and would anyway not be suitable in important situations
of high curvature. In fact, reformulations as free theories exist only
in special, non-dynamical backgrounds such as Minkowski space or planar
waves which, if used, immediately introduce a background.

If this is to be avoided in a background independent formulation, it
is necessary to deal with the full non-linear theory. This leads to
complicated expressions with factor ordering and other ambiguities
which are usually avoided in quantum field theory but not unfamiliar
from quantum theory in general. Sometimes it is said that such a
theory looses its predictive power or even suggested to stop working
on applications of the theory until all ambiguities are eliminated.
This view, of course, demonstrates a misunderstanding of the
scientific process where general effects play important roles even if
they can be quantified only at later stages. What is important is to
show that qualitative effects are robust enough such that their
implications do not crucially depend on one choice among many.

So far, applications of loop quantum gravity and cosmology are in
comparative stages where reliable effects can be derived from basic
properties and remaining ambiguities preclude sharp quantitative
predictions in general (notable exceptions are fundamental properties,
such as the computation of $\gamma$ through black hole entropy
\cite{ABCK:LoopEntro,IHEntro,Gamma,Gamma2}). These ambiguities have to
be constrained by further theoretical investigations of the overall
consistency, or by possible observations.

Ambiguities certainly mean that a theory cannot be formulated
uniquely, and uniqueness often plays a role in discussions of quantum
gravity. In the many approaches different kinds of uniqueness have
been advertised, most importantly the uniqueness of the whole theory,
or the uniqueness of a solution appropriate for the one universe we
can observe. Both expectations seem reasonable, though immodest. But
they are conceptually very different and even, maybe surprisingly,
inconsistent with each other as physical properties: For let us assume
that we have a theory from which we know that it has one and only one
solution. Provided that there is sufficient computational access to
that theory, it is falsifiable by comparing properties of the solution
with observations in the universe. Now, our observational access to
the universe will always be limited and so, even if the one solution
of our theory does agree with observations, we can always find ways to
change the theory without being in observational conflict. The theory
thus cannot be unique. Changing it in the described situation may only
violate other, external conditions which are not observable.

The converse, that a unique theory cannot have a unique solution,
follows by logically reversing the above argument. However, one has to
be careful about different notions of uniqueness of a theory. It is
clear from the above argument that uniqueness of a theory can be
realized only under external, such as mathematical, conditions which
always are a matter of taste and depend on existing knowledge.
Nevertheless, the statement seems to be supported by current
realizations of quantum gravity. String theory is one example where
the supposed uniqueness of the theory is far outweighed by the
non-uniqueness of its solutions. It should also be noted that the
uniqueness of a theory is not falsifiable, and therefore not a
scientific claim, unless its solutions are sufficiently restricted
within the theory.  Otherwise, one can always find new solutions if
one comes in conflict with observations. A theory itself, however, is
falsifiable if it implies characteristic effects for its solutions
even though it may otherwise be ambiguous.

\subsection{The role of time}

\begin{flushright}
\begin{quote}
 {\em Dies alles dauerte eine lange Zeit, oder eine kurze Zeit: denn,
  recht gesprochen, gibt es f\"ur dergleichen Dinge auf Erden {\em
    keine Zeit}.}
  
  ({\em All this took a long time, or a short time: for, strictly speaking,
  for such things {\em no time} on earth exists.})
\end{quote}

{\sc Friedrich Nietzsche}

 Thus Spoke Zarathustra
\end{flushright}

Often, time is intuitively viewed as coordinate time, i.e.\ one
direction of space-time. However, this does not have invariant
physical meaning in general relativity, and conceptually an internal
time is more appropriate. Evolution is then measured in a relational
manner of some degrees of freedom with respect to others
\cite{BergmannTime,RovelliTime,PartialCompleteObs}. In quantum
cosmology, as we have seen, this concept is even more general since
internal time keeps making sense at the quantum level also around
singularities where the classical space-time dissolves.

The wave function thus extends to a new branch beyond the classical
singularity, i.e.\ to a classically disconnected region. Intuitively
this leads to a picture of a collapsing universe preceding the big
bang, but one has to keep in mind that this is the picture obtained
from internal time where other time concepts are not available. In
such a situation it is not clear, intuitive pictures notwithstanding,
how this transition would be perceived by observers were they able to
withstand the extreme conditions. It can be said reliably that the
wave function is defined at both sides, ``before'' and ``after'', and
every computation of physical predictions, e.g.\ using observables, we
can do at ``our'' side can also be done at the other side. In this
sense, quantum gravity is free of singularities and provides a
transition between the two branches. The more complicated question is
what this means for evolution in a literal sense of our usual concept
of time (see also \cite{TimeBeforeTime}).

\begin{figure}[h]
  \centerline{\includegraphics[width=10cm,keepaspectratio]{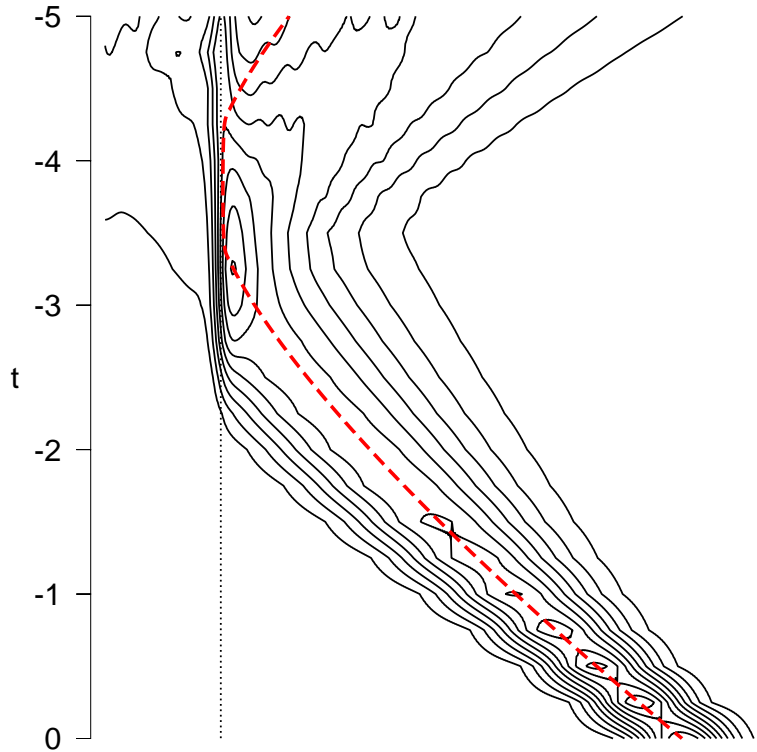}}
  \caption{\it Still of a Movie where the coordinate time evolution 
 \cite{Time} of a 
    wave packet starting at the bottom and moving toward the classical
    singularity (vertical dotted line) is shown for different values
    of an ambiguity parameter. Some part of the wave packet bounces back
    (and deforms) according to the effective classical solution
    (dashed), but other parts penetrate to negative $\mu$. The farther
    away from $a=0$ the effective bounce happens, depending on the
    ambiguity parameter, the smaller the part penetrating to negative
    $\mu$ is. The coordinate time evolution represents a physical
    state obtained after integrating over $t$ \cite{Time}. The movie is
  available from the online version \cite{LivRev} of this article at {\tt
  http://relativity.livingreviews.org/Articles/lrr-2005-11/}.}
  \label{Bounce}
\end{figure}

Effective equations displaying bounces in coordinate time evolution
indicate that indeed classical singularities are replaced by a
bouncing behavior. However, this does not occur completely generally
and does not say anything about the orientation reversal which is
characteristic for the quantum transition. In fact, effective
equations describe the motion of semiclassical wave packets, which
becomes less reliable at very small volume. And even if the effective
bounce happens far away from the classical singularity will there in
general be a part of the wave function splitting off and traversing to
the other orientation as can be seen in the example of Fig.~\ref{Bounce}.

It is not clear in general that a wave function penetrating a
classical singularity enters a new classical regime even if the volume
becomes large again. For instance, there can be oscillations on small
scales, i.e.\ violations of pre-classicality, picked up by the wave
function when it travels through the classical singularity. As
discussed in Sec.~\ref{s:PreClass}, the question of what conditions on
a wave function to require for a classical regime is still open, but
even if one can confidently say that there is such a new classical
region does the question arise if time continues during the transition
through the pure quantum regime. At least in the special model of a
free massless scalar in isotropic cosmology the answer to both
questions is affirmative, based on the availability of a physical
inner product and quantum observables in this model \cite{APS}.

Also related to this context is the question of unitary evolution.
Even if one uses a selfadjoint constraint operator, unitary evolution
is not guaranteed. First, the constraint splits into a time generator
part containing derivatives or difference operators with respect to
internal time and a source part containing, for instance, the matter
Hamiltonian. It is then not guaranteed that the time generator will
lead to unitary evolution. Secondly, it is not obvious in what inner
product to measure unitarity since the constraint is formulated in the
kinematical Hilbert space but the physical inner product is relevant
for its solutions. This shows that the usual expectation of unitary
evolution, commonly motivated by preservation of probability or
normalization of a wave function in an absolute time parameter, is not
reliable in quantum cosmology. It must be replaced by suitable
conditions on relational probabilities computed from physical wave
functions.

\subsection{Determinism}

\begin{flushright}
\begin{quote}
 {\em Hat die Zeit nicht Zeit?}
 ({\em Does time not have time?})
\end{quote}

{\sc Friedrich Nietzsche}

 Beyond Good and Evil
\end{flushright}

Loosely related to unitarity, but more general, is the concept of
determinism. This is usually weakened in quantum mechanics anyway
since in general one makes only probabilistic statements.
Nevertheless, the wave function is determined at all times by its
initial values, which is sometimes seen as the appropriate substitute
for deterministic behavior. In loop quantum cosmology the situation
again changes slightly since, as discussed in Sec.~\ref{s:DynIn}, the
wave function may not be determined by the evolution equation
everywhere, i.e.\ not at points of classical singularities, and
instead acquire new conditions on its initial values. This could be
seen as a form of indeterministic behavior, even though the values of
a wave function at classical singularities would not have any effect
on the behavior for non-degenerate configurations.\footnote{The author
  thanks Christian W\"uthrich for discussions.} (If they had such an
effect, the evolution would be singular.) In this situation one deals
with determinism in a background independent context, which requires a
new view.

In fact, rather than interpreting the freedom of choosing values at
classical singularities as indeterministic behavior, it seems more
appropriate to see this as an example for deterministic behavior in a
background independent theory. The internal time label $\mu$ first
appears as a kinematical object through the eigenvalues of the triad
operator (\ref{TriadOp}). It then plays a role in the constraint
equation (\ref{DiffIso}) when formulated in the triad representation.
Choosing internal time is just made for convenience, and it is the
constraint equation which must be used to see if this choice makes
sense in order to formulate evolution. This is indeed the case at
non-zero $\mu$ where we obtain a difference operator in the evolution
parameter. At zero $\mu$, however, the operator changes and does not
allow us to determine the wave function there from previous values.
Now, we can interpret this simply as a consequence of the constraint
equation rejecting the internal time value $\mu=0$. The background
independent evolution selects the values of internal time it needs to
propagate a wave function uniquely. As it turns out, $\mu=0$ is not
always necessary for this and thus simply decouples. In hindsight, one
could already have split off $|0\rangle$ from the kinematical Hilbert
space, thereby removing the classical singularity by hand.  Since we
did not do this, it is the evolution equation which tells us that this
is happening anyway.  Recall, however, that this is only one possible
scenario obtained from a non-symmetric constraint. For the evolution
(\ref{DiffIsoSymm}) following from the symmetric constraint, no
decoupling happens and $\mu=0$ is just like any other internal time
value.

\section{Research lines}

Currently, the development of loop quantum cosmology proceeds along
different lines, at all levels discussed before. We present here a
list of the main ones, ordered by topics rather than importance or
difficulty.

\subsection{Conceptual issues}

The list of conceptual issues is not much different from but equally
pressing as in quantum gravity in general. Here, mainly the issue of
time (its interpretation, different roles and explicit implementation
into physics), the interpretation of the wave function in quantum
theory, and technical as well as conceptual questions related to the
physical inner product need to be addressed.

\subsection{Mathematical development of models}

The main open issue, requiring new insights at all levels, is that of
inhomogeneities. While inhomogeneous models have been formulated and
partly analyzed, the following tasks are still to be completed:
\begin{description}
\item[Exact models:] In particular the dynamics of inhomogeneous
  models is much more complicated to analyze than in homogeneous ones.
  Understanding may be improved by an interesting cross-relation with
  black holes. This allows one to see if the different ingredients and
  effects of a loop quantization fit together in a complete picture,
  which so far seems to be the case
  \cite{BHInt,ModestoConn,BHPara,Horizon,SphSymmSing}. Moreover, the
  dynamics can possibly be simplified and understood better around
  slowly evolving horizons \cite{SlowHor,Horizon}. Other horizon
  conditions are also being studied in related approaches
  \cite{HWHorizon,TriangBH}.
 \item[Consistency:] Not directly related to physical applications but
   equally important is the issue of consistency of the constraints.
   The constraint algebra trivializes in homogeneous models, but is
   much more restrictive with inhomogeneities. Here, the feasibility
   of formulating a consistent theory of quantum gravity can be tested
   in a treatable situation. Related to consistency of the algebra, at
   least at a technical level, is the question of whether or not
   quantum gravity can predict initial conditions for a universe, or
   at least restrict its set of solutions.
 \item[Relation between models and the full theory:] By strengthening the
   relation between models and the full theory, ideally providing a
   complete derivation of models, physical applications will be put on
   a much firmer footing. This is also necessary to understand better
   effects of reductions such as degeneracies between different
   concepts or partial backgrounds. One aspect not realized in models
   so far is the large amount of non-Abelian effects in the full
   theory which can be significant also in models \cite{DegFull}.
 \item[Numerical quantum gravity:] Most systems of difference
   equations arising in loop quantum gravity are too complicated to
   solve exactly or even to analyze. Special techniques, such as those
   in \cite{Time,GenFunc,ContFrac,APS,PreClassBasis} have to be
   developed so as to apply to more general systems. In particular for
   including inhomogeneities, both for solving equations and
   interpreting solutions, a new area of numerical quantum gravity has
   to be developed.
 \item[Perturbations:] If the relation between different models is
   known, as presently realized for isotropic within homogeneous
   models \cite{AnisoPert}, one can formulate the less symmetric model
   perturbatively around the more symmetric one. This then provides a
   simpler formulation of the more complicated system, easing the
   analysis and uncovering new effects. In this context, also
   alternative methods to introduce approximate symmetries, based on
   coherent states as e.g.\ advocated in \cite{BoundCoh}, exist.
 \item[Effective equations:] Finding effective equations which capture
   the quantum behavior of basic difference equations at least in some
   regimes will be most helpful for a general analysis. However, their
   derivation is much more complicated for inhomogeneous systems owing
   to the consistency issue. On the other hand, trying to derive them
   will provide important tests for the framework, in addition to
   giving rise to new applications.
\end{description}

\subsection{Applications}

Once available, equations for inhomogeneous systems have the prospect
of applications such as
\begin{description}
 \item[Structure formation:] There are diverse scenarios for the early
   universe with a potential for viable structure formation, which can
   only be checked with a reliable handle on inhomogeneities. This
   applies to inflaton models with loop effects, inflation models
   without inflaton, and the generation of structure before and
   subsequent propagation through a bounce.
 \item[Robustness:] All results obtained so far have to be regarded as
   preliminary and their validity in the presence of perturbative
   inhomogeneities has to be established. A detailed analysis of their
   robustness to quantization freedom such as ambiguities or choosing
   matter fields is still to be undertaken.
 \item[New effects:] Some cosmological issues which have not been
   addressed so far from loop quantum gravity and which most likely
   require inhomogeneities are: the initial state of the inflaton
   (Gaussianity) or the present acceleration of cosmic expansion. The
   latter could be a result of small, local quantum corrections adding
   up to a sizeable effect on the whole universe. From a technical
   point of view, contact to quantum gravity phenomenology in a
   particle physics context can be made (as initially in
   \cite{CorrectScalar}).
 \item[Ans\"atze:] For the time being, those questions can be
   addressed preliminaryly by choosing suitable forms of inhomogeneous
   equations motivated by operators in full loop quantum gravity.
\end{description}

\subsection{Homogeneous models}

There are still several open areas in homogeneous models, which later
can be extended to inhomogeneous ones:
\begin{description}
 \item[Conceptual issues:] This has already been mentioned
   above. Isotropic models provide simpler settings to analyze,
   e.g., the physical inner product \cite{Golam,IsoSpinFoam,APS},
   observables, different interpretations of quantum aspects or the
   emergence of a classical world.
 \item[Effective equations:] Even in isotropic models effective
   equations have not yet been derived completely. A general scheme
   exists, shown to be analogous to standard effective action
   techniques \cite{EffAc}, but it remains to be applied completely to
   quantum cosmology. This will then lead to a complete set of
   correction terms and their ranges of validity and importance. Also
   the question of whether an effective action for quantum cosmology
   exists and what its form is can be addressed.
 \item[Matter systems:] Matter systems provide a rich source
   of diverse scenarios, but a full analysis is yet to be done. This
   includes adding different kinds of fluids \cite{LoopFluid},
   fermions or anisotropy parameters (shear term).
\end{description}

\subsection{Outlook}

All these developments will certainly also aid and suggest
developments in the full theory, and reciprocally be assisted by new
ideas realized there. At the other side, guidance as well as means for
testing can be expected from future observations.

\begin{appendix}

\section*{Appendix}

\section{Invariant connections}
\label{s:InvConn}

We first fix our notation by describing the additional structure
provided by a given action of a symmetry group on a space
manifold. This allows us to review the mathematical classification of
principal fiber bundles carrying an action of a symmetry group, and
their invariant connections.

\subsection{Partial backgrounds}

To describe a theory of connections we need to fix a principal fiber
bundle $P(\Sigma,G,\pi)$ over the analytic base manifold $\Sigma$ with
compact structure group $G$. Let $S<\mathrm{Aut}(P)$ be a Lie symmetry
subgroup of bundle automorphisms acting on the principal fiber bundle
$P$. Using the bundle projection $\pi\colon P\to\Sigma$ we get a symmetry
operation of $S$ on $\Sigma$. For simplicity we will assume that all
orbits of $S$ are of the same type. If necessary we will have to
decompose the base manifold into several orbit bundles
$\Sigma_{(F)}\subset\Sigma$, where $F\cong S_x$ is the isotropy
subgroup of $S$ consisting of elements fixing a point $x$ of the orbit
bundle $\Sigma_{(F)}$ (isotropy subgroups for different points in
$\Sigma_{(F)}$ are not identical but conjugate to each other). This
amounts to a special treatment of possible symmetry axes or centers.

By restricting ourselves to one fixed orbit bundle we fix an isotropy
subgroup $F\leq S$ up to conjugacy, and we require that the action of
$S$ on $\Sigma$ is such that the orbits are given by $S(x)\cong S/F$
for all $x\in\Sigma$. This will be the case if $S$ is compact but also
in most other cases of physical interest.  Moreover, we will have to
assume later on that the coset space $S/F$ is reductive
\cite{KobNom,KobNom2}, i.e.\ that ${\cal L}S$ can be written as a
direct sum ${\cal L}S={\cal L}F\oplus{\cal L}F_{\perp}$ with
$\mathrm{Ad}_F({\cal L}F_{\perp})\subset{\cal L}F_{\perp}$. If $S$ is
semisimple, ${\cal L}F_{\perp}$ is the orthogonal complement of ${\cal
  L}F$ with respect to the Cartan--Killing metric on ${\cal L}S$.
Further examples are provided by freely acting symmetry groups, in
which case we have $F=\{1\}$, and semidirect products of the form
$S=N\rtimes F$ where ${\cal L}F_{\perp}={\cal L}N$. The latter cases
are relevant for homogeneous and isotropic cosmological models.

The base manifold can be decomposed as $\Sigma\cong\Sigma/S\times S/F$
where $\Sigma/S\cong B\subset\Sigma$ is the base manifold of the orbit
bundle and can be realized as a submanifold $B$ of $\Sigma$ via a
section in this bundle.  As already noted in the main text, the action
of a symmetry group on space introduces a partial background into the
model. In particular, full diffeomorphism invariance is not preserved
but reduced to diffeomorphisms only on the reduced manifold $B$. To
see what kind of partial background we have in a model it is helpful
to contrast the mathematical definition of symmetry actions with the
physical picture.

To specify an action of a group on a manifold one has to give, for
each group element, a map between space points satisfying certain
conditions. Mathematically, each point is uniquely determined by
labels, usually by coordinates in a chosen (local) coordinate system.
The group action can then be written down in terms of maps of the
coordinate charts, and there are compatibility conditions for maps
expressed in different charts to ensure that the ensuing map on the
manifold is coordinate independent.  If we have active diffeomorphism
invariance, however, individual points in space are not well-defined.
This leads to the common view that geometrical observables such as the
area of a surface are, for physical purposes, not actually defined by
integrating over a submanifold simply in parameter form, but over
subsets of space defined by the values of matter fields
\cite{GeomObs1,GeomObs2}. Since matter fields are subject to
diffeomorphisms just as the metric, area defined in such a manner is
diffeomorphism invariant.

Similarly, orbits of the group action are not to be regarded as
fixed submanifolds, but as being deformed by diffeomorphisms. Fixing a
class of orbits filling the space manifold $\Sigma$ corresponds to
selecting a special coordinate system adapted to the symmetry. For
instance, in a spherically symmetric situation one usually chooses
spherical coordinates $(r,\vartheta,\varphi)$, where $r>0$ labels the
orbits and $\vartheta$ and $\varphi$ are angular coordinates and can
be identified with some parameters of the symmetry group SO(3). In a
Euclidean space the orbits can be embedded as spheres $S^2$ of
constant curvature. Applying a diffeomorphism, however, will deform
the spheres and they are in general only topological $S^2$.
Physically, the orbits can be specified as level surfaces of matter
fields, similar to specifying space points. This concept allows us to
distinguish in a diffeomorphism invariant manner between curves (such
as edges of spin networks) which are tangential and curves which are
transversal to the group orbits.

It is, however, not possible to label single points in a given orbit
in such a physical manner, simply because we could not introduce the
necessary matter fields without destroying the symmetry. Thus we have
to use the action of the symmetry group, which provides us with
additional structure, to label the points, e.g.\ by using the angular
coordinates in the example above. A similar role is played by the
embedding of the reduced manifold $B$ into $\Sigma$ by choosing a
section of the orbit bundle, which provides a base point for each
orbit (a north pole in the example of spherical symmetry). This
amounts to a partial fixing of the diffeomorphism invariance by
allowing only diffeomorphisms which respect the additional structure.
The reduced diffeomorphism constraint will then in general require
only invariance with respect to diffeomorphisms of the manifold $B$.

In a reduced model, a partial fixing of the diffeomorphism invariance
does not cause problems because all fields are constant along the
orbits anyway. However, if we study symmetric states as generalized
states of the full theory, as in Sec.~\ref{s:Link}, we inevitably have
to break partially the diffeomorphism invariance. The distributional
evaluation of symmetric states and the dual action of basic operators
thus depends on the partial background provided by the symmetry.

\subsection{Classification of symmetric principal fiber bundles}
\label{s:Class}

Fields which are invariant under the action of a symmetry group $S$ on
space $\Sigma$ are defined by a set of linear equations for invariant
field components. Nevertheless, finding invariant fields in gauge
theories is not always straightforward since, in general, fields need
to be invariant only up to gauge transformations which depend on the
symmetry transformation. An invariant connection, for instance,
satisfies the equation
\begin{equation}
 s^*A=g(s)^{-1}Ag(s)+g(s)^{-1}\mathrm{d} g(s)
\end{equation}
with a local gauge transformation $g(s)$ for each $s\in S$. These
gauge transformations are not arbitrary since two symmetry
transformations $s_1$ and $s_2$ applied one after another have to
imply a gauge transformation with $g(s_2s_1)$ related to $g(s_1)$ and
$g(s_2)$. However, this does not simply amount to a homomorphism
property and allowed maps $g\colon S\to G$ are not easily determined
by group theory. Thus, even though for a known map $g$ one simply has
to solve a system of linear equations for $A$, finding appropriate
maps $g$ can be difficult. In most cases, the equations would not have
any non-vanishing solution at all, which would certainly be
insufficient for interesting reduced field theories.

In the earlier physical literature, invariant connections and other
fields have indeed been determined by trial and error \cite{Cordero},
but the same problem has been solved in the mathematical literature
\cite{KobNom,KobNom2,Brodbeck} in impressive generality. This uses the
language of principal fiber bundles which already provides powerful
techniques. Moreover, the problem of solving one system of equations
for $A$ and $g(s)$ at the same time is split into two separate
problems which allows a more systematic approach. The first step is to
realize that a connection whose local 1-forms $A$ on $\Sigma$ are
invariant up to gauge is equivalent to a connection 1-form $\omega$
defined on the full fiber bundle $P$ which satisfies the simple
invariance conditions $s^*\omega=\omega$ for all $s\in S$. This is
indeed simpler to analyze since we now have a set of linear equations
for $\omega$ alone.  However, even though hidden in the notation, the
map $g\colon S\to G$ is still present. The invariance conditions for
$\omega$ defined on $P$ are well-defined only if we know a lift from
the original action of $S$ on the base manifold $\Sigma$ to the full
bundle $P$. As with maps $g\colon S\to G$, there are several
inequivalent choices for the lift which have to be determined. The
advantage of this procedure is that this can be done by studying
symmetric principal fiber bundles, i.e.\ principal fiber bundles
carrying the action of a symmetry group, independently of the behavior
of connections. In a second step, one can then ask what form invariant
connections on a given symmetric principal fiber bundle have.

We now discuss the first step of determining lifts of the symmetry
action of $S$ from $\Sigma$ to $P$.  Given a point $x\in\Sigma$, the
action of the isotropy subgroup $F$ yields a map
$F\colon\pi^{-1}(x)\to\pi^{-1}(x)$ of the fiber over $x$ which
commutes with the right action of $G$ on the bundle.  To each point
$p\in\pi^{-1}(x)$ we can assign a group homomorphism $\lambda_p\colon
F\to G$ defined by $f(p)=:p\cdot\lambda_p(f)$ for all $f\in F$. To
verify this we first note that commutativity of the action of
$S<\mathrm{Aut}(P)$ with right multiplication of $G$ on $P$ implies
that we have the conjugate homomorphism $\lambda_{p^{\prime}}=
\mathrm{Ad}_{g^{-1}}\circ\lambda_p$ for a different point
$p^{\prime}=p\cdot g$ in the same fiber:
\[
 p'\cdot\lambda_{p'}(f)=f(p\cdot g)=f(p)\cdot g= (p\cdot\lambda_p(f))\cdot
 g= p'\cdot\mathrm{Ad}_{g^{-1}}\lambda_p(f)\,.
\]
This yields
\[
 (f_1\circ f_2)(p)= f_1(p\cdot\lambda_p(f_2))=
 (p\cdot\lambda_p(f_2))\cdot\mathrm{Ad}_{\lambda_p(f_2)^{-1}}\lambda_p(f_1)=
p\cdot(\lambda_p(f_1)\cdot\lambda_p(f_2))
\]
demonstrating the homomorphism property. We thus obtain a map
$\lambda\colon P\times F\to G,(p,f)\mapsto\lambda_p(f)$ obeying the
relation $\lambda_{p\cdot g}=\mathrm{Ad}_{g^{-1}}\circ\lambda_p$.

Given a fixed homomorphism $\lambda\colon F\to G$, we can build
the principal fiber subbundle
\begin{equation}
 Q_{\lambda}(B,Z_{\lambda},\pi_Q):=\{p\in P_{|B}:\lambda_p=\lambda\}
\end{equation}
over the base manifold $B$ which as structure group has the
centralizer
\[
 Z_{\lambda}:=Z_G(\lambda(F))=\{g\in G: gf=fg \mbox{ for
  all }f\in \lambda(F)\}
\]
of $\lambda(F)$ in $G$.  $P_{|B}$ is the restricted fiber bundle over
$B$. A conjugate homomorphism
$\lambda^{\prime}=\mathrm{Ad}_{g^{-1}}\circ\lambda$ simply leads to an
isomorphic fiber bundle.

The structure elements $[\lambda]$ and $Q$ classify symmetric
principal fiber bundles according to the following theorem
\cite{Brodbeck}:

\begin{theo}\label{bundle}
  An $S$-symmetric principal fiber bundle $P(\Sigma,G,\pi)$ with
  isotropy subgroup $F\leq S$ of the action of $S$ on $\Sigma$ is
  uniquely characterized by a conjugacy class $[\lambda]$ of
  homomorphisms $\lambda\colon F\to G$ together with a {\em reduced bundle}
  $Q(\Sigma/S,Z_G(\lambda(F)),\pi_Q)$.
\end{theo}

Given two groups $F$ and $G$ we can make use of the relation
\cite{BroeckerDieck}
\begin{equation}\label{Hom}
 \mathrm{Hom}(F,G)/\mathrm{Ad}\cong\mathrm{Hom}(F,T(G))/W(G)
\end{equation}
in order to determine all conjugacy classes of homomorphisms
$\lambda\colon F\to G$. Here, $T(G)$ is a maximal torus and $W(G)$ the
Weyl group of $G$. Different conjugacy classes correspond to different
sectors of the theory which can be interpreted as having different
topological charge. In spherically symmetric electromagnetism, for
instance, this is just magnetic charge \cite{BFElec,SymmRed}.

\subsection{Classification of invariant connections}

Now let $\omega$ be an $S$-invariant connection on the symmetric
bundle $P$ classified by $([\lambda],Q)$, i.e.\ $s^*\omega=\omega$ for
any $s\in S$. After restriction, $\omega$ induces a connection
$\tilde{\omega}$ on the reduced bundle $Q$.  Because of $S$-invariance
of $\omega$ the reduced connection $\tilde{\omega}$ is a one-form on
$Q$ with values in the Lie algebra of the reduced structure group. To
see this, fix a point $p\in P$ and a vector $v$ in $T_p P$ such that
$\pi_*v\in\sigma_*T_{\pi(p)}B$ where $\sigma$ is the embedding of $B$
into $\Sigma$. Such a vector, which does not have components along
symmetry orbits, is fixed by the action of the isotropy group:
$\mathrm{d} f(v)=v$. The pull back of $\omega$ by $f\in F$ applied to
$v$ is by definition $f^{\star}\omega_p(v)=\omega_{f(p)}(\mathrm{d}
f(v))=\omega_{f(p)}(v)$. Now using the fact that $f$ acts as gauge
transformation in the fibers and observing the definition of
$\lambda_p$ and the adjoint transformation of $\omega$, we obtain
$\omega_{f(p)}(v)=\mathrm{Ad}_{\lambda_p(f)^{-1}}\omega_p(v)$. By
assumption the connection $\omega$ is $S$-invariant implying
$f^{\star}\omega_p(v)= \mathrm{Ad}_{\lambda_p(f)^{-1}}\omega_p(v)=
\omega_p(v)$ for all $f\in F$.  This shows that $\omega_p(v)\in {\cal
  L}Z_G(\lambda_p(F))$, and $\omega$ can be restricted to a connection
on the bundle $Q_{\lambda}$ with structure group $Z_{\lambda}$.

Furthermore, using $\omega$ we can construct the linear map
$\Lambda_p\colon{\cal L} S\to{\cal L} G,X\mapsto\omega_p(\tilde{X})$
for any $p\in P$. Here, $\tilde{X}$ is the vector field on $P$ given
by $\tilde{X}(h):= \mathrm{d} (\exp(tX)^{\star}h)/\mathrm{d} t|_{t=0}$
for any $X\in{\cal L} S$ and $h\in C^1(P,{\mathbb R})$.  For
$X\in{\cal L} F$ the vector field $\tilde{X}$ is a vertical vector
field, and we have $\Lambda_p(X)=\mathrm{d}\lambda_p(X)$ where
$\mathrm{d}\lambda\colon{\cal L} F\to{\cal L} G$ is the derivative of
the homomorphism defined above. This component of $\Lambda$ is
therefore already given by the classifying structure of the principal
fiber bundle. Using a suitable gauge, $\lambda$ can be held constant
along $B$. The remaining components $\Lambda_p|_{{\cal L}F_{\perp}}$
yield information about the invariant connection $\omega$. They are
subject to the condition
\begin{equation}\label{Higgs}
 \Lambda_p(\mathrm{Ad}_f(X))=\mathrm{Ad}_{\lambda_p(f)}(\Lambda_p(X))\quad
 \mbox{ for }f\in F,X\in{\cal L}S
\end{equation}
which follows from the transformation of $\omega$ under the
adjoint representation and which provides a set of equations
determining the form of the components $\Lambda$.

Keeping only the information characterizing $\omega$ we have, besides
$\tilde{\omega}$, the scalar field $\tilde\phi\colon Q\to{\cal L}
G\otimes{\cal L} F_{\perp}^{\star}$ which is determined by
$\Lambda_p|_{{\cal L}F_{\perp}}$ and can be regarded as having $\dim{\cal
  L}F_{\perp}$ components of ${\cal L}G$-valued scalar fields.  The
reduced connection and the scalar field suffice to characterize an
invariant connection \cite{Brodbeck}:

\begin{theo}[Generalized Wang Theorem]\label{connect}
  Let $P(\Sigma,G)$ be an $S$-symmetric principal fiber bundle
  classified by $([\lambda],Q)$ according to Theorem~\ref{bundle}, and
  let $\omega$ be an $S$-invariant connection on $P$.
  
  Then the connection $\omega$ is uniquely classified by a {\em
    reduced connection} $\tilde{\omega}$ on $Q$ and a {\em scalar
    field} $\tilde\phi\colon Q\times{\cal L} F_{\perp}\to{\cal L} G$
  obeying Eq.\ (\ref{Higgs}).
\end{theo}

In general, $\tilde\phi$ transforms under some representation of the
reduced structure group $Z_{\lambda}$: Its values lie in the subspace
of ${\cal L}G$ determined by Eq.\ (\ref{Higgs}) and form a
representation space for all group elements of $G$ (which act on
$\Lambda$) whose action preserves the subspace. These are by
definition precisely elements of the reduced group.

The connection $\omega$ can be reconstructed from its classifying
structure $(\tilde{\omega},\tilde{\phi})$ as follows: according to the
decomposition $\Sigma\cong B\times S/F$ we have
\begin{equation}\label{recons}
 \omega=\tilde{\omega}+\omega_{S/F}\,,
\end{equation}
where $\omega_{S/F}$ is given by
$\Lambda\circ\iota^{\star}\theta_{\mathrm{MC}}$ in a gauge depending
on the (local) embedding $\iota\colon S/F\hookrightarrow S$. Here
$\theta_{\mathrm{MC}}$ is the Maurer--Cartan form on $S$ taking values
in ${\cal L}S$. Through $\Lambda$, $\omega$ depends on $\lambda$ and
$\tilde{\phi}$.

\section{Examples}

With these general results we can now quickly derive the form of
invariant connections for the cases studied in the main text.

\subsection{Homogeneous models}
\label{s:Hom}

In Bianchi models the transitive symmetry group acts freely on
$\Sigma$, which implies that $\Sigma$ can locally be identified with
the group manifold $S$. The three generators of ${\cal L}S$ will be
denoted as $T_I$, $1\leq I\leq 3$, with relations
$[T_I,T_J]=C^K_{IJ}T_K$ where $C^K_{IJ}$ are the structure constants
of ${\cal L}S$ fulfilling $C^J_{IJ}=0$ for class A models by
definition. The Maurer--Cartan form on $S$ is given by
$\theta_{\mathrm{MC}}=\omega^IT_I$ with left invariant one-forms
$\omega^I$ on $S$ which fulfill the Maurer--Cartan equations
\begin{equation}\label{MC}
  \mathrm{d}\omega^I=-{\textstyle\frac{1}{2}}C^I_{JK}\omega^J\wedge\omega^K\,.
\end{equation}
Due to $F=\{1\}$ all homomorphisms $\lambda\colon F\to G$ are given by
$1\mapsto 1$, and we can use the embedding $\iota=\mathrm{id}\colon
S/F\hookrightarrow S$. An invariant connection then takes the form
$A=\tilde\phi\circ\theta_{\mathrm{MC}}=
\tilde\phi_I^i\tau_i\omega^I=A_a^i\tau_i\mathrm{d} x^a$ with matrices
$\tau_i$ generating ${\cal L}\mathrm{SU}(2)$. The scalar field is given by
$\tilde\phi\colon{\cal L}S\to{\cal
  L}G,T_I\mapsto\tilde\phi(T_I)=:\tilde\phi^i_I\tau_i$ already in its
final form, because condition (\ref{Higgs}) is empty for a trivial
isotropy group.

Using left invariant vector fields $X_I$ obeying
$\omega^I(X_J)=\delta^I_J$ and with Lie brackets
$[X_I,X_J]=C^K_{IJ}X_K$ the momenta canonically conjugate to
$A_a^i=\tilde\phi^i_I\omega^I_a$ can be written as
$E^a_i=\sqrt{g_0}\,\tilde{p}^I_iX^a_I$ with $\tilde{p}^I_i$ being canonically
conjugate to $\tilde\phi^i_I$. Here, $g_0=\det(\omega^I_a)^2$ is the
determinant of the left invariant metric
$(g_0)_{ab}:=\sum_I\omega^I_a\omega^I_b$ on $\Sigma$ which is used to
provide the density weight of $E^a_i$. The symplectic structure can be
derived from
\[
  \frac{1}{8\pi\gamma G}\int_{\Sigma}\mathrm{d}^3x\,\dot{A}^i_aE^a_i=
  \frac{1}{8\pi\gamma G}\int_{\Sigma}\mathrm{d}^3x\,\sqrt{g_0}\,
  \dot{\tilde\phi}{}^i_I \tilde{p}^J_i \omega^I(X_J)=
  \frac{V_0}{8\pi\gamma G}\dot{\tilde\phi}{}^i_I\tilde{p}^I_i\, ,
\]
to obtain
\begin{equation}
  \{\tilde\phi^i_I,\tilde{p}^J_j\}=8\pi\gamma G V_0\delta^i_j\delta^J_I
\end{equation}
with the volume $V_0:=\int_{\Sigma}\mathrm{d}^3x\sqrt{g_0}$ of $\Sigma$
measured with the invariant metric $g_0$. 

It is convenient to absorb the coordinate volume $V_0$ into the fields
by redefining $\phi^i_I:=V_0^{1/3}\tilde\phi^i_I$ and
$p_i^I:=V_0^{2/3}\tilde{p}^I_i$. This makes the symplectic structure
independent of $V_0$ in accordance with background independence. These
redefined variables automatically appear in holonomies and fluxes
through coordinate integrations.

\subsection{Isotropic models}
\label{s:Iso}

On Bianchi models additional symmetries can be imposed which
corresponds to a further symmetry reduction and introduces non-trivial
isotropy subgroups. These models with enhanced symmetry can be treated
on an equal footing by writing the symmetry group as a semidirect
product $S=N\rtimes_{\rho}F$, with the isotropy subgroup $F$ and the
translational subgroup $N$ which is one of the Bianchi groups.
Composition in this group is defined as
$(n_1,f_1)(n_2,f_2):=(n_1\rho(f_1)(n_2),f_1f_2)$ which depends on the
group homomorphism $\rho\colon F\to\mathrm{Aut} N$ into the
automorphism group of $N$ (which will be denoted
by the same letter as the representation on $\mathrm{Aut} {\cal L}N$
used below). Inverse elements are given by
$(n,f)^{-1}=(\rho(f^{-1})(n^{-1}),f^{-1})$. To determine the form of
invariant connections we have to compute the Maurer--Cartan form on
$S$ (using the usual notation):
\begin{eqnarray}\label{MaurerCartan}
  \theta^{(S)}_{\mathrm{MC}}(n,f) & = & (n,f)^{-1}\mathrm{d}
  (n,f)=(\rho(f^{-1})(n^{-1}),f^{-1})(\mathrm{d} n,\mathrm{d} f)\nonumber\\
  & = & (\rho(f^{-1})(n^{-1})\rho(f^{-1})(\mathrm{d} n),f^{-1}\mathrm{d} f)=
  (\rho(f^{-1})(n^{-1}\mathrm{d} n),f^{-1}\mathrm{d} f)\nonumber\\
  & = & \left(\rho(f^{-1})(\theta^{(N)}_{\mathrm{MC}}(n)),
    \theta^{(F)}_{\mathrm{MC}}(f)\right)\,.
\end{eqnarray}
Here the Maurer--Cartan forms $\theta^{(N)}_{\mathrm{MC}}$ on $N$ and
$\theta^{(F)}_{\mathrm{MC}}$ on $F$ appear. We then choose an
embedding $\iota\colon S/F=N\hookrightarrow S$, which can most easily
be done as $\iota\colon n\mapsto (n,1)$. Thus,
$\iota^*\theta^{(S)}_{\mathrm{MC}}=\theta^{(N)}_{\mathrm{MC}}$, and a
reconstructed connection takes the form
$\tilde\phi\circ\iota^*\theta^{(S)}_{\mathrm{MC}}=
\tilde\phi^i_I\omega^I\tau_i$ which is the same as for anisotropic
models before (where now $\omega^I$ are left invariant one-forms on
the translation group $N$).  However, here $\tilde{\phi}$ is
constrained by equation (\ref{Higgs}) and we get only a subset as
isotropic connections.

To solve equation (\ref{Higgs}) we have to treat LRS (locally rotationally
symmetric) models with a single rotational symmetry and isotropic
models separately. In the first case we choose ${\cal
  L}F=\langle\tau_3\rangle$, whereas in the second case we have ${\cal
  L}F=\langle\tau_1,\tau_2,\tau_3\rangle$ ($\langle\cdot\rangle$
denotes the linear span). Eq.\ (\ref{Higgs}) can be written
infinitesimally as
\[
\tilde\phi(\mathrm{ad}_{\tau_i}(T_I))=
\mathrm{ad}_{\mathrm{d}\lambda(\tau_i)}\tilde\phi(T_I)=
[\mathrm{d}\lambda(\tau_i),\tilde\phi(T_I)]
\]
($i=3$ for LRS, $1\leq i\leq 3$ for isotropy).  The $T_I$ are
generators of ${\cal L}N={\cal L}F_{\perp}$, on which the isotropy
subgroup $F$ acts by rotation,
$\mathrm{ad}_{\tau_i}(T_I)=\epsilon_{iIK}T_K$.  This is the derivative
of the representation $\rho$ defining the semidirect product $S$:
conjugation on the left hand side of (\ref{Higgs}) is
$\mathrm{Ad}_{(1,f)}(n,1)=(1,f)(n,1)(1,f^{-1})=(\rho(f)(n),1)$, which
follows from the composition in $S$.

Next, we have to determine the possible conjugacy classes of
homomorphisms $\lambda\colon F\to G$. For LRS models their
representatives are given by 
\[
 \lambda_k\colon \mathrm{U}(1)\to
 \mathrm{SU}(2),\exp t\tau_3\mapsto\exp kt\tau_3
\]
for $k\in{\mathbb N}_0=\{0,1,\ldots\}$ (as will be shown in detail
below for spherically symmetric connections).  For the components
$\tilde\phi^i_I$ of $\tilde\phi$ defined by
$\tilde\phi(T_I)=\tilde\phi^i_I\tau_i$, equation (\ref{Higgs}) takes
the form $\epsilon_{3IK}\tilde\phi^j_K=k\epsilon_{3lj}\tilde\phi^l_I$.
This has a non-trivial solution only for $k=1$, in which case
$\tilde\phi$ can be written as
\[
  \tilde\phi_1=\tilde{a}\tau_1+\tilde{b}\tau_2\quad,\quad
  \tilde\phi_2=-\tilde{b}\tau_1+\tilde{a}\tau_2\quad,
  \quad\tilde\phi_3=\tilde{c}\tau_3
\]
with arbitrary numbers $\tilde{a}$, $\tilde{b}$, $\tilde{c}$ (the
factors of $2^{-\frac{1}{2}}$ are introduced for the sake of
normalization). Their conjugate momenta take the form
\[
  \tilde{p}^1={\textstyle\frac{1}{2}} (\tilde{p}_a\tau_1+\tilde{p}_b\tau_2)
 \quad,\quad
  \tilde{p}^2={\textstyle\frac{1}{2}} (-\tilde{p}_b\tau_1+\tilde{p}_a\tau_2)
  \quad,
  \quad \tilde{p}^3=\tilde{p}_c\tau_3\,,
\]
and the symplectic structure is given by
\[
  \{\tilde{a},\tilde{p}_a\}=\{\tilde{b},\tilde{p}_b\}=
 \{\tilde{c},\tilde{p}_c\}=8\pi\gamma G V_0
\]
and vanishes in all other cases. There is remaining gauge freedom from
the reduced structure group $Z_{\lambda}\cong\mathrm{U}(1)$ which
rotates the pairs $(\tilde{a},\tilde{b})$ and
$(\tilde{p}_a,\tilde{p}_b)$. Gauge invariant are then only
$\sqrt{\tilde{a}^2+\tilde{b}^2}$ and its momentum
$(\tilde{a}\tilde{p}_a+\tilde{b}\tilde{p}_b)/\sqrt{\tilde{a}^2+\tilde{b}^2}$.

In the case of isotropic models we have only two
homomorphisms
 $\lambda_0\colon \mathrm{SU}(2)\to
 \mathrm{SU}(2),f\mapsto 1$
and $\lambda_1=\mathrm{id}$ up to conjugation (to simplify notation we
use the same letters for the homomorphisms as in the LRS case, which
is justified by the fact that the LRS homomorphisms are restrictions
of those appearing here).  Equation (\ref{Higgs}) takes the form
$\epsilon_{iIK}\tilde\phi^j_K=0$ for $\lambda_0$ without non-trivial
solutions, and $\epsilon_{iIK}\tilde\phi^j_K=
\epsilon_{ilj}\tilde\phi^l_I$ for $\lambda_1$. Each of the last
equations has the same form as for LRS models with $k=1$, and their
solution is $\tilde\phi^i_I=\tilde{c}\delta^i_I$ with an arbitrary
$\tilde{c}$.  In this case the conjugate momenta can be written as
$\tilde{p}^I_i=\tilde{p}\delta^I_i$, and we have the symplectic
structure $\{\tilde{c},\tilde{p}\}=\frac{8\pi}{3}G\gamma V_0$.

Thus, in both cases there is a unique non-trivial sector, and no
topological charge appears. The symplectic structure can again be made
independent of $V_0$ by redefining $a:=V_0^{1/3}\tilde{a}$,
$b:=V_0^{1/3}\tilde{b}$, $c:=V_0^{1/3}\tilde{c}$ and
$p_a:=V_0^{2/3}\tilde{p}_a$, $p_b:=V_0^{2/3}\tilde{p}_b$,
$p_c:=V_0^{2/3}\tilde{p}_c$, $p:=V_0^{2/3}\tilde{p}$. If one computes
the isotropic reduction of a Bianchi IX metric following from the
left-invariant 1-forms of SU(2), one obtains a closed
Friedmann--Robertson--Walker metric with scale factor
$a=2\tilde{a}=2\sqrt{|\tilde{p}|}$ (see, e.g., \cite{CosmoI} for the
calculation). Thus, we obtain the identification (\ref{paiso}) used in
isotropic loop cosmology. (Such a normalization can only be obtained
in curved models.)

\subsection{Spherical symmetry}
\label{s:SphSymm}

In the generic case (i.e., outside a symmetry center) of spherical
symmetry we have $S=\mathrm{SU}(2)$,
$F=\mathrm{U}(1)=\exp\langle\tau_3\rangle$ ($\langle\cdot\rangle$
denotes the linear span), and the connection form can be gauged to be
\begin{equation}\label{ASF}
 A_{S/F}=(\Lambda(\tau_2)\sin\vartheta+\Lambda(\tau_3)\cos\vartheta)
  \mathrm{d}\varphi+\Lambda(\tau_1)\mathrm{d}\vartheta\,.
\end{equation}
Here $(\vartheta,\varphi)$ are (local) coordinates on $S/F\cong S^2$
and as usually we use the basis elements $\tau_i$ of ${\cal L} S$.
$\Lambda(\tau_3)$ is given by $\mathrm{d}\lambda$, whereas
$\Lambda(\tau_{1,2})$ are the scalar field components.
Eq.\ (\ref{ASF}) contains as special cases the invariant connections
found in Ref.\ \cite{Cordero}.  These are gauge equivalent by gauge
transformations depending on the angular coordinates
$(\vartheta,\varphi)$, i.e.\ they correspond to homomorphisms
$\lambda$ which are not constant on the orbits of the symmetry group.

In order to specify the general form (\ref{ASF}) further, the first
step is again to find all conjugacy classes of homomorphisms
$\lambda\colon F=\mathrm{U}(1)\to \mathrm{SU}(2)=G$.  To do so we can
make use of Eq.\ (\ref{Hom}) to which end we need the following
information about SU(2) (see, e.g., Ref.\ \cite{BroeckerDieck}): The
standard maximal torus of SU(2) is given by
\[  
T(\mathrm{SU}(2))=\{\mathrm{diag}(z,z^{-1}):\, z\in
\mathrm{U}(1)\}\cong \mathrm{U}(1)
\]
and the Weyl group of SU(2) is the permutation group of two elements,
$W(\mathrm{SU}(2))\cong S_2$, its generator acting on
$T(\mathrm{SU}(2))$ by
$\mathrm{diag}(z,z^{-1})\mapsto\mathrm{diag}(z^{-1},z)$.

All homomorphisms in $\mathrm{Hom}(\mathrm{U}(1),T(\mathrm{SU}(2)))$
are given by
\[ 
  \lambda_k\colon z\mapsto\mathrm{diag}(z^k,z^{-k})
\]
for any $k\in{\mathbb Z}$, and we have to divide out the action of the
Weyl group leaving only the maps $\lambda_k$, $k\in{\mathbb N}_0$, as
representatives of all conjugacy classes of homomorphisms. We see that
spherically symmetric gravity has a topological charge taking values
in ${\mathbb N}_0$ (but only if degenerate configurations are allowed,
as we will see below).

We will represent $F$ as the subgroup
$\exp\langle\tau_3\rangle<\mathrm{SU}(2)$ of the symmetry group $S$,
and use the homomorphisms $\lambda_k\colon\exp t\tau_3\mapsto\exp
kt\tau_3$ out of each conjugacy class. This leads to a reduced
structure group $Z_G(\lambda_k(F))= \exp\langle\tau_3\rangle\cong
\mathrm{U}(1)$ for $k\not=0$ and $Z_G(\lambda_0(F))=\mathrm{SU}(2)$
($k=0$; this is the sector of manifestly invariant connections of
Ref.\ \cite{CorderoTeit}). The map $\Lambda|_{{\cal L} F}$ is given by
$\mathrm{d}\lambda_k\colon \langle\tau_3\rangle\to {\cal L} G, \tau_3\mapsto
k\tau_3$, and the remaining components of $\Lambda$, which give us the
scalar field, are determined by $\Lambda(\tau_{1,\,2})\in{\cal L} G$
subject to Eq.\ (\ref{Higgs}) which here can be written as
\[
\Lambda\circ\mathrm{ad}_{\tau_3}=\mathrm{ad}_{\mathrm{d}\lambda(\tau_3)}
\circ\Lambda\,.
\]
Using $\mathrm{ad}_{\tau_3}\tau_1=\tau_2$ and
$\mathrm{ad}_{\tau_3}\tau_2=-\tau_1$ we obtain
\[   
 \Lambda(a_0\tau_2-b_0\tau_1)=k(a_0[\tau_3,\Lambda(\tau_1)]+
  b_0[\tau_3,\Lambda(\tau_2)])\,,
\]
where $a_0\tau_1+b_0\tau_2$, $a_0,b_0\in{\mathbb R}$ is an arbitrary
element of ${\cal L} F_{\perp}$. Since $a_0$ and $b_0$ are arbitrary,
this is equivalent to the two equations
\[  
    k[\tau_3,\Lambda(\tau_1)]=\Lambda(\tau_2)\qquad\mbox{and}\qquad
    k[\tau_3,\Lambda(\tau_2)]=-\Lambda(\tau_1)\,.
\]
A general ansatz
\[   
  \Lambda(\tau_1)=a_1\tau_1+b_1\tau_2+c_1\tau_3\:,\quad
     \Lambda(\tau_2)=a_2\tau_1+b_2\tau_2+c_2\tau_3
\]
with arbitrary parameters $a_i,b_i,c_i\in{\mathbb R}$ yields
\begin{eqnarray*}
  k(a_1\tau_2-b_1\tau_1)&=& a_2\tau_1+b_2\tau_2+c_2\tau_3\,, \\
     k(-a_2\tau_2+b_2\tau_1)& =& a_1\tau_1+b_1\tau_2+c_1\tau_3
\end{eqnarray*}
which have non-trivial solutions only if $k=1$, namely 
\[
  b_2=a_1\:,\quad a_2 =-b_1\quad\mbox{ and }\quad  c_1=c_2=0\,.
\]

The configuration variables of the system are the above fields
$a,b,c\,\colon B\to{\mathbb R}$ of the $\mathrm{U}(1)$-connection form
$A=c(x)\,\tau_3\,\mathrm{d} x$ on the one hand and the two scalar field
components
\begin{eqnarray*}
 &&\Lambda|_{\langle\tau_1\rangle}\colon B \to {\cal L}\mathrm{SU}(2)
  ,x \mapsto a(x)\tau_1+b(x)\tau_2\\
 &&\qquad\qquad = \frac{1}{2}\left(
  \begin{array}{cc} 0 & -b(x)-i a(x)\\
  b(x)-i a(x) & 0\end{array}\right)
 =: \left(\begin{array}{cc} 0 & -\overline{w}(x)\\w(x) & 0
  \end{array}\right)
\end{eqnarray*}
on the other hand.  Under a local U(1)-gauge transformation
$z(x)=\exp (t(x)\tau_3)$ they transform as $c\mapsto c+\mathrm{d}
  t/\mathrm{d} x$ and $w(x)\mapsto\exp(-i t)w$ which can be read off from
\begin{eqnarray*} 
 A& \mapsto &z^{-1}Az+z^{-1}\mathrm{d} z=A+\tau_3\mathrm{d} t\,,\\ 
  \Lambda(\tau_1)&\mapsto & z^{-1}\Lambda(\tau_1)z=\left(
    \begin{array}{cc} 0 & -\exp(i t)\overline{w}\\ \exp(-i t)w & 0
    \end{array}\right)\,.
\end{eqnarray*}

In order to obtain a standard symplectic structure (see Eq.\ 
(\ref{sympl}) below), we reconstruct the general invariant connection
form
\begin{eqnarray} \label{InvConn}
 A(x,\vartheta,\varphi) & = & A_1(x)\tau_3\mathrm{d} x 
  +(A_2(x)\tau_1+A_3(x)\tau_2)\mathrm{d}\vartheta\\
  & & +(A_2(x)\tau_2-A_3(x)\tau_1)\sin\vartheta\mathrm{d}\varphi+
  \cos\vartheta\,\mathrm{d}\varphi\, \tau_3\nonumber\,.
\end{eqnarray}
An invariant densitized triad field is analogously given by
\begin{equation}\label{InvDrei}
 (E^x,E^{\vartheta},E^{\varphi})=(E^1\sin\vartheta\,\tau_3,
 {\textstyle\frac{1}{2}}  \sin\vartheta
 (E^2\tau_1+E^3\tau_2), {\textstyle\frac{1}{2}} (E^2\tau_2-E^3\tau_1))
\end{equation}
with coefficients $E^I$ canonically conjugate to $A_I$ ($E^2$ and
$E^3$ are non-vanishing only for $k=1$). The symplectic structure
\begin{equation}\label{sympl}
  \{A_I(x),E^J(y)\}=2\gamma G\delta^J_I\delta(x,y)
\end{equation}
can be derived by inserting the invariant expressions into
$(8\pi\gamma G)^{-1}\int_{\Sigma}\mathrm{d}^3x \dot{A}_a^i E_i^a$.

Information about the topological charge $k$ can be found
by expressing the volume in terms of the reduced triad coefficients
$E^I$: using
\begin{eqnarray}\label{DetE}
  \epsilon_{abc}\epsilon^{ijk}E_i^aE_j^bE_k^c & = &
  -2\epsilon_{abc}\mathrm{tr}(E^a[E^b,E^c])\\
  & = & \frac{3}{2} \sin^2\vartheta E^1\left((E^2)^2+(E^3)^2\right)\nonumber
\end{eqnarray}
we have
\begin{equation}\label{SphSymmVol}
 V=\int_{\Sigma}\mathrm{d}^3x \sqrt{{\textstyle\frac{1}{6}}
   \left|\epsilon_{abc}\epsilon^{ijk} E_a^iE_b^jE_c^k\right|}=
 2\pi \int_B\mathrm{d} x \sqrt{|E^1|((E^2)^2+(E^3)^2)}\,.
\end{equation}
We can now see that in all the sectors with $k\not=1$ the volume
vanishes because then $E^2=E^3=0$. All these degenerate sectors have
to be rejected on physical grounds and we arrive at a unique sector of
invariant connections given by the parameter $k=1$.

\end{appendix}


\end{document}